\documentclass[a4paper,11pt]{article}
\pdfoutput=1 % if your are submitting a pdflatex (i.e. if you have
\usepackage{jheppub} 
\usepackage{amsmath,amssymb,graphicx}
\allowdisplaybreaks[4]
\usepackage{tikz-cd}
\usetikzlibrary{arrows.meta,calc}
\usepackage{physics}
\usepackage{booktabs}
\usepackage{rotating}
\usepackage[T1]{fontenc} % if needed

\newcommand{\Ish}[1]{|#1\rangle\!\rangle}

\title{\boldmath Classification of Conformal Ising/Tricritical Ising Interfaces Preserving $\mathcal{W}_3^{\mathbb{Z}_2}$ Symmetry}

\author[a,b]{Bufan Zheng}

\affiliation[a]{Department of Physics, The University of Tokyo, Tokyo 113-0033, Japan}
\affiliation[b]{Kavli Institute for the Physics and Mathematics of the Universe (Kavli IPMU), UTIAS, The University of Tokyo, Kashiwa, Chiba 277-8583, Japan}

\emailAdd{bufan.zheng@ipmu.jp}

\abstract{
	We classify conformal interfaces between the Ising and
	tricritical Ising models that preserve the chiral algebra
	$\mathcal{W}_3^{\mathbb{Z}_2}$.
	Using the folding trick, we formulate the problem in terms
	of boundary states satisfying the Cardy condition and
	determine the unique NIM-rep compatible with the bulk
	partition function. We obtain a complete set of 24
	fundamental boundary states, of which 18 correspond to
	factorized interfaces preserving the full product Virasoro
	symmetry, while the remaining six break this symmetry.
	We construct these six interfaces explicitly and determine
	their annulus partition functions. They are related by
	topological defect line dressings, and one coincides with
	the RG domain wall proposed by Gaiotto.
}

\begin{document} 
\maketitle
\flushbottom

\section{Introduction}
\label{sec:intro}
Conformal interfaces in two dimensional conformal field theories have attracted considerable attention for their roles in a variety of contexts~\cite{Andrei:2018die}.  For example, recent work has explored their role in renormalization group flows~\cite{Gaiotto:2012np,Chen:2026byx}. A particular class of conformal interfaces arising in this context is known as RG domain walls. Moreover, the classification of conformal interfaces in conformal field theories has become a subject of considerable interest.

By employing the folding trick, the classification of conformal interfaces can be reformulated as the classification of conformal boundaries in the product of the two CFTs. Although the classification of conformal boundaries in rational conformal field theories (RCFTs) is well established~\cite{Cardy:1989ir,Behrend:1999bn}, the problem remains challenging even when both CFTs are Virasoro minimal models. In this case, factorized boundary states preserving the separate Virasoro algebras of the two factors can be constructed directly as tensor products of boundary states in the individual minimal models. These correspond to totally reflecting conformal interfaces. On the other hand, totally transmissive interfaces, i.e. topological defect lines, can be classified within the framework developed in~\cite{Frohlich:2006ch}.  However, more general boundary states preserve only the diagonal Virasoro algebra. Since the product theory is generally irrational with respect to this subalgebra, their classification becomes considerably more difficult.

Conformal interfaces in the Ising model were completely classified using an orbifold construction~\cite{Oshikawa:1996dj, Oshikawa:1996ww,Kormos:2009sk}. Nevertheless, even for the tricritical Ising model (TIM), which appears to be the next simplest case, a complete classification is still lacking, although partial results have been obtained in~\cite{Gang:2008sz, Makabe:2017ygy}. The RG domain wall between the Ising and tricritical Ising models was proposed in~\cite{Gaiotto:2012np} and has recently received numerical support in~\cite{Cogburn:2023xzw,Antunes:2026kqb}. In particular, the authors of~\cite{Antunes:2026kqb} pointed out the presence of an additional spin-three current at the interface, referred to as a ``phantom current'', suggesting a possible enhancement to $\mathcal{W}_3$ symmetry. Motivated by these observations, we aim to classify conformal
interfaces that break the full $\mathsf{Vir}_{c=1/2}\otimes\mathsf{Vir}_{c=7/10}$ chiral symmetry while preserving a smaller subalgebra, since the general classification of interfaces preserving only the diagonal Virasoro symmetry is considerably more difficult.

However, the phantom current is not local in the bulk theory, and $\mathcal{W}_3(c=6/5)$ is not a subalgebra of $\mathsf{Vir}_{c=1/2}\otimes\mathsf{Vir}_{c=7/10}$. Therefore, the preservation of $\mathcal{W}_3$ symmetry cannot be directly imposed through the standard gluing conditions. Instead, in this work, we identify the orbifold algebra $\mathcal{W}_3^{\mathbb{Z}_2}(c=6/5)$ as an appropriate chiral subalgebra and systematically classify all conformal interfaces preserving this symmetry. Interestingly, although our classification is based on $\mathcal{W}_3^{\mathbb{Z}_2}$ rather than $\mathcal{W}_3$, we find that the self-annulus partition functions of all six symmetry-breaking boundary states can be expressed in terms of $\mathcal{W}_3$ characters. This may indicate the presence of additional phantom currents at these boundaries, even though such currents are nonlocal in the bulk.

From the perspective of the folding trick, the idea of classifying symmetry breaking conformal boundaries can be traced back to~\cite{Fuchs:1998xx,Fuchs:1999zi,Fuchs:1999xn,Quella:2002ct}. In section~\ref{sec:2}, we first review some known results from the vertex operator algebra (VOA) literature that will be needed below. In section~\ref{sec:3}, we apply the folding trick and classify the resulting boundary states within the framework of~\cite{Petkova:2000dv}. We discuss their relations and identify one of them with the RG domain wall. However, this classification requires solving a complicated non-negative integer matrix representation (NIM-rep) problem. Its details are too technical to be included in the main text, so we present them in the appendix.

\section{Chiral algebras and modular data}
\label{sec:2}
\subsection{Chiral algebra extensions and fixed-point subalgebras}
\label{sec:2.1}
The Ising and TIM models correspond to the diagonal Virasoro minimal models $\mathcal{M}(3,4)$ and $\mathcal{M}(4,5)$ respectively. Their primaries are listed in Table~\ref{tab:Aprimaries}.

\begin{table}[htbp]
	\centering
	\renewcommand{\arraystretch}{1.3}
	\setlength{\tabcolsep}{4pt}
	\begin{tabular}{@{}c*{3}{c}@{}}
		\toprule
		Ising & $1$ & $\sigma$ & $\epsilon$ \\
		\midrule
		$h$ & $0$ & $\frac{1}{16}$ & $\frac{1}{2}$ \\
		$(r,s)$ & $(1,1)$ & $(1,2)$ & $(1,3)$ \\
		\bottomrule
	\end{tabular}%
	\hspace{16pt}%
	\begin{tabular}{@{}c*{6}{c}@{}}
		\toprule
		TIM & $1$ & $\sigma$ & $\epsilon$
		& $\sigma'$ & $\epsilon'$ & $\epsilon''$ \\
		\midrule
		$h$ & $0$ & $\frac{3}{80}$ & $\frac{1}{10}$
		& $\frac{7}{16}$ & $\frac{3}{5}$ & $\frac{3}{2}$ \\
		$(r,s)$
		& $(1,1)$ & $(2,2)$
		& $(1,2)$ & $(2,1)$
		& $(1,3)$ & $(3,1)$ \\
		\bottomrule
	\end{tabular}
	\caption{Primaries of the Ising and tricritical Ising models.}
	\label{tab:Aprimaries}
\end{table}

After folding, the bulk theory has $\mathrm{Vir}(c=\frac12)\times\mathrm{Vir}(c=\frac7{10})$ as its chiral algebra. In VOA notation, $L(c,h)$ denotes the irreducible highest weight module of the Virasoro algebra with central charge $c$ and highest weight $h$. In particular, $L(c,0)$ carries the structure of a simple Virasoro VOA~\cite{lepowsky2004introduction}. Thus, the chiral algebra of the folded theory can be written as~\cite{dong2003}
\begin{equation}
	\label{eq:Adef}
	\mathcal{A}:=L\left(\frac12,0\right)\otimes L\left(\frac7{10},0\right).
\end{equation}

It is well known that simple currents in CFT can be used to extend its chiral algebra through the simple current extension~\cite{Schellekens:1989am,Schellekens:1990xy,Fuchs:1996dd}. Any minimal model $\mathcal{M}(p,q)$ with $p > 2$ has a unique (non-trivial) simple current of conformal dimension $h_{p-1,1}=\frac{1}{4}(p-2)(q-2)$~\cite{Mathieu:2007fx}, so we can use the simple currents of Ising and TIM to construct the spin-2 current $J=\psi G$, which extends the chiral algebra of $\mathcal{M}(3,4)\times\mathcal{M}(4,5)$. $\psi$ and $G$ come from the $\mathbb{Z}_2$ twisted sectors of the Ising and TIM models and have conformal weights $\frac12$ and $\frac32$, respectively. This extension has been studied in detail in~\cite{dong2003}. Following the notation in the VOA literature~\cite{hohn2012mckay}, we denote it by
\begin{equation}
	U_{2A}:=L(\frac12,0)\otimes L(\frac{7}{10},0)\oplus L(\frac12,\frac12)\otimes L(\frac{7}{10},\frac32).
\end{equation}
By \cite[Theorem~2.1]{dong2003}, the outer automorphism group of this chiral algebra is 
\begin{equation}
	\label{eq:aut}
	\operatorname{Aut}(U_{2A})=\left\langle\sigma,\tau\mid\sigma^2=\tau^3=1,\, \sigma\tau\sigma=\tau^{-1}\right\rangle\cong S_3.
\end{equation}
For physical intuition, consider the following coset construction:
\begin{equation}
	\label{eq:coset}
	\mathcal{M}(k+1,k+2)\times\mathcal{M}(k+2,k+3)=\frac{\hat{\mathfrak{su}}(2)_{k-1}\times \hat{\mathfrak{su}}(2)_{1}}{\hat{\mathfrak{su}}(2)_{k}}\times \frac{\hat{\mathfrak{su}}(2)_{k}\times \hat{\mathfrak{su}}(2)_{1}}{\hat{\mathfrak{su}}(2)_{k+1}}.
\end{equation}
$U_{2A}$ can be realized as~\cite{Gaiotto:2012np}
\begin{equation}
	\label{eq:defB}
	\mathcal{B} = \frac{\hat{\mathfrak{su}}(2)_{1}\times \hat{\mathfrak{su}}(2)_{1}\times \hat{\mathfrak{su}}(2)_{1}}{\hat{\mathfrak{su}}(2)_{3}}\cong U_{2A}.
\end{equation}
This coset realization was also established mathematically in~\cite{lam2008faithful}. Thus, the $S_3$ symmetry of $U_{2A}$ can be easily understood as the permutation symmetry of the three $\hat{\mathfrak{su}}(2)_1$ factors in the numerator of $\mathcal{B}$. However, for general $k$, the coset in~\eqref{eq:defB} generalizes to
\begin{equation}
	\frac{\hat{\mathfrak{su}}(2)_{k-1}\times \hat{\mathfrak{su}}(2)_{1}\times \hat{\mathfrak{su}}(2)_{1}}{\hat{\mathfrak{su}}(2)_{k+1}}.
\end{equation}
Therefore, in general, the extended chiral algebra has only $\mathbb{Z}_2$ symmetry. We will see that this symmetry enhancement in $\mathcal{M}(3,4)\times\mathcal{M}(4,5)$ plays a very important role.

From the extended algebra $U_{2A}$, we can construct a fixed point subalgebra by projecting onto the subspace invariant under $g\in\operatorname{Aut}(U_{2A})\cong S_3$. First, consider the subgroup $\mathbb{Z}_3\subset \operatorname{Aut}(U_{2A})$ generated by $\tau$. It was proved in~\cite[Theorem~5.9]{dong2003} that $U_{2A}^{\mathbb{Z}_3}\cong \mathcal{W}_3(c=\frac65)$. The corresponding chiral theory is the $\mathcal{W}_3(5,6)$
minimal model. Its primary fields, listed in
Table~\ref{tab:W3primaries}, are labeled by four integers
$(r_1,s_1,r_2,s_2)$ satisfying
\begin{equation}
	r_i,s_i\geq 1,\quad r_1+r_2<5,\quad s_1+s_2<6,
\end{equation}
with conformal weights given by~\cite{Bouwknegt:1992wg}:
\begin{equation}
	h^{\mathcal W_3(5,6)}_{r_1,s_1,r_2,s_2}
	=\frac{
		3[6(r_1+r_2)-5(s_1+s_2)]^2
		+[6(r_1-r_2)-5(s_1-s_2)]^2-12
	}{360}.
\end{equation}
The labels are understood modulo the $\mathbb{Z}_3$ identification generated by:
\begin{equation}
	\begin{pmatrix}
		r_1 & s_1\\
		r_2 & s_2
	\end{pmatrix}
	\to
	\begin{pmatrix}
		5-(r_1+r_2) & 6-(s_1+s_2)\\
		r_1 & s_1
	\end{pmatrix}
\end{equation}
\begin{table}[htbp]
	\centering
	\renewcommand{\arraystretch}{1.3}
	\setlength{\tabcolsep}{2pt}
	\resizebox{\linewidth}{!}{%
		\begin{tabular}{@{}c*{4}{c}@{\hspace{9pt}}*{4}{c}@{\hspace{9pt}}*{12}{c}@{}}
			\toprule
			$\mathcal{W}_3(5,6)$
			& $1$ & $E$ & $F$ & $X$
			& $Z$ & $Z^\dagger$ & $D$ & $D^\dagger$
			& $\Phi_{\frac{2}{45}}$ & $\Phi_{\frac{2}{45}}^\dagger$
			& $\Phi_{\frac19}$ & $\Phi_{\frac19}^\dagger$
			& $\Phi_{\frac{17}{45}}$ & $\Phi_{\frac{17}{45}}^\dagger$
			& $\Phi_{\frac{32}{45}}$ & $\Phi_{\frac{32}{45}}^\dagger$
			& $\Phi_{\frac79}$ & $\Phi_{\frac79}^\dagger$
			& $\Phi_{\frac{13}{9}}$ & $\Phi_{\frac{13}{9}}^\dagger$ \\
			\midrule
			$h$
			& $0$ & $\frac1{10}$ & $\frac12$ & $\frac85$
			& $\frac35$ & $\frac35$ & $2$ & $2$
			& $\frac2{45}$ & $\frac2{45}$
			& $\frac19$ & $\frac19$
			& $\frac{17}{45}$ & $\frac{17}{45}$
			& $\frac{32}{45}$ & $\frac{32}{45}$
			& $\frac79$ & $\frac79$
			& $\frac{13}{9}$ & $\frac{13}{9}$ \\
			$w/q$
			& $0$ & $0$ & $0$ & $0$
			& $1$ & $-1$ & $6$ & $-6$
			& $\frac2{81}$ & $-\frac2{81}$
			& $\frac7{81}$ & $-\frac7{81}$
			& $\frac{11}{81}$ & $-\frac{11}{81}$
			& $\frac{88}{81}$ & $-\frac{88}{81}$
			& $\frac{119}{81}$ & $-\frac{119}{81}$
			& $\frac{187}{81}$ & $-\frac{187}{81}$ \\
			$\left(\begin{smallmatrix}r_1&s_1\\r_2&s_2\end{smallmatrix}\right)$
			& $\left(\begin{smallmatrix}1&1\\1&1\end{smallmatrix}\right)$
			& $\left(\begin{smallmatrix}2&2\\2&2\end{smallmatrix}\right)$
			& $\left(\begin{smallmatrix}1&2\\1&2\end{smallmatrix}\right)$
			& $\left(\begin{smallmatrix}2&1\\2&1\end{smallmatrix}\right)$
			& $\left(\begin{smallmatrix}2&1\\1&1\end{smallmatrix}\right)$
			& $\left(\begin{smallmatrix}1&1\\2&1\end{smallmatrix}\right)$
			& $\left(\begin{smallmatrix}3&1\\1&1\end{smallmatrix}\right)$
			& $\left(\begin{smallmatrix}1&1\\3&1\end{smallmatrix}\right)$
			& $\left(\begin{smallmatrix}2&2\\1&1\end{smallmatrix}\right)$
			& $\left(\begin{smallmatrix}1&1\\2&2\end{smallmatrix}\right)$
			& $\left(\begin{smallmatrix}1&1\\1&2\end{smallmatrix}\right)$
			& $\left(\begin{smallmatrix}1&2\\1&1\end{smallmatrix}\right)$
			& $\left(\begin{smallmatrix}1&2\\2&1\end{smallmatrix}\right)$
			& $\left(\begin{smallmatrix}2&1\\1&2\end{smallmatrix}\right)$
			& $\left(\begin{smallmatrix}2&1\\2&2\end{smallmatrix}\right)$
			& $\left(\begin{smallmatrix}2&2\\2&1\end{smallmatrix}\right)$
			& $\left(\begin{smallmatrix}1&1\\1&3\end{smallmatrix}\right)$
			& $\left(\begin{smallmatrix}1&3\\1&1\end{smallmatrix}\right)$
			& $\left(\begin{smallmatrix}1&2\\1&3\end{smallmatrix}\right)$
			& $\left(\begin{smallmatrix}1&3\\1&2\end{smallmatrix}\right)$ \\
			\bottomrule
		\end{tabular}%
	}
	\caption{Primaries of $\mathcal{W}_3(5,6)$,
		where $q=\sqrt{\frac{2}{35}}$.}
	\label{tab:W3primaries}
\end{table}

The $\mathcal{W}_3$ algebra is generated by the total stress tensor $T=T_{\text{Ising}}+T_{\text{TIM}}$ and an additional spin-3 current $W$, satisfying
\begin{equation}
	\label{eq:W3OPE}
	\begin{aligned}
		T(z)W(w)\sim{}&
		\frac{3W(w)}{(z-w)^2}
		+\frac{\partial W(w)}{z-w},
		\\[4pt]
		W(z)W(w)\sim{}&
		\frac{2/5}{(z-w)^6}
		+\frac{2T(w)}{(z-w)^4}
		+\frac{\partial T(w)}{(z-w)^3}
		\\
		&+\frac{
			\frac87\Lambda(w)
			+\frac3{10}\partial^2T(w)
		}{(z-w)^2}
		+\frac{
			\frac47\partial\Lambda(w)
			+\frac1{15}\partial^3T(w)
		}{z-w}.
	\end{aligned}
\end{equation}
Here,
\begin{equation}
	\Lambda(z)=:T(z)T(z):
	-\frac3{10}\partial^2T(z).
\end{equation}
The additional spin-3 current is expressed as
\begin{equation}
	\label{eq: Wdef}
	W(z)\propto\psi(z)\partial G(z)-3\partial\psi(z)G(z)=\left(L^{\text{TIM}}_{-1}-3L_{-1}^{\text{Ising}}\right)J.
\end{equation}

After gauging $\mathbb{Z}_3$, a $\mathbb{Z}_2$ symmetry generated by $\sigma$ still remains. It changes the sign of $J$, $\sigma J=-J$. It can be regarded as the $\mathbb{Z}_2$ charge conjugation symmetry of the $\mathcal{W}_3$ algebra. We can further gauge this $\mathbb{Z}_2$ symmetry to obtain another subalgebra $\mathcal{W}_3^{\mathbb{Z}_2}(c=\frac65)$, which we denote by $\mathcal{U}$. Since $\mathbb{Z}_3$ is a normal subgroup of $S_3$, $\mathcal{U}$ is just $(U_{2A}^{\mathbb{Z}_3})^{\mathbb{Z}_2}\cong U_{2A}^{S_3}$. 

On the other hand, we can first gauge the subgroup $\left\langle{\sigma}\right\rangle\cong \mathbb{Z}_2$. The simple current $J$ is then projected out, so we recover the original $\mathcal{A}$ in~\eqref{eq:Adef}. This can also be understood from the fact that the simple current $J$ satisfies the fusion rule $[J]\times[J]=[1]$, and thus generates the simple current group $\mathbb{Z}_2$. Thus, recovering $\mathcal{A}$ amounts to gauging the dual quantum $\mathbb Z_2$ symmetry~\cite{Vafa:1989ih,Bhardwaj:2017xup}. In this case, since $\mathbb{Z}_2$ is not a normal subgroup of $S_3$, there is no additional $\mathbb{Z}_3$ symmetry to gauge further. From~\eqref{eq: Wdef}, we know that $W(z)$ is projected out when gauging $\mathbb{Z}_2$ in $U_{2A}$ to obtain $\mathcal{A}$, since $\sigma W=-W$. Therefore, $\mathcal{W}_3$ is not a subalgebra of $\mathcal{A}$, so we cannot consider boundaries of $\mathcal{M}(3,4)\times \mathcal{M}(4,5)$ breaking $\mathcal{A}$ but preserving $\mathcal{W}_3$ in the bulk. 

However, since obtaining $\mathcal{U}$ from $U_{2A}$ requires gauging an additional $\mathbb{Z}_3$ symmetry compared to the construction of $\mathcal{A}$, the former can be regarded as a subalgebra of the latter. More interestingly, at the central charge $c=\frac{6}{5}$ considered here, this subalgebra is isomorphic to the $\mathcal{W}$-algebra $\mathcal{W}_{c=6/5}(2,6,8,10)$, generated by four currents of conformal weights $2$, $6$, $8$, and $10$~\cite{Blumenhagen:1994wg,milas2020permutation,Al-Ali:2016ndc}. Exploiting this identification, we now briefly argue that the outer automorphism group of $\mathcal{U}$ is trivial.

Following the argument in~\cite{Honecker:1992kz}, consider an automorphism
$g \in \operatorname{Aut}(\mathcal{U})$.
Since $g$ leaves the stress-energy tensor invariant, it must preserve
both the conformal spin and the Virasoro primary nature of the currents.
In the theory under consideration, the generating currents of the
chiral algebra have distinct spins. Consequently,
\begin{equation}
	g(W^{(s)}) = \lambda_s W^{(s)}, \quad \forall~W^{(s)}\in\mathcal{U}\cong \mathcal{W}_{c=6/5}(2,6,8,10),
\end{equation}
where $s$ denotes the spin and $\lambda_s \in \mathbb{C}$.
Using the procedure and normalization provided in~\cite{Blumenhagen:1994wg}, we find that at $c=6/5$, the following two-point functions are nonzero,
\begin{equation}
\begin{gathered}
	d_{6,6}=\frac{153292161318912}{625}\neq0,\quad d_{8,8}=\frac{429020048407233042579456}{625}\neq0,\\ d_{10,10}=\frac{8266133068627718675463285068934414336}{125}\neq0.
\end{gathered}
\end{equation}
Here $d_{s_1,s_2}:=\langle W^{(s_1)}(1)W^{(s_2)}(0)\rangle$. This implies $\lambda_s^2=1$, so that $\lambda_s=\pm 1$. Furthermore, the following OPE coefficients will be useful,
\begin{equation}
(C_{6,6}^6)^2=\frac{77775392}{2386545}\neq0,\quad (C_{6,6}^8)^2=\frac{1381024736}{3501915}\neq0, \quad (C_{6,6}^{10})^2=\frac{6387239404}{1780921315}\neq 0,
\end{equation}
which yield the following constraints on the coefficients $\lambda_s$:
\begin{equation}
	\lambda_6^3=\lambda_8\lambda_6^2=\lambda_{10}\lambda_6^2=1\Rightarrow \lambda_{6,8,10}=1.
\end{equation}
Since any VOA automorphism fixes the Virasoro subalgebra, we also have $\lambda_2=1$. Hence
\begin{equation}
	\operatorname{Aut}\left(\mathcal{U}\right)=1.
\end{equation}

We have established that $\mathcal{U}$ is a subalgebra of $\mathcal{A}$. Since its automorphism group is trivial, no smaller subalgebra can be obtained by gauging a nontrivial automorphism of $\mathcal{U}$. However, this does not rule out the existence of intermediate subalgebras between $\mathsf{Vir}_{c=6/5}^{\mathrm{diag}}$ and $\mathcal{U}$. A complete classification of the subalgebras of $\mathcal{U}$ is a difficult problem, which we shall not pursue here. In what follows, we therefore restrict our attention to conformal boundary states preserving $\mathcal{U}$.

On the other hand, the quantum Galois correspondence proved in~\cite[Theorem~1]{hanaki1997} allows us to show that there is no proper intermediate subalgebra between $\mathcal{U}$ and $\mathcal{A}$. Indeed, using the realizations
\begin{equation}
	\label{eq:no_inter}
	U_{2A}^{S_3}\cong \mathcal{U}
	\subsetneq 
	\mathcal{A}\cong U_{2A}^{\mathbb{Z}_2},
\end{equation}
the quantum Galois correspondence establishes a bijection between the intermediate subalgebras of $U_{2A}^{S_3}\subseteq U_{2A}$ and the subgroups of $S_3$. In particular, the existence of a proper intermediate subalgebra $\mathcal{V}$ satisfying $\mathcal{U}\subsetneq\mathcal{V}\subsetneq\mathcal{A}$ would imply the existence of a subgroup $K$ satisfying $\mathbb{Z}_2\subsetneq K\subsetneq S_3.$ However, $\mathbb{Z}_2$ is a maximal proper subgroup of $S_3$, so no such subgroup exists. Consequently, there is no proper intermediate subalgebra between $\mathcal{U}$ and $\mathcal{A}$.

The relations among the chiral algebras discussed in this section are summarized as follows:
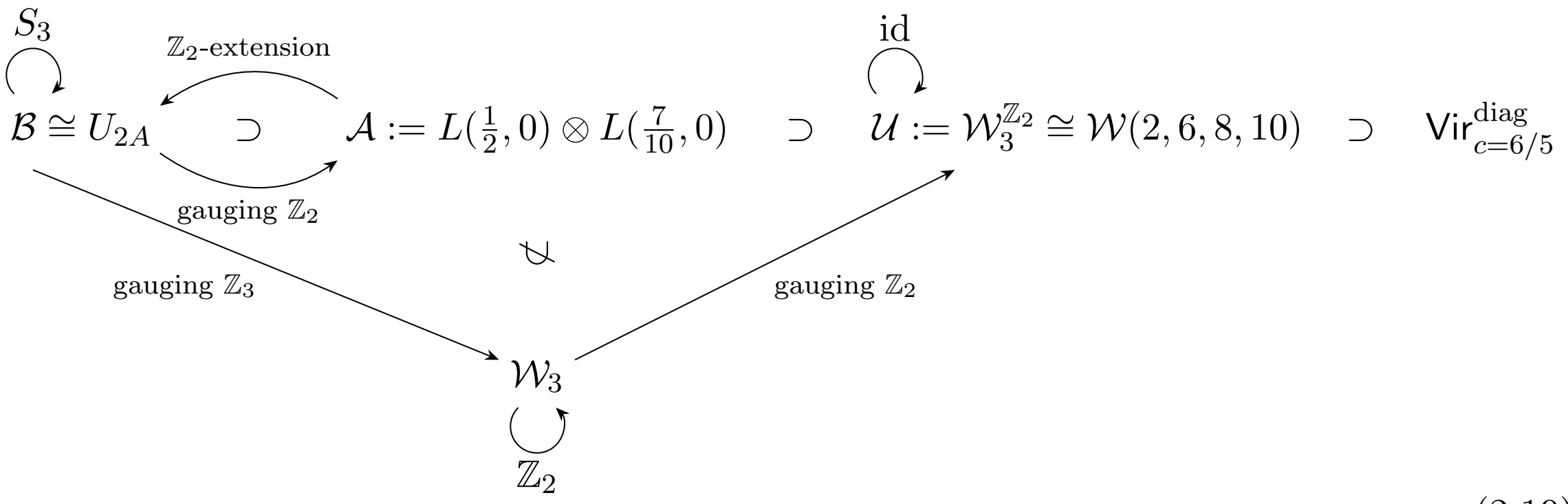
\begin{equation}
	\resizebox{\textwidth}{!}{%
	\begin{tikzpicture}[
		x=1em, y=1em,
		baseline=(current bounding box.center),
		>={Stealth[length=4pt,width=3.2pt]},
		every node/.style={inner sep=2pt},
		lab/.style={font=\scriptsize,inner sep=1.5pt}
		]
		\def\dleft{12}
		\def\dright{11}
		\def\dy{6.5}
		
		\node[anchor=base] (B) at (-\dleft,0)
		{$\mathcal B\cong U_{2A}$};
		
		\node[anchor=base] (A) at (0,0)
		{$\mathcal A:=
			L(\frac12,0)\otimes
			L(\frac{7}{10},0)$};
		
		\node[anchor=base] (U) at (\dright,0)
		{$\mathcal U:=\mathcal W_3^{\mathbb Z_2}$};
		
		\node[anchor=base] (W) at (0,-\dy)
		{$\mathcal W_3$};
		
		\node[anchor=base west,inner sep=0pt] (Utext)
		at (U.base east)
		{$\cong\mathcal W(2,6,8,10)$};
		
		\node[anchor=base west] (V)
		at ($(Utext.base east)+(3,0)$)
		{$\mathsf{Vir}_{c=6/5}^{\mathrm{diag}}$};
		
		\node[anchor=base]
		at ($(B.base east)!.5!(A.base west)$)
		{$\supset$};
		
		\node[anchor=base]
		at ($(A.base east)!.5!(U.base west)$)
		{$\supset$};
		
		\node[anchor=base]
		at ($(Utext.base east)!.5!(V.base west)$)
		{$\supset$};
		
		\node at ($(A.center)!.5!(W.center)$)
		{\rotatebox{90}{$\not\subset$}};
		
		\draw[->] (A.north west)
		to[out=145,in=35]
		node[lab,above=2pt]
		{$\mathbb Z_2\text{-extension}$}
		(B.north east);
		
		\draw[->] (B.south east)
		to[out=-35,in=-145]
		node[lab,below=2pt]
		{$\text{gauging }\mathbb Z_2$}
		(A.south west);
		
		\draw[->]
		($(B.base west)+(.8,-.8)$) --
		node[lab,midway,below left=1pt]
		{$\text{gauging }\mathbb Z_3$}
		($(W.base)+(-1,.7)$);
		
		\draw[->]
		($(W.base)+(1,.7)$) --
		node[lab,midway,below right=1pt]
		{$\text{gauging }\mathbb Z_2$}
		($(U.base)+(0,-.8)$);
		
		\coordinate (SB) at ($(B.base west)+(.8,1.7)$);
		\coordinate (SU) at ($(U.base west)+(.8,1.7)$);
		
		\foreach \P/\L in {SB/{S_3},SU/{\mathrm{id}}}{
			\draw[->]
			([shift={(225:.7em)}]\P)
			arc[
			start angle=225,
			end angle=-45,
			radius=.7em
			];
			\node[above=2pt,inner sep=0pt]
			at ($(\P)+(0,.7)$) {$\L$};
		}
		
		\coordinate (SW) at ($(W.south)+(0,-.7)$);
		
		\draw[->]
		([shift={(135:.7em)}]SW)
		arc[
		start angle=135,
		end angle=405,
		radius=.7em
		];
		
		\node[below=2pt,inner sep=0pt]
		at ($(SW)+(0,-.7)$) {$\mathbb Z_2$};
		
	\end{tikzpicture}
}
\end{equation}
We finally remark that~\eqref{eq:aut} can be obtained by embedding $U_{2A}$ into a larger chiral algebra $V_{\sqrt{2}A_2}$~\cite{dong1996algebraic,kitazume2000ternary}, which is the lattice VOA associated with $\sqrt{2}$ times the standard root lattice of type $A_2$:
\begin{equation}
	\alpha_1=(\sqrt{2},0),\quad \alpha_2=(-\frac{\sqrt{2}}{2},\sqrt{\frac{3}{2}}),\quad \alpha_0=-(\alpha_1+\alpha_2).
\end{equation}
In the physics language, it corresponds to the chiral theory of two compact free bosons $\boldsymbol{\phi}:=(\phi^1, \phi^2)$. Primary fields correspond to vertex operators $V_p:=\mathrm{e}^{ip\cdot\boldsymbol{\phi}}$, with momentum $p$ taking values in the dual lattice of $\sqrt{2}A_2$:
\begin{equation}
	\label{eq:plattice}
	p=
	\frac{(2m_1+m_2)\alpha_1+(m_1+2m_2)\alpha_2}
	{3\sqrt2},\quad m_1,m_2\in\mathbb{Z}.
\end{equation}
The two $\hat{\mathfrak{u}}(1)$ currents are $j^a=\mathrm{i}\partial \phi^a$. We define
\begin{equation}
	w(\alpha_i):=\frac{1}{2}:\left(\alpha_i\cdot \boldsymbol{j}\right)^2:-\left(:\mathrm{e}^{\mathrm{i}\sqrt{2}\alpha_i \cdot \boldsymbol{\phi}}:+:\mathrm{e}^{-\mathrm{i}\sqrt{2}\alpha_i \cdot \boldsymbol{\phi}}:\right).
\end{equation}
The embedding of $U_{2A}$ into $V_{\sqrt{2}A_2}$ is given by~\cite{dong2003}
\begin{equation}
	\label{eq:embbeding}
		T_{\mathrm{Ising}}
		=\frac14 w(\alpha_1),\quad
		T_{\mathrm{TIM}}
		=\frac15\bigl(w(\alpha_0)+w(\alpha_2)\bigr)
		-\frac1{20}w(\alpha_1),\quad
		J
		=6\bigl(w(\alpha_2)-w(\alpha_0)\bigr).
\end{equation}
The spin-3 current of the $\mathcal{W}_3$ algebra is
\begin{equation}
	W(z)=\frac{\mathrm{i}}{\sqrt{210}}
	\oint_z\frac{d\zeta}{2\pi\mathrm{i}}
	\left(
	[w(\alpha_1)](\zeta)[w(\alpha_2)](z)
	-[w(\alpha_2)](\zeta)[w(\alpha_1)](z)
	\right).
\end{equation}
These fields satisfy the following OPE relations, together with those in~\eqref{eq:W3OPE}:
\begin{align}
	\label{eq:TTOPE1}
	T_{\mathrm{Ising}}(z)T_{\mathrm{Ising}}(w)
	&\sim\frac{1}{4(z-w)^4}
	+\frac{2T_{\mathrm{Ising}}(w)}{(z-w)^2}
	+\frac{\partial T_{\mathrm{Ising}}(w)}{z-w},\\
	\label{eq:TTOPE2}
	T_{\mathrm{TIM}}(z)T_{\mathrm{TIM}}(w)
	&\sim\frac{7}{20(z-w)^4}
	+\frac{2T_{\mathrm{TIM}}(w)}{(z-w)^2}
	+\frac{\partial T_{\mathrm{TIM}}(w)}{z-w},\\
	\label{eq:TJOPE1}
	T_{\mathrm{Ising}}(z)J(w)
	&\sim\frac{J(w)}{2(z-w)^2}
	+\frac{\frac14\partial J(w)-\frac{3i\sqrt{210}}2W(w)}{z-w},\\
	\label{eq:TJOPE2}
	T_{\mathrm{TIM}}(z)J(w)
	&\sim\frac{3J(w)}{2(z-w)^2}
	+\frac{\frac34\partial J(w)+\frac{3i\sqrt{210}}2W(w)}{z-w},
\end{align}
\begin{equation}
	\label{eq:JJOPE}
	J(z)J(w)\sim
	\frac{252}{(z-w)^4}
	+\frac{504T_{\mathrm{Ising}}(w)+1080T_{\mathrm{TIM}}(w)}{(z-w)^2}
	+\frac{252\partial T_{\mathrm{Ising}}(w)+540\partial T_{\mathrm{TIM}}(w)}{z-w}.
\end{equation}
The $S_3$ symmetry can be identified as the Weyl group of $A_2$ generated by
\begin{equation}
	\tau:\alpha_1\to\alpha_2\to\alpha_0\to\alpha_1,\quad\rho:\alpha_1\leftrightarrow\alpha_2.
\end{equation}
Note that in~\eqref{eq:aut} we chose $\sigma=\rho\tau$ as the generator of the $\mathbb{Z}_2$ subgroup. We carefully chose the normalization in~\eqref{eq:embbeding} so that the transformation under $\rho$ agrees with that found in~\cite{Gaiotto:2012np}.\footnote{More precisely, we should use $\theta\rho$, with $\theta:\alpha_i\to-\alpha_i$. We will discuss this subtlety in section~\ref{sec:3.4}.} It is straightforward to check that $W(z)$ is invariant under $\tau$ but changes sign under $\rho$.

\subsection{Characters and modular properties of $\mathcal{W}_3(5,6)$}
\label{sec:2.2}
To study $\mathcal{U}$ in more detail, we first review some results for its parent algebra $\mathcal{W}_3(5,6)$. We begin by defining the twisted $\mathcal{W}_3$ characters as
\begin{equation}
	\hat{\chi}_{h}^{(g_1,g_2)}=\operatorname{Tr}_{\mathcal{H}^{g_1}_h} g_2 q^{L_0-\frac{c}{24}},\quad g_1,g_2\in\mathbb{Z}_2=\{1,\sigma\}.
\end{equation}
Here, a nontrivial $g_1$ corresponds to inserting a charge conjugation $\mathbb{Z}_2$ defect line along the time direction, so that the Hilbert space is in the corresponding twisted sector. Similarly, a nontrivial $g_2$ corresponds to inserting the $\mathbb{Z}_2$ defect line along the spatial direction. We will discuss their definitions and modular properties in detail, one case at a time.

For $g_1=g_2=1$, $\hat{\chi}^{(1,1)}$ is the ordinary $\mathcal{W}_3(5,6)$ character, whose expansion in $q=\mathrm{e}^{2\pi \mathrm{i}\tau}$ is given by~\cite{Bouwknegt:1992wg}
\begin{equation}
	\hat{\chi}_{\boldsymbol r,\boldsymbol s}^{(1,1)}(\tau)
	=\frac{1}{\eta(\tau)^2}
	\sum_{\substack{w\in W(A_2)\\ \beta\in \mathbb{Z}\alpha_1\oplus\mathbb{Z}\alpha_2}}
	\det(w)\,
	q^{\frac{|6w(\boldsymbol r)-5\boldsymbol s+30\beta|^2}{60}},\quad
	\eta(\tau)
	=
	q^{1/24}
	\prod_{m=1}^{\infty}(1-q^m).
\end{equation}
Here, we label the primary fields by $\boldsymbol{r}=r_1\omega^1+r_2\omega^2$ and $\boldsymbol{s}=s_1\omega^1+s_2\omega^2$, with the corresponding labels for each primary given in Table~\ref{tab:W3primaries}. We use the conventions
\begin{equation}
	\alpha_i\cdot\alpha_j=\begin{pmatrix}2&-1\\-1&2\end{pmatrix},\quad
	\omega^i\cdot\alpha_j=\delta^i_j,\quad
	W(A_2)=\langle s_{\alpha_1},s_{\alpha_2}\rangle\cong S_3,
	\quad
	s_{\alpha_i}(\lambda)
	=\lambda-(\alpha_i\cdot\lambda)\alpha_i
\end{equation}
Their modular $S$ transformations are given
by~\cite{Caldeira:2004jy}.
\begin{equation}
	\label{eq:SWfull}
		\begin{aligned}
			S_{(\boldsymbol r,\boldsymbol s),(\boldsymbol r',\boldsymbol s')}
			=\frac{1}{30\sqrt3}
			&
			\left[
			\sum_{w\in W(A_2)}\det(w)
			\exp\left(
			\frac{2\pi\mathrm{i}}{5}
			(6\boldsymbol r'-5\boldsymbol s')
			\cdot w(6\boldsymbol r-5\boldsymbol s)
			\right)
			\right]\\
			\times
			&\left[
			\sum_{w\in W(A_2)}\det(w)
			\exp\left(
			-\frac{\pi\mathrm{i}}{3}
			(6\boldsymbol r'-5\boldsymbol s')
			\cdot w(6\boldsymbol r-5\boldsymbol s)
			\right)
			\right]
		\end{aligned}
\end{equation}
Since charge conjugate representations have identical ordinary characters, the 20 irreducible representations give rise to only 12 linearly independent characters. For later convenience, we work directly in the reduced character basis
\begin{equation}
	\hat{\boldsymbol{\chi}}^{(1,1)}
	=\left(
	\hat{\chi}_0^{(1,1)},\hat{\chi}_{2/45}^{(1,1)},\hat{\chi}_{1/10}^{(1,1)},\hat{\chi}_{1/9}^{(1,1)},
	\hat{\chi}_{17/45}^{(1,1)},\hat{\chi}_{1/2}^{(1,1)},\hat{\chi}_{3/5}^{(1,1)},\hat{\chi}_{32/45}^{(1,1)},
	\hat{\chi}_{7/9}^{(1,1)},\hat{\chi}_{13/9}^{(1,1)},\hat{\chi}_{8/5}^{(1,1)},\hat{\chi}_2^{(1,1)}
	\right)^{\mathsf T},
\end{equation}
in which the reduced $S$ matrix takes the explicit form
\begin{equation}
	\label{eq:Sred}
	S_{\mathrm{red}}
	=\frac{\sin(\pi/5)}{3\sqrt{5}}
	\begin{pmatrix}
		1 & 4\varphi & 3\varphi & 4 & 4\varphi & 3
		& 2\varphi & 4\varphi & 4 & 4 & \varphi & 2 \\
		2\varphi &c_2 & 0 & \varphi c_1 & c_3 & 0
		& 2 & -c_1 & -\varphi c_3 & -\varphi c_2 & -2 & -2\varphi \\
		3\varphi & 0 & 3 & 0 & 0 & -3\varphi
		& -6 & 0 & 0 & 0 & -3 & 6\varphi \\
		2 & \varphi c_1 & 0 & -c_3 & -\varphi c_2 & 0
		& -2\varphi & -\varphi c_3 & -c_2 & c_1 & 2\varphi & -2 \\
		2\varphi & c_3 & 0 & -\varphi c_2 & -c_1 & 0
		& 2 & c_2 & \varphi c_1 & -\varphi c_3 & -2 & -2\varphi \\
		3 & 0 & -3\varphi & 0 & 0 & -3
		& 6\varphi & 0 & 0 & 0 & 3\varphi & 6 \\
		\varphi & 2 & -3 & -2\varphi & 2 & 3\varphi
		& -2 & 2 & -2\varphi & -2\varphi & -1 & 2\varphi \\
		2\varphi & -c_1 & 0 & -\varphi c_3 & c_2 & 0
		& 2 & c_3 & -\varphi c_2 & \varphi c_1 & -2 & -2\varphi \\
		2 & -\varphi c_3 & 0 & -c_2 & \varphi c_1 & 0
		& -2\varphi & -\varphi c_2 & c_1 & -c_3 & 2\varphi & -2 \\
		2 & -\varphi c_2 & 0 & c_1 & -\varphi c_3 & 0
		& -2\varphi & \varphi c_1 & -c_3 & -c_2 & 2\varphi & -2 \\
		\varphi & -4 & -3 & 4\varphi & -4 & 3\varphi
		& -2 & -4 & 4\varphi & 4\varphi & -1 & 2\varphi \\
		1 & -2\varphi & 3\varphi & -2 & -2\varphi & 3
		& 2\varphi & -2\varphi & -2 & -2 & \varphi & 2
	\end{pmatrix}
\end{equation}
The constants are defined by
\begin{equation}
	\label{eq:const}
	\varphi=\frac{1+\sqrt5}{2},\quad
	c_k=4\cos\left(\frac{2^{k-1}\pi}{9}\right),\quad k=1,2,3.
\end{equation}

For $g_1=\sigma$ and $g_2=1$, $\hat{\chi}^{(\sigma,1)}$ is the twisted sector character of $\mathcal{W}_3(5,6)$. For $g_1=1$ and $g_2=\sigma$, $\hat{\chi}^{(1, \sigma)}$ gives the ordinary $\mathcal{W}_3$ characters, $\hat{\chi}^{(1,1)}$, but with the signs of the terms corresponding to states created by an odd number of $W_{n\in\mathbb{Z}}$ modes flipped. By this definition, $\hat{\chi}^{(1, \sigma)}$ is defined only for the four charge conjugation invariant primaries of $\mathcal{W}_3(5,6)$, namely $1,\, E,\, F,\, X$. Since the modular $S$ transformation exchanges the time and space directions, $\hat{\chi}^{(1,\sigma)}$ and $\hat{\chi}^{(\sigma,1)}$ are related by the modular $S$ transformation:
\begin{equation}
	\label{eq:Stw}
	\hat{\boldsymbol{\chi}}^{(1,\sigma)}(-1/{\tau})=\hat{S}\hat{\boldsymbol{\chi}}^{(\sigma,1)}(\tau).
\end{equation}
Therefore, the four charge conjugation invariant primaries,
characterized by $r_1=r_2=r$ and $s_1=s_2=s$ as shown in
Table~\ref{tab:W3primaries}, correspond to four distinct
$\mathbb{Z}_2$ twisted sectors of $\mathcal{W}_3(5,6)$.
Their conformal weights are given by~\cite{Ho-Kim:1988ast,Honecker:1992kz}
\begin{equation}
	\label{eq:twistedW}
	\hat{h}_{r,s}
	=\frac{(6r-5s)^2}{120}+\frac7{240},
	\qquad
	\left\{\hat{h}_{r,s}\right\}_{r,s=1,2}
	=\left\{\frac{3}{80},\frac{1}{16},
	\frac{13}{80},\frac{7}{16}\right\}.
\end{equation}
Labeling the charge conjugation invariant primaries and their corresponding $\mathbb{Z}_2$ twisted fields by $(r,s)$, we write $\hat{\chi}^{(\sigma,1)}$ and $\hat{\chi}^{(1,\sigma)}$ as~\cite{Ho-Kim:1988ast,Caldeira:2004jy}
\begin{align}
		\hat{\chi}^{(\sigma,1)}_{r,s}(\tau)
		&=\frac{1}{\eta(\tau/2)}
		\sum_{n\in\mathbb Z}\left[
		q^{\frac{[30n+6r-5s]^2}{120}}
		-q^{\frac{[30n+6r+5s]^2}{120}}
		\right],\\
		\hat{\chi}^{(1,\sigma)}_{r,s}(\tau)
		&=\frac{1}{\eta(2\tau)}
		\sum_{n\in\mathbb Z}
		\left[
		q^{\frac{(30n+6r-5s)^2}{30}}
		-q^{\frac{(30n-6r-5s)^2}{30}}
		\right],
\end{align}
with the following modular S matrix
\begin{equation}
	\label{eq:hatSgen}
	\hat{S}_{(r,s),(r',s')}\\
	=\frac{4}{\sqrt{30}}
	\sin\left[
	\frac{2\pi}{5}(6r-5s)(6r'-5s')
	\right]
	\sin\left[
	\frac{2\pi}{6}(6r-5s)(6r'-5s')
	\right].
\end{equation}
 We choose the bases of $\hat{\boldsymbol{\chi}}^{(1,\sigma)}$ and $\hat{\boldsymbol{\chi}}^{(\sigma,1)}$ as follows:
\begin{equation}
	\hat{\boldsymbol{\chi}}^{(1,\sigma)}
	=
	\left(
	\hat{\chi}_0^{(1,\sigma)},
	\hat{\chi}_{1/10}^{(1,\sigma)},
	\hat{\chi}_{1/2}^{(1,\sigma)},
	\hat{\chi}_{8/5}^{(1,\sigma)}
	\right)^{\mathsf T},\quad
	\hat{\boldsymbol{\chi}}^{(\sigma,1)}
	=
	\left(
	\hat{\chi}_{3/80}^{(\sigma,1)},
	\hat{\chi}_{1/16}^{(\sigma,1)},
	\hat{\chi}_{13/80}^{(\sigma,1)},
	\hat{\chi}_{7/16}^{(\sigma,1)}
	\right)^{\mathsf T}.
\end{equation}
Using~\eqref{eq:hatSgen}, we obtain the matrix $\hat{S}$ in~\eqref{eq:Stw}, in agreement with~\cite{Antunes:2026kqb}:
\begin{equation}
	\label{eq:hatS}
	\hat S
	=\frac{1}{\sqrt{2(2+\varphi)}}
	\begin{pmatrix}
		\varphi & 1 & \varphi & 1\\
		1 & \varphi & -1 & -\varphi\\
		\varphi & -1 & -\varphi & 1\\
		1 & -\varphi & 1 & -\varphi
	\end{pmatrix},
	\quad \hat S^{-1}=\hat S.
\end{equation}
Here, the golden ratio $\varphi$ is defined in~\eqref{eq:const}. 

For $g_1=g_2=\sigma$, $\hat{\chi}_{\hat h}^{(\sigma,\sigma)}$ is obtained from the twisted sector character $\hat{\chi}_{\hat h}^{(\sigma,1)}$ by flipping the signs of terms corresponding to states created by an odd number of $W_{r\in\mathbb{Z}+1/2}$ modes. Equivalently, the insertion of $\sigma$ along the time direction in the trace over the twisted sector $\mathcal H_{\hat h}^\sigma$ can be understood as acting with $\hat \sigma=e^{2\pi\mathrm{i}(L_0-\hat h)}$. With this convention, the modular T transformation of $\hat{\chi}^{\sigma,\sigma}$ is
\begin{equation}
	\begin{aligned}
		\hat{\chi}_{\hat h}^{(\sigma,\sigma)}(\tau+1)
		&=\operatorname{Tr}_{\mathcal H_{\hat h}^{\sigma}}
		\left[
		e^{2\pi\mathrm{i}(L_0-\hat h)}
		e^{2\pi\mathrm{i}(L_0-c/24)}
		q^{L_0-c/24}
		\right]\\
		&=e^{2\pi\mathrm{i}(\hat h-c/24)}
		\operatorname{Tr}_{\mathcal H_{\hat h}^{\sigma}}
		\left[
		e^{4\pi\mathrm{i}(L_0-\hat h)}
		q^{L_0-c/24}
		\right]\\
		&=e^{2\pi\mathrm{i}(\hat h-c/24)}
		\hat{\chi}_{\hat h}^{(\sigma,1)}(\tau)
	\end{aligned}
\end{equation}
Here we used the fact that the twisted sector $\mathcal H_{\hat h}^{\sigma}$ is generated by integer moded Virasoro modes $L_{n\in\mathbb{Z}}$ and half-integer moded $W_{r\in\mathbb{Z}+1/2}$ modes, so the spectrum of $L_0-\hat h$ lies in $\frac12\mathbb{Z}$. Consequently, $e^{4\pi \mathrm{i}(L_0-\hat h)}$ acts as the identity on $\mathcal H_{\hat h}^{\sigma}$. Similarly, the modular T transformation of $\hat{\chi}^{(\sigma,1)}_{\hat h}$ is
\begin{equation}
	\begin{aligned}
		\hat{\chi}^{(\sigma,1)}_{\hat h}(\tau+1)
		&=\operatorname{Tr}_{\mathcal{H}^\sigma_{\hat h} }\left[\mathrm{e}^{2\pi\mathrm{i}(L_0-c/24)} q^{L_0-c/24}\right]\\
		&=\mathrm{e}^{2\pi\mathrm{i}(\hat h-c/24)}\operatorname{Tr}_{\mathcal{H}^\sigma_{\hat h} }\left[\mathrm{e}^{2\pi\mathrm{i}(L_0-\hat h)}q^{L_0-c/24}\right]\\
		&=\mathrm{e}^{2\pi\mathrm{i}(\hat h-c/24)}\hat{\chi}_{\hat h}^{(\sigma,\sigma)}(\tau)
	\end{aligned}
\end{equation}

The modular S transformation of $\hat{\chi}^{(\sigma,\sigma)}_{\hat h}$ is more involved. It can be derived using the following chain of transformations:~\cite{Birke:1999ik}
\begin{equation}
	\begin{aligned}
	\hat{\boldsymbol{\chi}}^{(\sigma,\sigma)}(\tau)\xrightarrow{T}\hat{\boldsymbol{\chi}}^{(\sigma,1)}(\tau+1)\xrightarrow{\hat S} 	&\hat{\boldsymbol{\chi}}^{(1,\sigma)}(-\frac{1}{\tau+1})
	\\
	\xrightarrow{T}&\hat{\boldsymbol{\chi}}^{(1,\sigma)}(\frac{\tau}{\tau+1})
	\xrightarrow{\hat S^{-1}}\hat{\boldsymbol{\chi}}^{(\sigma,1)}(-1-\frac{1}{\tau})\xrightarrow{T}	\hat{\boldsymbol{\chi}}^{(\sigma,\sigma)}(-\frac1\tau),
	\end{aligned}
\end{equation}
which gives
\begin{equation}
	\label{eq:P}
	\hat{\boldsymbol{\chi}}^{(\sigma,\sigma)}(-1/\tau)=P\hat{\boldsymbol{\chi}}^{(\sigma,\sigma)} (\tau),\quad
			P=\frac{1}{\sqrt{\varphi+2}}
			\begin{pmatrix}
				0&\varphi&1&0\\
				\varphi&0&0&1\\
				1&0&0&-\varphi\\
				0&1&-\varphi&0
			\end{pmatrix}.
\end{equation}
Here, we chose the basis
\begin{equation}
	\hat{\boldsymbol{\chi}}^{(\sigma,\sigma)}
	=
	\left(
	\hat{\chi}_{3/80}^{(\sigma,\sigma)},
	\hat{\chi}_{1/16}^{(\sigma,\sigma)},
	\hat{\chi}_{13/80}^{(\sigma,\sigma)},
	\hat{\chi}_{7/16}^{(\sigma,\sigma)}
	\right)^{\mathsf T}.
\end{equation}
We summarize the modular S transformation as
\begin{equation}
	\label{eq:SW}
	\begin{pmatrix}
		\hat{\boldsymbol{\chi}}^{(1,1)}\\
		\hat{\boldsymbol{\chi}}^{(1,\sigma)}\\
		\hat{\boldsymbol{\chi}}^{(\sigma,1)}\\
		\hat{\boldsymbol{\chi}}^{(\sigma,\sigma)}
	\end{pmatrix}(-1/\tau)=
	\begin{pmatrix}
		S_{\mathrm{red}} & 0 & 0 & 0\\
		0& 0 & \hat S & 0\\
		0 & {\hat S}^{-1} & 0 & 0\\
		0& 0 & 0 & P
	\end{pmatrix}
	\begin{pmatrix}
		\hat{\boldsymbol{\chi}}^{(1,1)}\\
		\hat{\boldsymbol{\chi}}^{(1,\sigma)}\\
		\hat{\boldsymbol{\chi}}^{(\sigma,1)}\\
		\hat{\boldsymbol{\chi}}^{(\sigma,\sigma)}
	\end{pmatrix}(\tau)
\end{equation}
where $S_{\mathrm{red}}$, $\hat S$ and $P$ are given in \eqref{eq:Sred}, \eqref{eq:hatS}, and \eqref{eq:P}, respectively.

\subsection{Primaries and fusion rules of $\mathcal{W}_3^{\mathbb{Z}_2}$}
To determine the primary field content of $\mathcal{U}$, we need to study its irreducible modules, which can be constructed from the untwisted irreducible modules of $\mathcal{W}_3$ in Table~\ref{tab:W3primaries} and the twisted irreducible modules in~\eqref{eq:twistedW}, following~\cite{Birke:1999ik,dong2017orbifold}.

Two $\mathcal{W}_3$-modules exchanged under charge conjugation belong to the same $\mathbb{Z}_2$ orbit and give rise to a single irreducible $\mathcal{U}$-module up to isomorphism. Schematically, for example,
\begin{equation}
	| D\rangle,|{D^\dagger}\rangle
	\xrightarrow{\text{gauging }\mathbb{Z}_2}
	\frac{1}{\sqrt{2}}\left(\ket{D}\pm|{D^\dagger}\rangle\right),
\end{equation}
where the two choices of sign correspond to isomorphic $\mathcal{U}$-modules and hence do not give rise to distinct irreducible representations.

On the other hand, a $\mathcal{W}_3$-module invariant under charge conjugation splits into two eigenspaces with eigenvalues $\sigma=+1$ and $\sigma=-1$, which furnish two inequivalent irreducible $\mathcal{U}$-modules. For generic conformal weight $h$, the corresponding primary states can be identified with $\ket{h}$ and $W_{-1}\ket{h}$, respectively. Similarly, all the twisted modules are fixed under charge conjugation and therefore split into two irreducible $\mathcal{U}$-modules. Since the spin-3 current $W$ has a half-integer mode expansion in the twisted sector, the corresponding primary states are generically given by $|{\hat h}\rangle$ and $W_{-1/2}|{\hat h}\rangle$.

However, there are two exceptional cases: the vacuum module $\ket{0}$ and the twisted module $ \ket{1/16}$. Using the commutation relations derived from the OPEs in~\eqref{eq:W3OPE}, we find
\begin{equation}
		\left\|W_{-1}|h\rangle\right\|^2
		=\frac{h(32h+2-c)}{22+5c},\quad
		\left\|W_{-1/2}|h\rangle\right\|^2
		=\frac{c}{256}-\frac{9h}{40}
		+\frac{16h(h+\frac15)}{22+5c}.
\end{equation}
These norms vanish in the respective exceptional cases. Thus, the corresponding $\sigma=-1$ primary states must be constructed from nonvanishing descendants at higher levels. More precisely, they are given by $W_{-3}\ket{0}$ and $W_{-3/2}\ket{\frac{1}{16}}$, respectively.

Based on the above analysis, we find that $\mathcal{U}$ has a total of 24 primary fields, whose conformal weights are given by
\begin{equation}
	\begin{aligned}
		h^{\text{orbit}}&=
		\left\{
		\frac{2}{45},\frac{1}{9},\frac{17}{45},\frac{3}{5},
		\frac{32}{45},\frac{7}{9},\frac{13}{9},2
		\right\},\\
		\quad
		\left(h^{\text{split}}_+,h^{\text{split}}_-\right)&=
		\left\{
		\left(0,3\right),
		\left(\frac{1}{10},\frac{11}{10}\right),
		\left(\frac{1}{2},\frac32\right),
		\left(\frac{8}{5},\frac{13}{5}\right)
		\right\},
		\\
		\left(\hat h^{\text{split}}_+,\hat h^{\text{split}}_-\right)&=
		\left\{
		\left(\frac{3}{80},\frac{43}{80}\right),
		\left(\frac{1}{16},\frac{25}{16}\right),
		\left(\frac{13}{80},\frac{53}{80}\right),
		\left(\frac{7}{16},\frac{15}{16}\right)
		\right\}.
	\end{aligned}
\end{equation}

For a non-fixed orbit, the character of the resulting $\mathcal{U}$-module coincides with the ordinary $\mathcal{W}_3$ character $\hat{\chi}_h^{(1,1)}$. In contrast, for modules invariant under charge conjugation, the characters of the two irreducible $\mathcal{U}$-modules obtained by splitting are given by projections onto the two eigenspaces of the $\mathbb{Z}_2$ action, schematically:
\begin{equation}
	\chi^{\mathcal{U},\text{split}}_{g,\pm}=\operatorname{Tr}_{{\mathcal{H}}^g}\left[\frac{1\pm\hat\sigma}{2}q^{L_0-c/24}\right]=\frac{1}{2}\left(\hat{\chi}^{(g,1)}\pm\hat{\chi}^{(g,\sigma)}\right),\quad g\in\{1,\sigma\}.
\end{equation}
Thus, the characters of $\mathcal{U}$ (for general $h$) can be expressed in terms of the characters of $\mathcal{W}_3$ as follows:
\begin{equation}
	\label{UtoW}
	\begin{gathered}
		\chi^{\mathcal{U}}_h=\hat{\chi}^{(1,1)}_h,\quad h\in h^{\text{orbit}},
		\\
		\begin{aligned}
			\chi_{h_+}^{\mathcal{U}}&=\frac{1}{2}\left(\hat{\chi}_{h_+}^{(1,1)}+\hat{\chi}_{h_+}^{(1,\sigma)}\right),\quad h_+\in h^{\text{split}}_+,
			\quad
			&\chi_{h_-}^{\mathcal{U}}&=\frac{1}{2}\left(\hat{\chi}_{{h_-}-1}^{(1,1)}-\hat{\chi}_{{h_-}-1}^{(1,\sigma)}\right),\quad h_-\in 		h^{\text{split}}_-,
			\\
			\chi_{\hat h_+}^{\mathcal{U}}&=\frac{1}{2}\left(\hat{\chi}_{\hat h_+}^{(\sigma,1)}+\hat{\chi}_{\hat h_+}^{(\sigma,\sigma)}\right),\quad \hat 	h_+\in \hat h^{\text{split}}_+,\quad
		   &\chi_{\hat h_-}^{\mathcal{U}}&=\frac{1}{2}\left(\hat{\chi}_{\hat h_--\frac{1}{2}}^{(\sigma,1)}-\hat{\chi}_{\hat h_--\frac{1}{2}}^{(\sigma,\sigma)}\right),\quad \hat h_-\in \hat h^{\text{split}}_-.
		\end{aligned}
	\end{gathered}
\end{equation}
For the two exceptional cases, we have
\begin{equation}
	\chi_{3}^{\mathcal{U}}	= \frac{1}{2}\left(\hat{\chi}_0^{(1,1)}-\hat{\chi}_0^{(1,\sigma)}\right),\quad
	\chi_{\frac{25}{16}}^{\mathcal{U}}	= \frac{1}{2}\left(\hat{\chi}^{(\sigma,1)}_{\frac{1}{16}}-\hat{\chi}^{(\sigma,\sigma)}_{\frac{1}{16}}\right).
\end{equation}

Using~\eqref{eq:SW} and the above character relations, we obtain the modular S matrix of $\mathcal{U}$, which factorizes as
\begin{equation}
	S^{\mathcal{U}}=S^M\otimes S^{\mathsf{Fib}},
\end{equation}
where the constants $c_k$ are defined in \eqref{eq:const} and
\begin{equation}
	\label{eq:SM}
	S^M=
	\frac{1}{12}\begin{pmatrix}
		1&3\sqrt2&4&3\sqrt2&3&4&3\sqrt2&4&3&3\sqrt2&2&1\\
		3\sqrt2&0&0&6&-3\sqrt2&0&-6&0&3\sqrt2&0&0&-3\sqrt2\\
		4&0&-2 c_3&0&0&-2 c_2&0&2 c_1&0&0&-4&4\\
		3\sqrt2&6&0&0&3\sqrt2&0&0&0&-3\sqrt2&-6&0&-3\sqrt2\\
		3&-3\sqrt2&0&3\sqrt2&-3&0&3\sqrt2&0&-3&-3\sqrt2&6&3\\
		4&0&-2 c_2&0&0&2 c_1&0&-2 c_3&0&0&-4&4\\
		3\sqrt2&-6&0&0&3\sqrt2&0&0&0&-3\sqrt2&6&0&-3\sqrt2\\
		4&0&2 c_1&0&0&-2 c_3&0&-2 c_2&0&0&-4&4\\
		3&3\sqrt2&0&-3\sqrt2&-3&0&-3\sqrt2&0&-3&3\sqrt2&6&3\\
		3\sqrt2&0&0&-6&-3\sqrt2&0&6&0&3\sqrt2&0&0&-3\sqrt2\\
		2&0&-4&0&6&-4&0&-4&6&0&4&2\\
		1&-3\sqrt2&4&-3\sqrt2&3&4&-3\sqrt2&4&3&-3\sqrt2&2&1
	\end{pmatrix}.
\end{equation}
Here, $S^{\mathsf{Fib}}$ is the Fibonacci modular S matrix, with the golden ratio $\varphi$ defined in \eqref{eq:const}:
\begin{equation}
	S^{\mathsf{Fib}}=\frac{1}{\sqrt{\varphi+2}}
	\begin{pmatrix}
		 1    & \varphi\\
		\varphi & -1
	\end{pmatrix},\quad  \varphi= \frac{1+\sqrt{5}}{2}.
\end{equation}
Here, let $m=1,\ldots,12$ label the indices of $S^M$, and let
$\alpha\in\{1,\tau\}$ label those of $S^{\mathsf{Fib}}$.
We adopt the ordering $(m,1),(m,\tau)$ for each $m$,
with the corresponding conformal weights arranged as
\begin{equation}
	\label{eq:fibbasis}
	\begin{aligned}
		h_{(m,1)}
		&=\left\{0,\frac{1}{16},\frac19,\frac7{16},\frac{1}{2},\frac{7}{9},\frac{15}{16},\frac{13}{9},\frac32,\frac{25}{16},2,3\right\},
		\\
		h_{(m,\tau)}
		&=\left\{\frac{13}{5},\frac{53}{80},\frac{32}{45},\frac{3}{80},\frac{11}{10},\frac{17}{45},\frac{43}{80},\frac{2}{45},\frac1{10},\frac{13}{80},\frac35,\frac85\right\}.
	\end{aligned}
\end{equation}
Using the Verlinde formula, the fusion algebra of $\mathcal{W}_{3}^{\mathbb{Z}_2}$ can be expressed as
\begin{equation}
	\label{eq:fusion}
	{\mathcal{N}_{ij}}^k=\sum_l\frac{S_{il}^{\mathcal{U}}S_{jl}^{\mathcal{U}}S_{kl}^{\mathcal{U}}}{S_{0l}^{\mathcal{U}}}=M\otimes \mathsf{Fib},\quad \mathsf{Fib}_1=\mathbb{I}_{2\times 2},\quad
	\mathsf{Fib}_\tau=
	\begin{pmatrix}
		0&1\\
		1&1
	\end{pmatrix}.
\end{equation}
The full fusion rules are too lengthy to display explicitly. However, we will see in the appendix that this factorization is very useful for finding the NIM-rep.
\subsection{Branching rules and the bulk partition function}
Using the relations between the $\mathcal{U}$ and $\mathcal{W}_3$ characters derived in~\eqref{UtoW}, we explicitly compute the $q$-expansions of the $\mathcal{U}$ characters to sufficiently high orders, following the method described in Section~\ref{sec:2.2}. The Ising and TIM characters are given by~\cite{DiFrancesco:1997nk}
\begin{equation}
	\chi_{r,s}^{\mathcal M(p,p')}(\tau)
	=\frac{1}{\eta(\tau)}
	\sum_{n\in\mathbb Z}
	\left[
	q^{\frac{(2pp'n+p'r-ps)^2}{4pp'}}
	-
	q^{\frac{(2pp'n+p'r+ps)^2}{4pp'}}
	\right].
\end{equation}
By matching the coefficients in the $q$-expansions of the $\mathcal U$ characters and the products of the Ising and TIM characters, we obtain the following character identities:
\begin{equation}
	\label{eq:branch rule}
		\begin{alignedat}{3}
			\chi^{\mathrm{Ising}}_{0}\chi^{\mathrm{TIM}}_{3/80}
			&=\chi^{\mathcal U}_{3/80},
			&\qquad
			\chi^{\mathrm{Ising}}_{0}\chi^{\mathrm{TIM}}_{1/10}
			&=\chi^{\mathcal U}_{1/10},
			&\qquad
			\chi^{\mathrm{Ising}}_{0}\chi^{\mathrm{TIM}}_{0}
			&=\chi^{\mathcal U}_{0}+\chi^{\mathcal U}_{2},
			\\[4pt]
			\chi^{\mathrm{Ising}}_{0}\chi^{\mathrm{TIM}}_{7/16}
			&=\chi^{\mathcal U}_{7/16},
			&\qquad
			\chi^{\mathrm{Ising}}_{0}\chi^{\mathrm{TIM}}_{3/2}
			&=\chi^{\mathcal U}_{3/2},
			&\qquad
			\chi^{\mathrm{Ising}}_{0}\chi^{\mathrm{TIM}}_{3/5}
			&=\chi^{\mathcal U}_{3/5}+\chi^{\mathcal U}_{13/5},
			\\[4pt]
			\chi^{\mathrm{Ising}}_{1/16}\chi^{\mathrm{TIM}}_{0}
			&=\chi^{\mathcal U}_{1/16},
			&\qquad
			\chi^{\mathrm{Ising}}_{1/16}\chi^{\mathrm{TIM}}_{1/10}
			&=\chi^{\mathcal U}_{13/80},
			&\qquad
			\chi^{\mathrm{Ising}}_{1/16}\chi^{\mathrm{TIM}}_{3/80}
			&=\chi^{\mathcal U}_{1/10}+\chi^{\mathcal U}_{11/10},
			\\[4pt]
			\chi^{\mathrm{Ising}}_{1/16}\chi^{\mathrm{TIM}}_{3/5}
			&=\chi^{\mathcal U}_{53/80},
			&\qquad
			\chi^{\mathrm{Ising}}_{1/16}\chi^{\mathrm{TIM}}_{3/2}
			&=\chi^{\mathcal U}_{25/16},
			&\qquad
			\chi^{\mathrm{Ising}}_{1/16}\chi^{\mathrm{TIM}}_{7/16}
			&=\chi^{\mathcal U}_{1/2}+\chi^{\mathcal U}_{3/2},
			\\[4pt]
			\chi^{\mathrm{Ising}}_{1/2}\chi^{\mathrm{TIM}}_{0}
			&=\chi^{\mathcal U}_{1/2},
			&\qquad
			\chi^{\mathrm{Ising}}_{1/2}\chi^{\mathrm{TIM}}_{3/80}
			&=\chi^{\mathcal U}_{43/80},
			&\qquad
			\chi^{\mathrm{Ising}}_{1/2}\chi^{\mathrm{TIM}}_{1/10}
			&=\chi^{\mathcal U}_{3/5}+\chi^{\mathcal U}_{8/5},
			\\[4pt]
			\chi^{\mathrm{Ising}}_{1/2}\chi^{\mathrm{TIM}}_{7/16}
			&=\chi^{\mathcal U}_{15/16},
			&\qquad
			\chi^{\mathrm{Ising}}_{1/2}\chi^{\mathrm{TIM}}_{3/5}
			&=\chi^{\mathcal U}_{11/10},
			&\qquad
			\chi^{\mathrm{Ising}}_{1/2}\chi^{\mathrm{TIM}}_{3/2}
			&=\chi^{\mathcal U}_{2}+\chi^{\mathcal U}_{3}.
		\end{alignedat}
\end{equation}
Since the irreducible $\mathcal U$-modules have distinct conformal weights, their characters have distinct leading powers of $q$ and are therefore linearly independent. The branching rules for the decomposition of $\mathcal A$-modules into $\mathcal U$-modules can thus be read off unambiguously from the above character identities.

We now consider the full bulk CFT. From the field content in Table~\ref{tab:Aprimaries}, the bulk partition function is
\begin{equation}
		\begin{aligned}
			Z=&Z^{\text{Ising}}Z^{\text{TIM}}\\
			=&
			\left|\chi^{\mathrm{Ising}}_0\chi^{\mathrm{TIM}}_0\right|^2
			+\left|\chi^{\mathrm{Ising}}_0\chi^{\mathrm{TIM}}_{3/80}\right|^2
			+\left|\chi^{\mathrm{Ising}}_0\chi^{\mathrm{TIM}}_{1/10}\right|^2
			+\left|\chi^{\mathrm{Ising}}_0\chi^{\mathrm{TIM}}_{7/16}\right|^2
			\\
			+&\left|\chi^{\mathrm{Ising}}_0\chi^{\mathrm{TIM}}_{3/5}\right|^2
			+\left|\chi^{\mathrm{Ising}}_0\chi^{\mathrm{TIM}}_{3/2}\right|^2
			+\left|\chi^{\mathrm{Ising}}_{1/16}\chi^{\mathrm{TIM}}_0\right|^2
			+\left|\chi^{\mathrm{Ising}}_{1/16}\chi^{\mathrm{TIM}}_{3/80}\right|^2
			+\left|\chi^{\mathrm{Ising}}_{1/16}\chi^{\mathrm{TIM}}_{1/10}\right|^2
			\\
			+&\left|\chi^{\mathrm{Ising}}_{1/16}\chi^{\mathrm{TIM}}_{7/16}\right|^2
			+\left|\chi^{\mathrm{Ising}}_{1/16}\chi^{\mathrm{TIM}}_{3/5}\right|^2
			+\left|\chi^{\mathrm{Ising}}_{1/16}\chi^{\mathrm{TIM}}_{3/2}\right|^2
			+\left|\chi^{\mathrm{Ising}}_{1/2}\chi^{\mathrm{TIM}}_0\right|^2
			\\
			+&\left|\chi^{\mathrm{Ising}}_{1/2}\chi^{\mathrm{TIM}}_{3/80}\right|^2
			+\left|\chi^{\mathrm{Ising}}_{1/2}\chi^{\mathrm{TIM}}_{1/10}\right|^2
			+\left|\chi^{\mathrm{Ising}}_{1/2}\chi^{\mathrm{TIM}}_{7/16}\right|^2
			+\left|\chi^{\mathrm{Ising}}_{1/2}\chi^{\mathrm{TIM}}_{3/5}\right|^2
			+\left|\chi^{\mathrm{Ising}}_{1/2}\chi^{\mathrm{TIM}}_{3/2}\right|^2
		\end{aligned}
\end{equation}
Note that when the chiral algebra is restricted to $\mathcal{U}$, the bulk CFT becomes a non-diagonal CFT with the following partition function:
\begin{equation}
	\label{eq:bulk_partition_U}
		\begin{aligned}
			Z={}&
			2\Bigl(
			|\chi^{\mathcal U}_{\frac{1}{10}}|^2
			+|\chi^{\mathcal U}_{\frac{1}{2}}|^2
			+|\chi^{\mathcal U}_{\frac{3}{5}}|^2
			+|\chi^{\mathcal U}_{\frac{11}{10}}|^2
			+|\chi^{\mathcal U}_{\frac{3}{2}}|^2
			+|\chi^{\mathcal U}_{2}|^2
			\Bigr)
			\\
			&+|\chi^{\mathcal U}_0|^2
			+|\chi^{\mathcal U}_{\frac{3}{80}}|^2
			+|\chi^{\mathcal U}_{\frac{1}{16}}|^2
			+|\chi^{\mathcal U}_{\frac{13}{80}}|^2
			+|\chi^{\mathcal U}_{\frac{7}{16}}|^2
			+|\chi^{\mathcal U}_{\frac{43}{80}}|^2
			\\
			&+|\chi^{\mathcal U}_{\frac{53}{80}}|^2
			+|\chi^{\mathcal U}_{\frac{15}{16}}|^2
			+|\chi^{\mathcal U}_{\frac{25}{16}}|^2
			+|\chi^{\mathcal U}_{\frac{8}{5}}|^2
			+|\chi^{\mathcal U}_{\frac{13}{5}}|^2
			+|\chi^{\mathcal U}_{3}|^2
			\\
			&+\Bigl(
			\chi^{\mathcal U}_0\bar{\chi}^{\mathcal U}_2
			+\bar{\chi}^{\mathcal U}_0\chi^{\mathcal U}_2
			\Bigr)
			+\Bigl(
			\chi^{\mathcal U}_{\frac{1}{10}}\bar{\chi}^{\mathcal U}_{\frac{11}{10}}
			+\bar{\chi}^{\mathcal U}_{\frac{1}{10}}\chi^{\mathcal U}_{\frac{11}{10}}
			\Bigr)
			+\Bigl(
			\chi^{\mathcal U}_{\frac{1}{2}}\bar{\chi}^{\mathcal U}_{\frac{3}{2}}
			+\bar{\chi}^{\mathcal U}_{\frac{1}{2}}\chi^{\mathcal U}_{\frac{3}{2}}
			\Bigr)
			\\
			&+\Bigl(
			\chi^{\mathcal U}_{\frac{3}{5}}\bar{\chi}^{\mathcal U}_{\frac{8}{5}}
			+\bar{\chi}^{\mathcal U}_{\frac{3}{5}}\chi^{\mathcal U}_{\frac{8}{5}}
			\Bigr)
			+\Bigl(
			\chi^{\mathcal U}_{\frac{3}{5}}\bar{\chi}^{\mathcal U}_{\frac{13}{5}}
			+\bar{\chi}^{\mathcal U}_{\frac{3}{5}}\chi^{\mathcal U}_{\frac{13}{5}}
			\Bigr)
			+\Bigl(
			\chi^{\mathcal U}_2\bar{\chi}^{\mathcal U}_3
			+\bar{\chi}^{\mathcal U}_2\chi^{\mathcal U}_3
			\Bigr)
		\end{aligned}
\end{equation}
Consequently, the boundary states preserving $\mathcal{U}$ cannot be obtained directly through the standard Cardy construction for diagonal CFT. However, since the bulk theory remains an RCFT with respect to this subalgebra, we instead classify these boundary states by solving the corresponding NIM-rep problem directly in what follows.
\section{Classification of symmetry breaking interfaces}
\label{sec:3}
In this section, we present a complete classification of conformal boundary states preserving $\mathcal{U}=\mathcal{W}_3^{\mathbb{Z}_2}$. We begin with a brief review of the classification of boundary states in RCFT, following~\cite{Behrend:1999bn, Petkova:2000dv}.
\subsection{The Cardy condition and NIM-rep}
\label{sec:3.1}
Since $\operatorname{Aut}(\mathcal{U})=1$, boundaries preserving $\mathcal{U}$ should satisfy
\begin{equation}
	\label{eq:gluing}
	\left(W_n^{(s)}-(-1)^s\bar W_{-n}^{(s)}\right)\ket{B}=0.
\end{equation}
Here $W^{(s)}$ is the spin-$s$ current of $\mathcal{U}\cong\mathcal{W}(2,6,8,10)$. Let $\mathcal{E}$ denote the set of diagonal sectors in the bulk CFT spectrum, counted with multiplicity. Each $(j,a)\in\mathcal{E}$ corresponds to an Ishibashi state $\Ish{j;a}$, where $j$ labels an irreducible module and $a$ is a multiplicity index. These Ishibashi states span the space of solutions to the above gluing conditions~\cite{Ishibashi:1988kg}. Once the normalization of the bulk fields is fixed, the Ishibashi states can be explicitly constructed as follows:
\begin{equation}
	\label{eq:ishibashi}
	\Ish{j;a}=\sum_{N}
	\left|j;a,N\right\rangle
	\otimes U\left|\overline{j,a,N}\right\rangle.
\end{equation}
Here $\left|j;a,N\right\rangle$ runs over an orthonormal basis of $\mathcal{H}_j$, and $U:\mathcal{H}_j\to\mathcal{H}_{j^\vee}$ ($j^\vee$ denotes the charge conjugate representation of $j$, for $\mathcal{U}$, $j^\vee=j$) is an antiunitary operator satisfying
\begin{equation}
	U\overline{W}_n^j U^{-1}
	=(-1)^{h_j}(\overline{W}_{-n}^j)^\dagger.
\end{equation}
In what follows, we adopt the following normalization convention:
\begin{equation}
	\langle\!\langle i;a|q^{\frac{1}{2}\left(L_0+\bar{L}_0-\frac{c}{12}\right)}\Ish{j;b}=\delta_{ij}\delta_{ab}\chi_j(q).
\end{equation}
However, this normalization leaves a phase ambiguity, or more generally, a unitary transformation ambiguity in sectors with multiplicities. When the chiral algebra admits nontrivial automorphisms,
we may choose an automorphism $\rho$ and generalize
the gluing condition to
\begin{equation}
		\left(W_n^{(s)}-(-1)^s \rho \left(\bar W_{-n}^{(s)}\right)\right)\ket{B}=0.
\end{equation}
 For an irreducible module
$\mathcal{H}_i$ carrying a representation $\pi_i$, the automorphism $\rho$ induces a new representation $\pi_i^\rho:=\pi_i\circ\rho$, and hence a permutation $i\mapsto\omega(i)$
of the irreducible modules, such that $\mathcal{H}_i\curvearrowleft\pi_i^\rho$ is equivalent to $\mathcal{H}_{\omega(i)}\curvearrowleft\pi_{\omega(i)}$.
We denote the isomorphism by
$\Omega_\rho:\mathcal{H}_{\omega(i)}\to\mathcal{H}_i$,
which satisfies
\begin{equation}
	\pi_i^\rho(J)
	=\Omega_\rho\pi_{\omega(i)}(J)\Omega_\rho^{-1},
	\qquad [\Omega_\rho,U]=0.
\end{equation}
Consequently, each bulk sector satisfying
$i=\omega(i^\vee)$ admits a $\rho$-twisted
Ishibashi state of the form
\begin{equation}
	\label{eq:twish}
	\Ish{i;\Omega}=\mathrm{id}\otimes\Omega_\rho\Ish{i}.
\end{equation}

Only appropriate linear combinations of Ishibashi states correspond to physical boundary states. We consider the following ansatz:
\begin{equation}
	\label{eq:bdy}
	\ket{\alpha}=\sum_{(j;a)\in\mathcal{E}}\frac{\psi_\alpha^{(j;a)}}{\sqrt{S_{0j}}}\Ish{j;a},
\end{equation}
and we define the outgoing boundary states using the antiunitary CPT operator $\Theta$, normalized so that $\Theta\Ish{j;a}=\Ish{j^\vee;a^\vee}$~\cite{Recknagel:2013uja}:
\begin{equation}
	\bra{\alpha}:=\left(\Theta\ket{\alpha}\right)^\dagger=\sum_{(j;a)\in\mathcal{E}}\langle\!\langle j;a|\frac{\psi_\alpha^{(j;a)^\vee}}{\sqrt{S_{0j}}}.
\end{equation}
Following~\cite{Behrend:1999bn}, we assume $\psi_\alpha^{(j;a)^\vee}=(\psi_\alpha^{(j;a)})^*$~\footnote{In the standard RCFT setting, this reality condition may be viewed, at least partly, as a choice of an appropriate basis of Ishibashi states. For more general CFT, however, this condition may not hold, see, e.g., \cite{Gaberdiel:2001xm, Hoback:2026yqj}.}. With this definition, the closed channel partition function between two boundaries can be computed as
\begin{equation}
	\label{eq:clPF}
	\mathcal{Z}^{\text{cl}}_{\beta\mid \alpha}:=\bra{\beta}\tilde{q}^{\frac12(L_0+\bar{L}_0-\frac{c}{12})}\ket{\alpha}=\sum_{(j;a)\in\mathcal{E}}\psi^{(j;a)}_\alpha\left(\psi^{(j;a)}_\beta\right)^*\frac{\chi_j(\tilde q)}{S_{0j}}.
\end{equation}
Here $\tilde{q}=\mathrm{e}^{-2\pi\mathrm{i}/\tau}$. On the other hand, the open channel partition function is
\begin{equation}
	\label{eq:opPF}
		\mathcal{Z}^{\text{op}}_{\beta\mid \alpha}:=\operatorname{Tr}_{\mathcal{H}_{\beta\mid \alpha}}q^{L_0-\frac{c}{24}}=\sum_{i\in \mathcal{I}}{({n_i})_{\alpha}}^\beta\chi_i(q),\quad {({n_i})_{\alpha}}^\beta\in \mathbb{Z}_{\geq 0},
\end{equation}
with $q=\mathrm{e}^{2\pi\mathrm{i}\tau}$ and $\mathcal{I}$ denoting the set of irreducible modules. The open and closed channel partition functions should be related by the modular S transformation, $\tau\mapsto -1/\tau$. Comparing the coefficients of the characters after the modular $S$ transformation, we obtain the Cardy condition:~\footnote{A subtlety arises from the possible linear dependence of characters, which may obstruct a direct comparison of their coefficients. This issue is avoided in the TFT construction of RCFT \cite{Fuchs:2002cm}. It does not arise in our case.}
\begin{equation}
	\label{eq:nidecomp}
	{({n_i})_{\alpha}}^\beta =\sum_{(j;a)\in\mathcal{E}}\left(\psi^{(j;a)}_\beta\right)^*\frac{S_{ij}}{S_{0j}}\psi^{(j;a)}_\alpha\in \mathbb{Z}_{\geq 0},\quad \forall i\in\mathcal{I}.
\end{equation}
We are looking for a complete set of boundary states satisfying
\begin{equation}
	\label{eq:cond}
	{(n_0)_{\alpha}}^\beta=\sum_{(j;a)\in\mathcal{E}}\psi_\alpha^{(j;a)}(\psi_\beta^{(j;a)})^*=\delta_{\alpha\beta},\quad
	\sum_{\alpha}\psi_\alpha^{(i;a)}(\psi_\alpha^{(j;b)})^*=\delta_{ab}\delta_{ij}.
\end{equation}

The first condition expresses the requirement that these boundary states are fundamental, i.e., the open-channel spectrum between identical boundaries contains exactly one copy of the vacuum representation. Since these fundamental boundary states form a complete set, more general boundary states can be constructed as appropriate linear combinations of them. For a more detailed discussion of these conditions in RCFT, we refer the reader to~\cite{Fuchs:2002cm}. The above conditions immediately imply that the number of fundamental boundary states equals the number of Ishibashi states.

The Verlinde formula implies that $S_{ij}/S_{0j}$ form a one-dimensional representation of the fusion algebra,
\begin{equation}
	\frac{S_{il}}{S_{0l}}\frac{S_{jl}}{S_{0l}}=\sum_{k\in\mathcal{I}}{\mathcal{N}_{ij}}^k\frac{S_{kl}}{S_{0l}},\quad \forall i,j,l\in\mathcal{I}.
\end{equation}
Hence, by~\eqref{eq:nidecomp}and~\eqref{eq:cond}, the matrices $n_i$, of size $|\mathcal{E}|\times|\mathcal{E}|$, also form a representation of the fusion algebra,
\begin{equation}
	n_in_j=\sum_{k\in\mathcal{I}}{\mathcal{N}_{ij}}^k n_k,\quad n_0=\mathbb{I}_{|\mathcal{E}|\times|\mathcal{E}|}, \, n_i^\mathsf{T}=n_{i^\vee}.
\end{equation}

Furthermore, since $n_i$ are $\mathbb{Z}_{\geq 0}$-valued matrices, this representation is called a non-negative integer matrix representation (NIM-rep)~\cite{Gannon:2001ki}. Once such a representation is found, we simultaneously diagonalize the matrices $n_i$ according to~\eqref{eq:nidecomp} to obtain $\psi$, thereby constructing a complete and orthonormal solution to the Cardy condition, with the boundary states given by~\eqref{eq:bdy}.
\subsection{Boundaries preserving $\mathcal{W}_3^{\mathbb{Z}_2}$}
From the diagonal part of \eqref{eq:bulk_partition_U}, we find 24 Ishibashi states in our theory,\footnote{In what follows, we omit the chiral algebra labels of Ishibashi states whenever no confusion arises.}
\begin{equation}
	\label{eq:ishibashis}
	\begin{aligned}
			\mathcal{E}^\mathcal{U}=\Bigl\{
		&\Ish{0},
		\Ish{\frac{3}{80}},
		\Ish{\frac{1}{16}},
		\Ish{\frac{13}{80}},
		\Ish{\frac{7}{16}},
		\Ish{\frac{43}{80}},
		\Ish{\frac{53}{80}},
		\Ish{\frac{15}{16}},
		\Ish{\frac{25}{16}},
		\Ish{\frac{8}{5}},
		\Ish{\frac{13}{5}},
		\Ish{3},
		\\
		&\Ish{\frac{1}{10};(0,\frac{1}{10})},
		\Ish{\frac{1}{10};(\frac{1}{16},\frac{3}{80})},
		\Ish{\frac{1}{2};(\frac{1}{2},0)},
		\Ish{\frac{1}{2};(\frac{1}{16},\frac{7}{16})},
		\Ish{\frac{3}{5};(0,\frac{3}{5})},
		\Ish{\frac{3}{5};(\frac{1}{2},\frac{1}{10})},
		\\
		&\Ish{\frac{11}{10};(\frac{1}{16},\frac{3}{80})},
		\Ish{\frac{11}{10};(\frac{1}{2},\frac{3}{5})},
		\Ish{\frac{3}{2};(0,\frac{3}{2})},
		\Ish{\frac{3}{2};(\frac{1}{16},\frac{7}{16})},
		\Ish{2;(0,0)},
		\Ish{2;(\frac{1}{2},\frac{3}{2})}
		\Bigr\}.
	\end{aligned}
\end{equation}
Here, we use $(h_{\text{Ising}},h_{\text{TIM}})$ to label the origin of each module under the branching rules in~\eqref{eq:branch rule}, thereby distinguishing different copies when multiplicities occur.

Hence, as discussed in section~\ref{sec:3.1}, there are 24 fundamental boundary states preserving $\mathcal{U}$ (denoted by $\mathcal{B}^\mathcal{U}$). To determine them, we need to find the NIM-rep of the fusion algebra in~\eqref{eq:fusion}. We give the details in the appendix. Before presenting the result, we discuss one subtlety. Since the Ishibashi states in~\eqref{eq:ishibashis} occur with multiplicities, simultaneous diagonalization of the matrices $n_i$ determines the eigenvectors $\psi^{(j;a)}$ only up to $U(1)$ phases and, in subspaces with multiplicity two, additional $U(2)$ rotations. For a fixed normalization of the bulk fields, only a particular choice corresponds to the physical solution~\cite{Sousa:2002te}. However, before such a normalization is specified, this ambiguity amounts to a choice of convention.

For convenience in the following discussion, we assume that the field normalizations in the Ising and TIM models have been fixed and choose a basis naturally adapted to the branching rules. More precisely, we require that the decomposition of the Ising $\times$ TIM Ishibashi states into $\mathcal{U}$ Ishibashi states follows directly from the module decompositions given by the branching rules in~\eqref{eq:branch rule}. For example, we demand
\begin{equation}
	\label{eq:ishibashi_convention}
	\Ish{0}_{\text{Ising}}\otimes\Ish{0}_{\text{TIM}}=\Ish{0}_{\mathcal{U}}+\Ish{2;(0,0)}_{\mathcal{U}},\quad
	\Ish{\frac12}_{\text{Ising}}\otimes\Ish{\frac32}_{\text{TIM}}=\Ish{3}_{\mathcal{U}}+\Ish{2;(\frac12,\frac32)}_{\mathcal{U}}.
\end{equation}

To obtain the correct $\psi$ in our chosen basis, we first take an arbitrary $\tilde{\psi}$ obtained by simultaneously diagonalizing the matrices $n_i$. Recall that the 18 product Cardy states preserving the full chiral algebra $\mathcal{A}$ are given by
\begin{equation}
	\label{eq:prodstate}
	\ket{h_{\text{Ising}},h_{\text{TIM}}}_{\mathcal{A}}=\left(\sum_{i\in\mathcal{E}^{\text{Ising}}}\frac{S^{(3,4)}_{h_{\text{Ising}},i}}{\sqrt{S^{(3,4)}_{0,i}}}\Ish{i}_{\text{Ising}}\right)\otimes\left(\sum_{j\in\mathcal{E}^{\text{TIM}}}\frac{S^{(4,5)}_{h_{\text{TIM}},j}}{\sqrt{S^{(4,5)}_{0,j}}}\Ish{j}_{\text{TIM}}\right),
\end{equation}
where the modular S matrix is given by the general formula for $\mathcal{M}(p,q)$:
\begin{equation}
	\label{eq:Sminimal}
	S_{(r,s),(r',s')}^{(p,q)}
	=
	2\sqrt{\frac{2}{pq}}\,
	(-1)^{1+rs'+sr'}
	\sin\left(\frac{\pi qrr'}{p}\right)
	\sin\left(\frac{\pi pss'}{q}\right),
\end{equation}
and $\mathcal{E}^{\text{Ising, TIM}}$ are the same as the sets of primaries listed in Table~\ref{tab:Aprimaries}, since the corresponding bulk theories are diagonal. Since these Cardy states (denoted by $\mathcal{B}^\mathcal{A}$) also preserve the subalgebra $\mathcal{U}$, we expect them to be among the 24 boundary states $\mathcal{B}^\mathcal{U}$. This can be verified by computing the closed channel partition functions in~\eqref{eq:clPF}, expressing them in terms of the Ising and TIM characters using~\eqref{eq:branch rule}, and comparing them with the closed channel partition functions of the product Cardy states in~\eqref{eq:prodstate}. 

We can thus identify the 18 rows of the desired matrix $\psi$ corresponding to these product Cardy states, denoted by $\psi_{\alpha}^{(j;a)}$, $\alpha\in\mathcal{B}^\mathcal{A}$. They are determined by~\eqref{eq:prodstate} and our conventions for the Ishibashi states in~\eqref{eq:ishibashi_convention}. Let ${P_{\alpha}}^\beta$ be an $18\times24$ partial permutation matrix that selects and relabels the corresponding rows of $\tilde\psi^{(j;a)}_{\beta}$, $\beta\in\mathcal{B}^{\mathcal{U}}$. As discussed above, $P$ can be determined by comparing the closed channel partition functions, which are independent of the choice of Ishibashi basis. 

Let $Q$ denote the unitary transformation relating the two choices of Ishibashi basis. By construction, $Q$ is block diagonal, with $U(1)$ blocks for the multiplicity free states in $\mathcal{E}^{\mathcal{U}}$ and $U(2)$ blocks for the multiplicity two subspaces. Next, we need to find this transformation matrix satisfying
\begin{equation}
	\psi_{\alpha}^{(j;a)}=\sum_{\beta\in\mathcal{B}^\mathcal{U}}\sum_{(j';a')\in\mathcal{E}^\mathcal{U}}{P_{\alpha}}^\beta\tilde\psi^{(j';a')}_{\beta} {Q_{(j';a')}}^{(j;a)},\quad \forall \alpha\in\mathcal{B}^\mathcal{A}.
\end{equation}
The final result is given in~\eqref{eq:psi}, with the column labels $(j;a)$ ordered as in~\eqref{eq:ishibashis}.
\begin{sidewaysfigure}[p]
	\centering
\begin{equation}
		\label{eq:psi}
\begin{aligned}
		&B_\alpha^{(j;a)}=\frac{1}{2\varphi^{\frac{1}{2}}(\varphi+2)^{\frac{1}{4}}}\times\\[4pt]
		&\resizebox{\linewidth}{!}{$\displaystyle
		\begin{pmatrix}
			\varphi^{\frac{1}{2}} & 2^{\frac{1}{4}} \varphi & 2^{\frac{1}{4}} \varphi^{\frac{1}{2}} & 2^{\frac{1}{4}} \varphi & 2^{\frac{1}{4}} \varphi^{\frac{1}{2}} & 2^{\frac{1}{4}} \varphi & 2^{\frac{1}{4}} \varphi & 2^{\frac{1}{4}} \varphi^{\frac{1}{2}} & 2^{\frac{1}{4}} \varphi^{\frac{1}{2}} & \varphi & \varphi & \varphi^{\frac{1}{2}} & \varphi & 2^{\frac{1}{2}} \varphi & \varphi^{\frac{1}{2}} & 2^{\frac{1}{2}} \varphi^{\frac{1}{2}} & \varphi & \varphi & 2^{\frac{1}{2}} \varphi & \varphi & \varphi^{\frac{1}{2}} & 2^{\frac{1}{2}} \varphi^{\frac{1}{2}} & \varphi^{\frac{1}{2}} & \varphi^{\frac{1}{2}} \\
			2^{\frac{1}{2}} \varphi^{\frac{3}{2}} & 0 & 2^{\frac{3}{4}} \varphi^{\frac{3}{2}} & 2^{\frac{3}{4}} & 0 & 0 & -2^{\frac{3}{4}} & 0 & -2^{\frac{3}{4}} \varphi^{\frac{3}{2}} & 2^{\frac{1}{2}} & -2^{\frac{1}{2}} & -2^{\frac{1}{2}} \varphi^{\frac{3}{2}} & 2^{\frac{1}{2}} & 0 & 2^{\frac{1}{2}} \varphi^{\frac{3}{2}} & 0 & -2^{\frac{1}{2}} & 2^{\frac{1}{2}} & 0 & -2^{\frac{1}{2}} & -2^{\frac{1}{2}} \varphi^{\frac{3}{2}} & 0 & 2^{\frac{1}{2}} \varphi^{\frac{3}{2}} & -2^{\frac{1}{2}} \varphi^{\frac{3}{2}} \\
			\varphi^{\frac{3}{2}} & 2^{\frac{1}{4}} & 2^{\frac{1}{4}} \varphi^{\frac{3}{2}} & -2^{\frac{1}{4}} & -2^{\frac{1}{4}} \varphi^{\frac{3}{2}} & 2^{\frac{1}{4}} & -2^{\frac{1}{4}} & -2^{\frac{1}{4}} \varphi^{\frac{3}{2}} & 2^{\frac{1}{4}} \varphi^{\frac{3}{2}} & -1 & -1 & \varphi^{\frac{3}{2}} & -1 & 2^{\frac{1}{2}} & \varphi^{\frac{3}{2}} & -2^{\frac{1}{2}} \varphi^{\frac{3}{2}} & -1 & -1 & 2^{\frac{1}{2}} & -1 & \varphi^{\frac{3}{2}} & -2^{\frac{1}{2}} \varphi^{\frac{3}{2}} & \varphi^{\frac{3}{2}} & \varphi^{\frac{3}{2}} \\
			2^{\frac{1}{2}} \varphi^{\frac{1}{2}} & 0 & 2^{\frac{3}{4}} \varphi^{\frac{1}{2}} & -2^{\frac{3}{4}} \varphi & 0 & 0 & 2^{\frac{3}{4}} \varphi & 0 & -2^{\frac{3}{4}} \varphi^{\frac{1}{2}} & -2^{\frac{1}{2}} \varphi & 2^{\frac{1}{2}} \varphi & -2^{\frac{1}{2}} \varphi^{\frac{1}{2}} & -2^{\frac{1}{2}} \varphi & 0 & 2^{\frac{1}{2}} \varphi^{\frac{1}{2}} & 0 & 2^{\frac{1}{2}} \varphi & -2^{\frac{1}{2}} \varphi & 0 & 2^{\frac{1}{2}} \varphi & -2^{\frac{1}{2}} \varphi^{\frac{1}{2}} & 0 & 2^{\frac{1}{2}} \varphi^{\frac{1}{2}} & -2^{\frac{1}{2}} \varphi^{\frac{1}{2}} \\
			\varphi^{\frac{3}{2}} & -2^{\frac{1}{4}} & 2^{\frac{1}{4}} \varphi^{\frac{3}{2}} & -2^{\frac{1}{4}} & 2^{\frac{1}{4}} \varphi^{\frac{3}{2}} & -2^{\frac{1}{4}} & -2^{\frac{1}{4}} & 2^{\frac{1}{4}} \varphi^{\frac{3}{2}} & 2^{\frac{1}{4}} \varphi^{\frac{3}{2}} & -1 & -1 & \varphi^{\frac{3}{2}} & -1 & -2^{\frac{1}{2}} & \varphi^{\frac{3}{2}} & 2^{\frac{1}{2}} \varphi^{\frac{3}{2}} & -1 & -1 & -2^{\frac{1}{2}} & -1 & \varphi^{\frac{3}{2}} & 2^{\frac{1}{2}} \varphi^{\frac{3}{2}} & \varphi^{\frac{3}{2}} & \varphi^{\frac{3}{2}} \\
			\varphi^{\frac{1}{2}} & -2^{\frac{1}{4}} \varphi & 2^{\frac{1}{4}} \varphi^{\frac{1}{2}} & 2^{\frac{1}{4}} \varphi & -2^{\frac{1}{4}} \varphi^{\frac{1}{2}} & -2^{\frac{1}{4}} \varphi & 2^{\frac{1}{4}} \varphi & -2^{\frac{1}{4}} \varphi^{\frac{1}{2}} & 2^{\frac{1}{4}} \varphi^{\frac{1}{2}} & \varphi & \varphi & \varphi^{\frac{1}{2}} & \varphi & -2^{\frac{1}{2}} \varphi & \varphi^{\frac{1}{2}} & -2^{\frac{1}{2}} \varphi^{\frac{1}{2}} & \varphi & \varphi & -2^{\frac{1}{2}} \varphi & \varphi & \varphi^{\frac{1}{2}} & -2^{\frac{1}{2}} \varphi^{\frac{1}{2}} & \varphi^{\frac{1}{2}} & \varphi^{\frac{1}{2}} \\
			2^{\frac{1}{2}} \varphi^{\frac{1}{2}} & 2^{\frac{3}{4}} \varphi & 0 & 0 & 2^{\frac{3}{4}} \varphi^{\frac{1}{2}} & -2^{\frac{3}{4}} \varphi & 0 & -2^{\frac{3}{4}} \varphi^{\frac{1}{2}} & 0 & -2^{\frac{1}{2}} \varphi & 2^{\frac{1}{2}} \varphi & -2^{\frac{1}{2}} \varphi^{\frac{1}{2}} & 2^{\frac{1}{2}} \varphi & 0 & -2^{\frac{1}{2}} \varphi^{\frac{1}{2}} & 0 & 2^{\frac{1}{2}} \varphi & -2^{\frac{1}{2}} \varphi & 0 & -2^{\frac{1}{2}} \varphi & 2^{\frac{1}{2}} \varphi^{\frac{1}{2}} & 0 & 2^{\frac{1}{2}} \varphi^{\frac{1}{2}} & -2^{\frac{1}{2}} \varphi^{\frac{1}{2}} \\
			2 \varphi^{\frac{3}{2}} & 0 & 0 & 0 & 0 & 0 & 0 & 0 & 0 & -2 & -2 & 2 \varphi^{\frac{3}{2}} & 2 & 0 & -2 \varphi^{\frac{3}{2}} & 0 & -2 & -2 & 0 & 2 & -2 \varphi^{\frac{3}{2}} & 0 & 2 \varphi^{\frac{3}{2}} & 2 \varphi^{\frac{3}{2}} \\
			2^{\frac{1}{2}} \varphi^{\frac{3}{2}} & 2^{\frac{3}{4}} & 0 & 0 & -2^{\frac{3}{4}} \varphi^{\frac{3}{2}} & -2^{\frac{3}{4}} & 0 & 2^{\frac{3}{4}} \varphi^{\frac{3}{2}} & 0 & 2^{\frac{1}{2}} & -2^{\frac{1}{2}} & -2^{\frac{1}{2}} \varphi^{\frac{3}{2}} & -2^{\frac{1}{2}} & 0 & -2^{\frac{1}{2}} \varphi^{\frac{3}{2}} & 0 & -2^{\frac{1}{2}} & 2^{\frac{1}{2}} & 0 & 2^{\frac{1}{2}} & 2^{\frac{1}{2}} \varphi^{\frac{3}{2}} & 0 & 2^{\frac{1}{2}} \varphi^{\frac{3}{2}} & -2^{\frac{1}{2}} \varphi^{\frac{3}{2}} \\
			2 \varphi^{\frac{1}{2}} & 0 & 0 & 0 & 0 & 0 & 0 & 0 & 0 & 2 \varphi & 2 \varphi & 2 \varphi^{\frac{1}{2}} & -2 \varphi & 0 & -2 \varphi^{\frac{1}{2}} & 0 & 2 \varphi & 2 \varphi & 0 & -2 \varphi & -2 \varphi^{\frac{1}{2}} & 0 & 2 \varphi^{\frac{1}{2}} & 2 \varphi^{\frac{1}{2}} \\
			2^{\frac{1}{2}} \varphi^{\frac{3}{2}} & -2^{\frac{3}{4}} & 0 & 0 & 2^{\frac{3}{4}} \varphi^{\frac{3}{2}} & 2^{\frac{3}{4}} & 0 & -2^{\frac{3}{4}} \varphi^{\frac{3}{2}} & 0 & 2^{\frac{1}{2}} & -2^{\frac{1}{2}} & -2^{\frac{1}{2}} \varphi^{\frac{3}{2}} & -2^{\frac{1}{2}} & 0 & -2^{\frac{1}{2}} \varphi^{\frac{3}{2}} & 0 & -2^{\frac{1}{2}} & 2^{\frac{1}{2}} & 0 & 2^{\frac{1}{2}} & 2^{\frac{1}{2}} \varphi^{\frac{3}{2}} & 0 & 2^{\frac{1}{2}} \varphi^{\frac{3}{2}} & -2^{\frac{1}{2}} \varphi^{\frac{3}{2}} \\
			2^{\frac{1}{2}} \varphi^{\frac{1}{2}} & -2^{\frac{3}{4}} \varphi & 0 & 0 & -2^{\frac{3}{4}} \varphi^{\frac{1}{2}} & 2^{\frac{3}{4}} \varphi & 0 & 2^{\frac{3}{4}} \varphi^{\frac{1}{2}} & 0 & -2^{\frac{1}{2}} \varphi & 2^{\frac{1}{2}} \varphi & -2^{\frac{1}{2}} \varphi^{\frac{1}{2}} & 2^{\frac{1}{2}} \varphi & 0 & -2^{\frac{1}{2}} \varphi^{\frac{1}{2}} & 0 & 2^{\frac{1}{2}} \varphi & -2^{\frac{1}{2}} \varphi & 0 & -2^{\frac{1}{2}} \varphi & 2^{\frac{1}{2}} \varphi^{\frac{1}{2}} & 0 & 2^{\frac{1}{2}} \varphi^{\frac{1}{2}} & -2^{\frac{1}{2}} \varphi^{\frac{1}{2}} \\
			\varphi^{\frac{1}{2}} & 2^{\frac{1}{4}} \varphi & -2^{\frac{1}{4}} \varphi^{\frac{1}{2}} & -2^{\frac{1}{4}} \varphi & 2^{\frac{1}{4}} \varphi^{\frac{1}{2}} & 2^{\frac{1}{4}} \varphi & -2^{\frac{1}{4}} \varphi & 2^{\frac{1}{4}} \varphi^{\frac{1}{2}} & -2^{\frac{1}{4}} \varphi^{\frac{1}{2}} & \varphi & \varphi & \varphi^{\frac{1}{2}} & \varphi & -2^{\frac{1}{2}} \varphi & \varphi^{\frac{1}{2}} & -2^{\frac{1}{2}} \varphi^{\frac{1}{2}} & \varphi & \varphi & -2^{\frac{1}{2}} \varphi & \varphi & \varphi^{\frac{1}{2}} & -2^{\frac{1}{2}} \varphi^{\frac{1}{2}} & \varphi^{\frac{1}{2}} & \varphi^{\frac{1}{2}} \\
			2^{\frac{1}{2}} \varphi^{\frac{3}{2}} & 0 & -2^{\frac{3}{4}} \varphi^{\frac{3}{2}} & -2^{\frac{3}{4}} & 0 & 0 & 2^{\frac{3}{4}} & 0 & 2^{\frac{3}{4}} \varphi^{\frac{3}{2}} & 2^{\frac{1}{2}} & -2^{\frac{1}{2}} & -2^{\frac{1}{2}} \varphi^{\frac{3}{2}} & 2^{\frac{1}{2}} & 0 & 2^{\frac{1}{2}} \varphi^{\frac{3}{2}} & 0 & -2^{\frac{1}{2}} & 2^{\frac{1}{2}} & 0 & -2^{\frac{1}{2}} & -2^{\frac{1}{2}} \varphi^{\frac{3}{2}} & 0 & 2^{\frac{1}{2}} \varphi^{\frac{3}{2}} & -2^{\frac{1}{2}} \varphi^{\frac{3}{2}} \\
			\varphi^{\frac{3}{2}} & 2^{\frac{1}{4}} & -2^{\frac{1}{4}} \varphi^{\frac{3}{2}} & 2^{\frac{1}{4}} & -2^{\frac{1}{4}} \varphi^{\frac{3}{2}} & 2^{\frac{1}{4}} & 2^{\frac{1}{4}} & -2^{\frac{1}{4}} \varphi^{\frac{3}{2}} & -2^{\frac{1}{4}} \varphi^{\frac{3}{2}} & -1 & -1 & \varphi^{\frac{3}{2}} & -1 & -2^{\frac{1}{2}} & \varphi^{\frac{3}{2}} & 2^{\frac{1}{2}} \varphi^{\frac{3}{2}} & -1 & -1 & -2^{\frac{1}{2}} & -1 & \varphi^{\frac{3}{2}} & 2^{\frac{1}{2}} \varphi^{\frac{3}{2}} & \varphi^{\frac{3}{2}} & \varphi^{\frac{3}{2}} \\
			2^{\frac{1}{2}} \varphi^{\frac{1}{2}} & 0 & -2^{\frac{3}{4}} \varphi^{\frac{1}{2}} & 2^{\frac{3}{4}} \varphi & 0 & 0 & -2^{\frac{3}{4}} \varphi & 0 & 2^{\frac{3}{4}} \varphi^{\frac{1}{2}} & -2^{\frac{1}{2}} \varphi & 2^{\frac{1}{2}} \varphi & -2^{\frac{1}{2}} \varphi^{\frac{1}{2}} & -2^{\frac{1}{2}} \varphi & 0 & 2^{\frac{1}{2}} \varphi^{\frac{1}{2}} & 0 & 2^{\frac{1}{2}} \varphi & -2^{\frac{1}{2}} \varphi & 0 & 2^{\frac{1}{2}} \varphi & -2^{\frac{1}{2}} \varphi^{\frac{1}{2}} & 0 & 2^{\frac{1}{2}} \varphi^{\frac{1}{2}} & -2^{\frac{1}{2}} \varphi^{\frac{1}{2}} \\
			\varphi^{\frac{3}{2}} & -2^{\frac{1}{4}} & -2^{\frac{1}{4}} \varphi^{\frac{3}{2}} & 2^{\frac{1}{4}} & 2^{\frac{1}{4}} \varphi^{\frac{3}{2}} & -2^{\frac{1}{4}} & 2^{\frac{1}{4}} & 2^{\frac{1}{4}} \varphi^{\frac{3}{2}} & -2^{\frac{1}{4}} \varphi^{\frac{3}{2}} & -1 & -1 & \varphi^{\frac{3}{2}} & -1 & 2^{\frac{1}{2}} & \varphi^{\frac{3}{2}} & -2^{\frac{1}{2}} \varphi^{\frac{3}{2}} & -1 & -1 & 2^{\frac{1}{2}} & -1 & \varphi^{\frac{3}{2}} & -2^{\frac{1}{2}} \varphi^{\frac{3}{2}} & \varphi^{\frac{3}{2}} & \varphi^{\frac{3}{2}} \\
			\varphi^{\frac{1}{2}} & -2^{\frac{1}{4}} \varphi & -2^{\frac{1}{4}} \varphi^{\frac{1}{2}} & -2^{\frac{1}{4}} \varphi & -2^{\frac{1}{4}} \varphi^{\frac{1}{2}} & -2^{\frac{1}{4}} \varphi & -2^{\frac{1}{4}} \varphi & -2^{\frac{1}{4}} \varphi^{\frac{1}{2}} & -2^{\frac{1}{4}} \varphi^{\frac{1}{2}} & \varphi & \varphi & \varphi^{\frac{1}{2}} & \varphi & 2^{\frac{1}{2}} \varphi & \varphi^{\frac{1}{2}} & 2^{\frac{1}{2}} \varphi^{\frac{1}{2}} & \varphi & \varphi & 2^{\frac{1}{2}} \varphi & \varphi & \varphi^{\frac{1}{2}} & 2^{\frac{1}{2}} \varphi^{\frac{1}{2}} & \varphi^{\frac{1}{2}} & \varphi^{\frac{1}{2}} \\
			2^{\frac{3}{2}} \varphi^{\frac{1}{2}} & 0 & 0 & 0 & 0 & 0 & 0 & 0 & 0 & -2^{\frac{3}{2}} \varphi & 2^{\frac{3}{2}} \varphi & -2^{\frac{3}{2}} \varphi^{\frac{1}{2}} & 0 & -2 \varphi & 0 & 2 \varphi^{\frac{1}{2}} & -2^{\frac{1}{2}} \varphi & 2^{\frac{1}{2}} \varphi & 2 \varphi & 0 & 0 & -2 \varphi^{\frac{1}{2}} & -2^{\frac{1}{2}} \varphi^{\frac{1}{2}} & 2^{\frac{1}{2}} \varphi^{\frac{1}{2}} \\
			2^{\frac{3}{2}} \varphi^{\frac{1}{2}} & 0 & 0 & 0 & 0 & 0 & 0 & 0 & 0 & -2^{\frac{3}{2}} \varphi & 2^{\frac{3}{2}} \varphi & -2^{\frac{3}{2}} \varphi^{\frac{1}{2}} & 0 & 2 \varphi & 0 & -2 \varphi^{\frac{1}{2}} & -2^{\frac{1}{2}} \varphi & 2^{\frac{1}{2}} \varphi & -2 \varphi & 0 & 0 & 2 \varphi^{\frac{1}{2}} & -2^{\frac{1}{2}} \varphi^{\frac{1}{2}} & 2^{\frac{1}{2}} \varphi^{\frac{1}{2}} \\
			4 \varphi^{\frac{1}{2}} & 0 & 0 & 0 & 0 & 0 & 0 & 0 & 0 & 4 \varphi & 4 \varphi & 4 \varphi^{\frac{1}{2}} & 0 & 0 & 0 & 0 & -2 \varphi & -2 \varphi & 0 & 0 & 0 & 0 & -2 \varphi^{\frac{1}{2}} & -2 \varphi^{\frac{1}{2}} \\
			2^{\frac{3}{2}} \varphi^{\frac{3}{2}} & 0 & 0 & 0 & 0 & 0 & 0 & 0 & 0 & 2^{\frac{3}{2}} & -2^{\frac{3}{2}} & -2^{\frac{3}{2}} \varphi^{\frac{3}{2}} & 0 & 2 & 0 & 2 \varphi^{\frac{3}{2}} & 2^{\frac{1}{2}} & -2^{\frac{1}{2}} & -2 & 0 & 0 & -2 \varphi^{\frac{3}{2}} & -2^{\frac{1}{2}} \varphi^{\frac{3}{2}} & 2^{\frac{1}{2}} \varphi^{\frac{3}{2}} \\
			2^{\frac{3}{2}} \varphi^{\frac{3}{2}} & 0 & 0 & 0 & 0 & 0 & 0 & 0 & 0 & 2^{\frac{3}{2}} & -2^{\frac{3}{2}} & -2^{\frac{3}{2}} \varphi^{\frac{3}{2}} & 0 & -2 & 0 & -2 \varphi^{\frac{3}{2}} & 2^{\frac{1}{2}} & -2^{\frac{1}{2}} & 2 & 0 & 0 & 2 \varphi^{\frac{3}{2}} & -2^{\frac{1}{2}} \varphi^{\frac{3}{2}} & 2^{\frac{1}{2}} \varphi^{\frac{3}{2}} \\
			4 \varphi^{\frac{3}{2}} & 0 & 0 & 0 & 0 & 0 & 0 & 0 & 0 & -4 & -4 & 4 \varphi^{\frac{3}{2}} & 0 & 0 & 0 & 0 & 2 & 2 & 0 & 0 & 0 & 0 & -2 \varphi^{\frac{3}{2}} & -2 \varphi^{\frac{3}{2}}
		\end{pmatrix}
	$}
\end{aligned}
\end{equation}
\end{sidewaysfigure}
Here, $B_\alpha^{(j;a)}=\psi_{\alpha}^{(j;a)}/\sqrt{S_{0j}}$ is the matrix of boundary state coefficients. Hence, the 24 boundary states can be obtained from this matrix as
\begin{equation}
	\ket{\alpha}=\sum_{(j;a)\in\mathcal{E}^\mathcal{U}}B_\alpha^{(j;a)}\Ish{j;a},\quad \alpha\in\mathcal{B}^\mathcal{U}.
\end{equation}
The first 18 rows of $B_\alpha^{(j;a)}$ correspond to the product Cardy states $\mathcal{B}^\mathcal{A}$, ordered as follows:
\begin{equation}
\begin{aligned}
	\mathcal{B}^\mathcal{A}=
	\Big\{
	&|0,0\rangle ,|0,\frac{3}{80}\rangle ,|0,\frac{1}{10}\rangle ,|0,\frac{7}{16}\rangle ,|0,\frac{3}{5}\rangle ,|0,\frac{3}{2}\rangle ,|\frac{1}{16},0\rangle ,|\frac{1}{16},\frac{3}{80}\rangle ,|\frac{1}{16},\frac{1}{10}\rangle ,
	\\
	&|\frac{1}{16},\frac{7}{16}\rangle ,|\frac{1}{16},\frac{3}{5}\rangle ,|\frac{1}{16},\frac{3}{2}\rangle ,|\frac{1}{2},0\rangle ,|\frac{1}{2},\frac{3}{80}\rangle ,|\frac{1}{2},\frac{1}{10}\rangle ,|\frac{1}{2},\frac{7}{16}\rangle ,|\frac{1}{2},\frac{3}{5}\rangle ,|\frac{1}{2},\frac{3}{2}\rangle
	\Big\}.
\end{aligned}
\end{equation}
The last six rows of $B_\alpha^{(j;a)}$ correspond to symmetry breaking boundaries. Since we argued around~\eqref{eq:no_inter} that there is no intermediate algebra between $\mathcal{A}$ and $\mathcal{U}$, these boundaries cannot preserve a larger symmetry within the bulk chiral algebra. We write down these states explicitly for later use. They constitute the main result of this paper:
\begin{align}
	\lvert B_1\rangle ={}&
	\frac{1}{(2+\varphi)^{1/4}}\Bigl[
	\sqrt2\bigl(\Ish{0}-\Ish{3}\bigr)
	-\sqrt{2\varphi}\bigl(\Ish{\frac85}-\Ish{\frac{13}{5}}\bigr)
	\notag\\
	&{}-\sqrt{\varphi}\bigl(
	\Ish{\frac{1}{10};(\frac{1}{16},\frac{3}{80})}
	-\Ish{\frac{11}{10};(\frac{1}{16},\frac{3}{80})}
	\bigr)
	+\Ish{\frac12;(\frac{1}{16},\frac{7}{16})}
	-\Ish{\frac32;(\frac{1}{16},\frac{7}{16})}
	\notag\\
	&{}-\sqrt{\frac{\varphi}{2}}\bigl(
	\Ish{\frac35;(0,\frac35)}
	-\Ish{\frac35;(\frac12,\frac{1}{10})}
	\bigr)
	-\frac{1}{\sqrt2}\bigl(
	\Ish{2;(0,0)}-\Ish{2;(\frac12,\frac32)}
	\bigr)
	\Bigr].
	\label{eq:B1}\\
	\lvert B_2\rangle ={}&
	\frac{1}{(2+\varphi)^{1/4}}\Bigl[
	\sqrt2\bigl(\Ish{0}-\Ish{3}\bigr)
	-\sqrt{2\varphi}\bigl(\Ish{\frac85}-\Ish{\frac{13}{5}}\bigr)
	\notag\\
	&{}+\sqrt{\varphi}\bigl(
	\Ish{\frac{1}{10};(\frac{1}{16},\frac{3}{80})}
	-\Ish{\frac{11}{10};(\frac{1}{16},\frac{3}{80})}
	\bigr)
	-\Ish{\frac12;(\frac{1}{16},\frac{7}{16})}
	+\Ish{\frac32;(\frac{1}{16},\frac{7}{16})}
	\notag\\
	&{}-\sqrt{\frac{\varphi}{2}}\bigl(
	\Ish{\frac35;(0,\frac35)}
	-\Ish{\frac35;(\frac12,\frac{1}{10})}
	\bigr)
	-\frac{1}{\sqrt2}\bigl(
	\Ish{2;(0,0)}-\Ish{2;(\frac12,\frac32)}
	\bigr)
	\Bigr].
	\label{eq:B2}\\
	\lvert B_3\rangle ={}&
	\frac{1}{(2+\varphi)^{1/4}}\Bigl[
	2\bigl(\Ish{0}+\Ish{3}\bigr)
	+2\sqrt{\varphi}\bigl(
	\Ish{\frac85}+\Ish{\frac{13}{5}}
	\bigr)
	\notag\\
	&{}-\sqrt{\varphi}\bigl(
	\Ish{\frac35;(0,\frac35)}
	+\Ish{\frac35;(\frac12,\frac{1}{10})}
	\bigr)
	-\bigl(
	\Ish{2;(0,0)}+\Ish{2;(\frac12,\frac32)}
	\bigr)
	\Bigr].
	\label{eq:B3}\\
	\lvert B_4\rangle ={}&
	\frac{1}{(2+\varphi)^{1/4}}\Bigl[
	\sqrt2\varphi\bigl(\Ish{0}-\Ish{3}\bigr)
	+\sqrt{\frac{2}{\varphi}}\bigl(
	\Ish{\frac85}-\Ish{\frac{13}{5}}
	\bigr)
	\notag\\
	&{}+\sqrt{\frac{1}{\varphi}}\bigl(
	\Ish{\frac{1}{10};(\frac{1}{16},\frac{3}{80})}
	-\Ish{\frac{11}{10};(\frac{1}{16},\frac{3}{80})}
	\bigr)
	+\varphi\bigl(
	\Ish{\frac12;(\frac{1}{16},\frac{7}{16})}
	-\Ish{\frac32;(\frac{1}{16},\frac{7}{16})}
	\bigr)
	\notag\\
	&{}+\sqrt{\frac{1}{2\varphi}}\bigl(
	\Ish{\frac35;(0,\frac35)}
	-\Ish{\frac35;(\frac12,\frac{1}{10})}
	\bigr)
	-\frac{\varphi}{\sqrt2}\bigl(
	\Ish{2;(0,0)}-\Ish{2;(\frac12,\frac32)}
	\bigr)
	\Bigr].
	\label{eq:B4}\\
	\lvert B_5\rangle ={}&
	\frac{1}{(2+\varphi)^{1/4}}\Bigl[
	\sqrt2\varphi\bigl(\Ish{0}-\Ish{3}\bigr)
	+\sqrt{\frac{2}{\varphi}}\bigl(
	\Ish{\frac85}-\Ish{\frac{13}{5}}
	\bigr)
	\notag\\
	&{}-\sqrt{\frac{1}{\varphi}}\bigl(
	\Ish{\frac{1}{10};(\frac{1}{16},\frac{3}{80})}
	-\Ish{\frac{11}{10};(\frac{1}{16},\frac{3}{80})}
	\bigr)
	-\varphi\bigl(
	\Ish{\frac12;(\frac{1}{16},\frac{7}{16})}
	-\Ish{\frac32;(\frac{1}{16},\frac{7}{16})}
	\bigr)
	\notag\\
	&{}+\sqrt{\frac{1}{2\varphi}}\bigl(
	\Ish{\frac35;(0,\frac35)}
	-\Ish{\frac35;(\frac12,\frac{1}{10})}
	\bigr)
	-\frac{\varphi}{\sqrt2}\bigl(
	\Ish{2;(0,0)}-\Ish{2;(\frac12,\frac32)}
	\bigr)
	\Bigr].
	\label{eq:B5}\\
	\lvert B_6\rangle ={}&
	\frac{1}{(2+\varphi)^{1/4}}\Bigl[
	2\varphi\bigl(\Ish{0}+\Ish{3}\bigr)
	-2\sqrt{\frac{1}{\varphi}}\bigl(
	\Ish{\frac85}+\Ish{\frac{13}{5}}
	\bigr)
	\notag\\
	&{}+\sqrt{\frac{1}{\varphi}}\bigl(
	\Ish{\frac35;(0,\frac35)}
	+\Ish{\frac35;(\frac12,\frac{1}{10})}
	\bigr)
	-\varphi\bigl(
	\Ish{2;(0,0)}+\Ish{2;(\frac12,\frac32)}
	\bigr)
	\Bigr].
	\label{eq:B6}
\end{align}
Here, $\varphi=\frac{1+\sqrt{5}}{2}$. Their closed channel partition functions with identical boundaries are:
	\begin{align}
		\mathcal{Z}^{\text{cl}}_{B_1|B_1}
		=\mathcal{Z}^{\text{cl}}_{B_2|B_2}
		=&\frac{1}{\sqrt{2+\varphi}}
		\Bigl[
		\varphi\bigl(
		\chi^{\mathcal W}_{1/10}
		+\chi^{\mathcal W}_{3/5}
		+2\chi^{\mathcal W}_{8/5}
		\bigr)
		+2\chi^{\mathcal W}_0
		+\chi^{\mathcal W}_{1/2}
		+\chi^{\mathcal W}_2
		\Bigr],
		\\[6pt]
		\mathcal{Z}^{\text{cl}}_{B_3|B_3}
		=&\frac{2}{\sqrt{2+\varphi}}
		\Bigl[
		\varphi\bigl(
		\chi^{\mathcal W}_{3/5}
		+2\chi^{\mathcal W}_{8/5}
		\bigr)
		+2\chi^{\mathcal W}_0
		+\chi^{\mathcal W}_2
		\Bigr],
		\\[6pt]
		\mathcal{Z}^{\text{cl}}_{B_4|B_4}
		=\mathcal{Z}^{\text{cl}}_{B_5|B_5}
		=&\frac{1}{\sqrt{2+\varphi}}
		\Bigl[
		\frac{1}{\varphi}\bigl(
		\chi^{\mathcal W}_{1/10}
		+\chi^{\mathcal W}_{3/5}
		+2\chi^{\mathcal W}_{8/5}
		\bigr)
		+\varphi^2\bigl(
		2\chi^{\mathcal W}_0
		+\chi^{\mathcal W}_{1/2}
		+\chi^{\mathcal W}_2
		\bigr)
		\Bigr],
		\\[6pt]
		\mathcal{Z}^{\text{cl}}_{B_6|B_6}
		=&\frac{2}{\sqrt{2+\varphi}}
		\Bigl[
		\frac{1}{\varphi}\bigl(
		\chi^{\mathcal W}_{3/5}
		+2\chi^{\mathcal W}_{8/5}
		\bigr)
		+\varphi^2\bigl(
		2\chi^{\mathcal W}_0
		+\chi^{\mathcal W}_2
		\bigr)
		\Bigr],
	\end{align}
and their open channel counterparts are:
	\begin{align}
		\mathcal{Z}^{\text{op}}_{B_1|B_1}
		=\mathcal{Z}^{\text{op}}_{B_2|B_2}
		=&\chi_0^{\mathcal W}
		+\chi_{1/9}^{\mathcal W}
		+\chi_{1/2}^{\mathcal W}
		+\chi_{7/9}^{\mathcal W}
		+\chi_{13/9}^{\mathcal W}
		+2\chi_2^{\mathcal W},
		\\
		\mathcal{Z}^{\text{op}}_{B_3|B_3}
		=&\chi_0^{\mathcal W}
		+2\chi_{1/9}^{\mathcal W}
		+3\chi_{1/2}^{\mathcal W}
		+2\chi_{7/9}^{\mathcal W}
		+2\chi_{13/9}^{\mathcal W}
		+2\chi_2^{\mathcal W},
		\\
		\mathcal{Z}^{\text{op}}_{B_4|B_4}
		=\mathcal{Z}^{\text{op}}_{B_5|B_5}
		=&\chi_0^{\mathcal W}
		+\chi_{2/45}^{\mathcal W}
		+\chi_{1/10}^{\mathcal W}
		+\chi_{1/9}^{\mathcal W}
		+\chi_{17/45}^{\mathcal W}
		+\chi_{1/2}^{\mathcal W}
		\notag
		\\
		&
		+2\chi_{3/5}^{\mathcal W}
		+\chi_{32/45}^{\mathcal W}
		+\chi_{7/9}^{\mathcal W}
		+\chi_{13/9}^{\mathcal W}
		+\chi_{8/5}^{\mathcal W}
		+2\chi_2^{\mathcal W},
		\\
		\mathcal{Z}^{\text{op}}_{B_6|B_6}
		=&\chi_0^{\mathcal W}
		+2\chi_{2/45}^{\mathcal W}
		+3\chi_{1/10}^{\mathcal W}
		+2\chi_{1/9}^{\mathcal W}
		+2\chi_{17/45}^{\mathcal W}
		+3\chi_{1/2}^{\mathcal W}
		\notag
		\\
		&
		+2\chi_{3/5}^{\mathcal W}
		+2\chi_{32/45}^{\mathcal W}
		+2\chi_{7/9}^{\mathcal W}
		+2\chi_{13/9}^{\mathcal W}
		+\chi_{8/5}^{\mathcal W}
		+2\chi_2^{\mathcal W}.
	\end{align}
	
Here, using~\eqref{UtoW}, we express all these partition functions solely in terms of ordinary $\mathcal{W}_3$ characters, $\hat{\chi}^{(1,1)}$, denoted by $\chi^{\mathcal{W}}$ in the equations above. Therefore, although these boundaries preserve only the $\mathcal{U}$ chiral symmetry generated by local bulk currents, rather than $\mathcal{W}_3$ symmetry, $\mathcal{W}_3$ still plays some role in their spectra. It was suggested in~\cite{Antunes:2026kqb,Chen:2026byx} that phantom currents may be present on these boundaries while being nonlocal in the bulk. However, such currents cannot enter the gluing condition~\eqref{eq:gluing}, which is the starting point for our classification.

\subsection{Transmission and topological defect dressings}
\label{sec:3.3}
The reflection and transmission coefficients are among the most important physical quantities characterizing a conformal interface~\cite{Quella:2006de}. Consider two CFTs with stress-energy tensors $T^1$ and $T^2$ and central charges $c_1$ and $c_2$, respectively. Let $\alpha$ denote the boundary state corresponding to a conformal interface between these two theories. We define the following matrix:
\begin{equation}
	\label{eq:R}
	R_{ij}
	=\frac{\langle 0|L_2^i\bar{L}_2^j|\alpha\rangle}
	{\langle 0|\alpha\rangle}.
\end{equation}
From $T^1$ and $T^2$, we can construct two spin-two currents:
\begin{equation}
	T=T^1+T^2,\quad T'=c_2T^1-c_1T^2.
\end{equation}
Using the Virasoro algebra, it is straightforward to verify that the corresponding states at conformal weight $h=2$ are orthonormal:
\begin{equation}
	\ket{T}=\sqrt{\frac{2}{c_1+c_2}}\left(L_{-2}^1+L^2_{-2}\right)\ket{0},\quad \ket{T'}=\sqrt{\frac{2}{c_1c_2(c_1+c_2)}}T'_{-2}\ket{0}.
\end{equation}
The anti-holomorphic counterparts are defined in the same way. It can be shown that the matrix in~\eqref{eq:R} is completely determined by a single parameter $\omega_\alpha$~\cite{Quella:2006de}:
\begin{equation}
	R=\frac{c_1c_2}{2(c_1+c_2)}
	\left[
	\begin{pmatrix}
		\frac{c_1}{c_2} & 1\\
		1 & \frac{c_2}{c_1}
	\end{pmatrix}
	+\omega_\alpha
	\begin{pmatrix}
		1 & -1\\
		-1 & 1
	\end{pmatrix}
	\right],
\end{equation}
where
\begin{equation}
	\omega_\alpha
	=
	\frac{2}{c_1c_2(c_1+c_2)}
	\frac{\langle 0|T'_2\bar{T'}_2|\alpha\rangle}
	{\langle 0|\alpha\rangle}.
\end{equation}
The reflection and transmission coefficients are then defined as
\begin{equation}
	\label{eq:RT}
	\begin{aligned}
		\mathcal{R}
		&=\frac{2}{c_1+c_2}(R_{11}+R_{22})
		=\frac{c_1^2+2c_1c_2\omega_\alpha+c_2^2}
		{(c_1+c_2)^2},\\
		\mathcal{T}
		&=\frac{2}{c_1+c_2}(R_{12}+R_{21})
		=\frac{2c_1c_2(1-\omega_\alpha)}
		{(c_1+c_2)^2}.
	\end{aligned}
\end{equation}

To determine the transmission coefficients, we need to calculate the overlap between $\ket{\alpha}$ and $\ket{T'}\otimes\ket{\bar T'}$. Among the Ishibashi states of $\mathcal{U}$, only $\Ish{0}$, $\Ish{2;(0,0)}$, and $\Ish{2;(\frac12,\frac32)}$ contain states of conformal weights $(2,2)$ and hence can potentially have nonvanishing overlaps with $\ket{T'}\otimes\ket{\bar T'}$. However, the state of conformal weights $(2,2)$ in $\Ish{\frac12}\otimes\Ish{\frac32}$ is generated by the simple current $J$. Using \eqref{eq:TJOPE1} and \eqref{eq:TJOPE2}, one can verify that the OPE between $J$ and $T'$ has no fourth order pole, implying that their two point function vanishes, $\braket{T'}{J}=0$.
\begin{equation}
	T'(z)J(w)\sim
	-\frac{2J(w)}{5(z-w)^2}
	-\frac{\frac{9i\sqrt{210}}5W(w)+\frac15\partial J(w)}{z-w}.
\end{equation}
Since $\Ish{3}$ contains no state of conformal weights $(2,2)$, it follows from our convention in~\eqref{eq:ishibashi_convention} that the overlap between $\Ish{2;(\frac12,\frac32)}$ and $\ket{T'}\otimes\ket{\bar T'}$ vanishes. Using the explicit construction of Ishibashi states in~\eqref{eq:ishibashi}, we obtain:
\begin{equation}
	\label{eq:inclusion}
	\Ish{0}_{\text{Ising}}\otimes\Ish{0}_{\text{TIM}}\supset\ket{T}\otimes U\ket{\bar T}+\ket{T'}\otimes U\ket{\bar T'},\quad
	\Ish{0}_{\mathcal{U}}\supset \ket{T}\otimes U\ket{\bar T}.
\end{equation}
The first inclusion follows from the fact that the vacuum module of $\mathcal{A}$ is generated by the stress-energy tensors $T^{\mathrm{Ising}}$ and $T^{\mathrm{TIM}}$, while the second follows from the fact that the vacuum module of $\mathcal{U}$ is generated by the total stress-energy tensor $T$ and the higher-spin currents. Furthermore, our convention in~\eqref{eq:ishibashi_convention} implies that the overlap between $\Ish{2;(0,0)}$ and $\ket{T'}\otimes\ket{\bar T'}$ is unity. Since $\braket{0}{\alpha}$ is simply given by the coefficient of $\Ish{0}$ in the boundary state $\ket{\alpha}$, it follows that, in our case, $\omega_\alpha$ can be directly obtained from the coefficient of $\Ish{2;(0,0)}$, normalized by that of $\Ish{0}$:
\begin{equation}
	\label{eq:omega}
	\omega_\alpha =\frac{B_\alpha^{\left(2;(0,0)\right)}}{B_\alpha^0}.
\end{equation}
Applying~\eqref{eq:RT} and~\eqref{eq:omega} to the symmetry breaking boundaries in~\eqref{eq:B1}--\eqref{eq:B6}, we obtain:
\begin{equation}
	\label{eq:transmission}
	\mathcal{T}_1=\mathcal{T}_2=\mathcal{T}_3=\mathcal{T}_4=\mathcal{T}_5=\mathcal{T}_6=\frac{35}{48}.
\end{equation}

This result strongly suggests that these boundaries can be related by dressing them with topological defect lines. Recall that in a diagonal CFT, a topological defect line can be constructed as
\begin{equation}
	\mathcal{D}_{i}=\sum_{j}\frac{S_{ij}}{S_{0j}}\|j\|.
\end{equation}
Here, $\|j\|$ denotes the projection operator onto the diagonal bulk sector $\mathcal{H}_j\otimes\overline{\mathcal{H}}_j$. As discussed around~\eqref{eq:ishibashi_convention}, we have chosen an Ishibashi basis compatible with the branching rules. The advantage of this choice is that the topological defect lines of both the Ising and TIM models act diagonally on the Ishibashi states.~\footnote{For notational simplicity, we have suppressed the source labels $(h_{\mathrm{Ising}},h_{\mathrm{TIM}})$ for multiplicity free modules, which can be recovered from the branching rules in~\eqref{eq:branch rule}.}
\begin{equation}
	\begin{aligned}
		\mathcal{D}^{\mathrm{Ising}}_i\boxtimes
		\mathcal{D}^{\mathrm{TIM}}_j
		\Ish{k;(h_{\mathrm{Ising}},h_{\mathrm{TIM}})}
		&=
		\sum_{h_1,h_2}
		\frac{S^{(3,4)}_{i,h_1}}{S^{(3,4)}_{0,h_1}}
		\frac{S^{(4,5)}_{j,h_2}}{S^{(4,5)}_{0,h_2}}
		\bigl\|(h_1,h_2)\bigr\|
		k;(h_{\mathrm{Ising}},h_{\mathrm{TIM}})\rangle\!\rangle
		\\
		&=
		\frac{S^{(3,4)}_{i,h_{\mathrm{Ising}}}}
		{S^{(3,4)}_{0,h_{\mathrm{Ising}}}}
		\frac{S^{(4,5)}_{j,h_{\mathrm{TIM}}}}
		{S^{(4,5)}_{0,h_{\mathrm{TIM}}}}
		\Ish{k;(h_{\mathrm{Ising}},h_{\mathrm{TIM}})}.
	\end{aligned}
\end{equation}

Using the above formula and the modular $S$ matrices computed from~\eqref{eq:Sminimal}, we find that all symmetry breaking boundary states in~\eqref{eq:B1}--\eqref{eq:B6} are related by the following topological defect line dressings, which explains why they have the same transmission coefficient:
\begin{equation}
\begin{gathered}
	\ket{B_2}=	\mathcal{D}^{\mathrm{Ising}}_{\epsilon} \ket{B_1},\quad
	\ket{B_3}=	\mathcal{D}^{\mathrm{Ising}}_{\sigma} \ket{B_1},\quad
	\ket{B_4}=	\mathcal{D}^{\mathrm{TIM}}_{\epsilon'} \ket{B_1},
	\\
	\ket{B_5}=	\mathcal{D}^{\mathrm{Ising}}_{\epsilon}\boxtimes \mathcal{D}^{\mathrm{TIM}}_{\epsilon'}\ket{B_1},\quad
	\ket{B_6}=	\mathcal{D}^{\mathrm{Ising}}_{\sigma}\boxtimes \mathcal{D}^{\mathrm{TIM}}_{\epsilon'}\ket{B_1}.
\end{gathered}
\end{equation}
Here $\mathcal{D}_{\sigma}^{\text{Ising}}$, $\mathcal{D}_{\epsilon}^{\text{Ising}}$ and $\mathcal{D}_{\epsilon'}^{\text{TIM}}$ are known as Kramers-Wannier duality, $\mathbb{Z}_2$ symmetry and Fibonacci defect line respectively~\cite{Chang:2018iay}. Note that the transmission coefficients in~\eqref{eq:transmission} agree with those of the RG domain wall studied in~\cite{Furuta:2025ahl}. In the next section, we will identify one of the boundary states in~\eqref{eq:B1}--\eqref{eq:B6} with the RG domain wall.
\subsection{Identification with the RG domain wall}
\label{sec:3.4}
Following~\cite{Gaiotto:2012np}, we first introduce the notation used in the following discussion. According to the coset construction of Virasoro minimal models in~\eqref{eq:coset}, each primary field in $\mathcal{M}(k+1,k+2)$ can be labeled by three $\widehat{\mathfrak{su}}(2)$ representations as $[r,d;s]$, modulo the field identification $[r,d;s]\sim[k+1-r,\,3-d;\,k+2-s]$. Here $r$ and $s$ are Kac labels, with $1\leq r\leq k$ and $1\leq s\leq k+1$. The label $d\in\{1,2\}$ is determined by the selection rule $r+d+s=1\mod 2$. The conformal weights are given by
\begin{equation}
		h^{(k+1,k+2)}_{(r,s)}
		=\frac{\bigl((k+2)r-(k+1)s\bigr)^2-1}{4(k+1)(k+2)}.
\end{equation}
Similarly, from the coset construction of $\mathcal{B}$ in~\eqref{eq:defB}, its primary fields can be labeled by four $\widehat{\mathfrak{su}}(2)$ representations as $[t,d,\tilde d;s]$. The branching rules from $\mathcal{B}$ to $\mathcal{A}$ are then given by:
\begin{equation}
	\phi^\mathcal{B}_{[t,d,\tilde{d};s]}=\bigoplus_{\substack{1\le r\le k+1\\
			t+d+r=1\ (\mathrm{mod}\ 2)}}\phi^{(k+1,k+2)}_{[t,d;r]}\otimes\phi^{(k+2,k+3)}_{[r,\tilde d;s]}.
\end{equation}

The Ising and TIM models are adjacent Virasoro minimal models
connected by Zamolodchikov's RG flow~\cite{Zamolodchikov:1987ti,Zamolodchikov:1991vx}. Starting from TIM in the UV,
we turn on the relevant perturbation only in one half plane.
In the IR limit, the perturbed half plane flows to the Ising model,
while the other half remains TIM. This naturally gives rise to
a conformal interface between the two theories, known as the
RG domain wall. For adjacent minimal models, Gaiotto proposed
the following boundary state describing the RG domain wall
in the folded picture:
\begin{equation}
	\label{eq:RG}
	\begin{aligned}
		|RG\rangle
		=\sum_{\substack{t,r,s\\s-t\in2\mathbb Z}}
		\frac{\sqrt{S_{1,t}^{(k-1)}S_{1,s}^{(k+1)}}}{S_{1,r}^{(k)}}
		\left(\|\mathcal{A},t,r,s\| t,1,1,s;\mathcal B,\mathbb{Z}_2\rangle\!\rangle
		+\|\mathcal{A},t,r,s\| t,2,2,s;\mathcal B,\mathbb{Z}_2\rangle\!\rangle\right),
	\end{aligned}
\end{equation}
where
\begin{equation}
	\label{eq:SWZW}
	S_{n,m}^{(k)}=\sqrt{\frac{2}{k+2}}\sin\frac{\pi nm}{k+2},
\end{equation}
and $\|\mathcal{A},t,r,s\| $ projects onto the $[t,d;r]\times [r,\tilde d;s]$ sector of $\mathcal{A}$. Here, $\Ish{t,d,\tilde{d},s;\mathcal B,\mathbb{Z}_2}$ denotes the $\mathbb{Z}_2$-twisted Ishibashi state associated with the automorphism exchanging the last two $\widehat{\mathfrak{su}}(2)$ factors in the numerator of the coset construction of $\mathcal{B}$. We emphasize that the Ishibashi states are defined with respect to the diagonal CFT with chiral algebra $\mathcal{B}$. Thus, the $\mathbb{Z}_2$-twisted Ishibashi states arise only from $\mathcal{B}$-modules invariant under $d\leftrightarrow\tilde d$, which explains the restriction $d=\tilde d=1,2$ in~\eqref{eq:RG}.

In what follows, we mainly use the free-boson realization introduced in Section~\ref{sec:2.1} to study the action of the twist on states. According to the embedding \eqref{eq:embbeding}, the three dimensional weight-two subspace of $U_{2A}$ is spanned by $T_{\mathrm{Ising}}$, $T_{\mathrm{TIM}}$, and the simple current $J$. $V_{\sqrt{2}A_2}$ also contains the stress-energy tensor of an auxiliary Potts theory~\cite{dong2003},
\begin{equation}
	T_{\mathrm{Potts}}
	=T_{\mathrm{bos}}
	-T_{\mathrm{Ising}}-T_{\mathrm{TIM}},
	\qquad
	T_{\mathrm{bos}}
	=\frac12\sum_{a=1}^{2}:j^aj^a:.
\end{equation}
These three stress-energy tensors have mutually regular OPEs and central charges $1/2$, $7/10$, and $4/5$, respectively. Thus, this embedding can be described using the following coset construction~\cite{lam2008faithful}
\begin{equation}
	\mathcal{B}\cong U_{2A}\cong\frac{V_{\sqrt{2}A_2}}{\mathcal{W}_3(c=4/5)},
\end{equation}
and any state constructed from $V_{\sqrt{2}A_2}$ can be represented as
\begin{equation}
	\ket{h_{\text{Ising}}}\otimes\ket{h_{\text{TIM}} }\otimes\ket{h_{\text{Potts}}}\sim j^{a_1}_{-n_1}\cdots j^{a_r}_{-n_r}V_p|0\rangle,
	\qquad
	h_{\rm bos}=\frac{p^2}{2}+\sum_{\ell=1}^r n_\ell=h_{\mathcal{B}}+h_{\text{Potts}}.
\end{equation}
Here, the values of $p$ are given in~\eqref{eq:plattice}. Since the three stress-energy tensors mutually commute, acting with Potts descendants preserves the Ising and TIM highest-weight conditions. This leads to a first subtlety: a given primary with weights $(h_{\mathrm{Ising}},h_{\mathrm{TIM}})$ admits infinitely many realizations in the lattice embedding, differing in their auxiliary Potts states. As mentioned in Section~\ref{sec:2.1}, the $\mathbb Z_2$ automorphism of $\mathcal B$ is induced by the lattice transformation $\rho:\alpha_1\leftrightarrow\alpha_2$. This transformation preserves $T_{\mathrm{Potts}}$ and hence the Potts conformal weights. We may therefore perform the calculation with a convenient allowed Potts factor. In what follows, we choose this factor to be $\ket{0}$ or $\ket{\frac25}$.
 
 However, there is another subtlety. The lattice VOA
$V_{\sqrt{2}A_2}$ admits a nontrivial symmetry
$\theta:\alpha_i\mapsto-\alpha_i$, which acts trivially on
$\mathcal{B}$. Consequently, $\rho$ and $\theta\rho$
induce the same automorphism of $\mathcal{B}$ but act
differently on $V_{\sqrt{2}A_2}$. Therefore, there is an
ambiguity in lifting $\rho$ from $\mathcal{B}$ to
$V_{\sqrt{2}A_2}$, which requires a choice of lift. The convention in~\cite{Gaiotto:2012np} is that the $\mathbb{Z}_2$ twist acts on $\ket{\frac{1}{16}}\otimes\ket{\frac{7}{16}}$\footnote{This field is denoted by $\psi$ in~\cite{Gaiotto:2012np}, whereas $\psi$ in this paper refers to the $h=\frac12$ field of the Ising model.} by flipping its sign. Using the following OPE between $w(\alpha_i)$ and the physical vertex operator $V_p:=\mathrm{e}^{ip\cdot\boldsymbol{\phi}}$,
\begin{equation}
	\label{eq:wVOPE}
	\begin{aligned}
		w(\alpha_i)(z)V_p(w)\sim{}&
		\frac{(\alpha_i\cdot p)^2}{2(z-w)^2}V_p(w)
		+\frac{\alpha_i\cdot p}{z-w}
		:(\alpha_i\cdot\boldsymbol{j})V_p:(w)\\
		&-(z-w)^{\sqrt{2}\alpha_i\cdot p}
		:e^{i\sqrt{2}\alpha_i\cdot\boldsymbol{\phi}(z)
			+ip\cdot\boldsymbol{\phi}(w)}:\\
		&-(z-w)^{-\sqrt{2}\alpha_i\cdot p}
		:e^{-i\sqrt{2}\alpha_i\cdot\boldsymbol{\phi}(z)
			+ip\cdot\boldsymbol{\phi}(w)}:,
	\end{aligned}
\end{equation}
one can check that for $i=0,2$,
\begin{equation}
		T_{\mathrm{Ising}}(z)\Psi_i(w)\sim\frac{\Psi_i(w)}{16(z-w)^2}+\mathcal{O}\bigl((z-w)^{-1}\bigr),\quad
		T_{\mathrm{TIM}}(z)\Psi_i(w)\sim\frac{7\Psi_i(w)}{16(z-w)^2}+\mathcal{O}\bigl((z-w)^{-1}\bigr).
\end{equation}
Here $\Psi_i:=V_{\alpha_i/\sqrt{2}}-V_{-\alpha_i/\sqrt{2}}$, and we choose the auxiliary Potts factor to be $\ket{0}$. However, $\Psi_2$ is not mapped to itself under $\rho$, which identifies $\Psi_0$ as the correct realization of $\ket{\frac{1}{16}}\otimes\ket{\frac{7}{16}}$. If we act only with $\rho:\alpha_1\leftrightarrow\alpha_2$, $\Psi_0$ is unchanged, so we must also apply $\theta: \alpha_i\mapsto-\alpha_i$. Thus, the $\mathbb{Z}_2$ twist of $\mathcal{B}$ should be implemented on $V_{\sqrt{2}A_2}$ by $\hat\rho=\theta\rho$. It is precisely this choice that ensures that the following spin-3 current of the auxiliary Potts model remains invariant ($\alpha_{i+3}:=\alpha_i$):
\begin{equation}
			Q={}\frac19:\prod_{i=0}^{2}\bigl((\alpha_{i+1}-\alpha_{i+2})\cdot\boldsymbol{j}\bigr):
			-\frac12\sum_{i=0}^{2}:\bigl((\alpha_{i+1}-\alpha_{i+2})\cdot\boldsymbol{j}\bigr)
			\left(V_{\sqrt2\alpha_i}+V_{-\sqrt2\alpha_i}\right):.
\end{equation}

To compare the RG domain wall with our boundaries, we need to decompose $\mathbb{Z}_2$-twisted $\mathcal{B}$ Ishibashi states into $\mathcal{U}$ Ishibashi states. We first generalize the definition in~\eqref{eq:ishibashi} to
\begin{equation}
	\Ish{i;a,b}=\sum_{N}\ket{i,a,N}\otimes U \ket{\overline{i,b,N}},
\end{equation}
which allows mixing between different copies of the same representation. Although such mixed terms appear in the decomposition of $\mathbb{Z}_2$-twisted $\mathcal{B}$ Ishibashi states, the projector $\|\mathcal{A},t,r,s\|$ removes these contributions from the final state $\ket{RG}$, since they are absent from the diagonal bulk spectrum, even though they satisfy the gluing condition. 

Consider a $\mathcal{B}$-module decomposed into $\mathcal{U}$-modules as $\phi_\mu^{\mathcal{B}}=\bigoplus_i M_{\mu i}\otimes \phi_i^\mathcal{U}$. Using the definition in~\eqref{eq:twish} and noting that the restriction of $\hat\rho$ to $\mathcal{U}$ is trivial, we find that $\Omega_{\hat\rho}$ preserves the $\mathcal{U}$-modules. Its action on descendants is completely determined by its action on primary states, i.e. it only acts on the multiplicity space $M_{\mu i}$. We obtain:
\begin{equation}
\begin{aligned}
		\Ish{\mu;\mathcal{B},\mathbb{Z}_2}&=\mathrm{id}\otimes \Omega_{\hat\rho}\Ish{\mu;\mathcal{B}}=\left(\mathrm{id}\otimes \Omega_{\hat\rho}\right)\sum_i \sum_{a\in M_{\mu i}}\Ish{i;a}\\
	&=\sum_{i,a,b,N}\ket{i;a,N}\otimes U {\left(\Omega_{\hat\rho}^{\mu i}\right)_a}^b\ket{\overline{i;b,N}}
	=\sum_i\sum_{a,b}{\left(\Omega_{\hat\rho}^{\mu i}\right)_a}^b\Ish{i;a,b}.
\end{aligned}
\end{equation}
Thus, the key is to determine the matrix representing the action of $\Omega_{\hat\rho}$ on the multiplicity spaces of the Ishibashi states, which is completely determined by its action on the corresponding lowest weight states. In our case, setting $k=2$ in~\eqref{eq:RG} gives
\begin{equation}
	\ket{RG}=\sum_{r=1}^{3}\frac{\sqrt{S^{(1)}_{1,1}}}{S^{(2)}_{1,r}}
	\left(
		\sqrt{S^{(3)}_{1,1}}\|\mathcal{A},1,r,1\|1,1;\tilde{\mathcal{B}},\mathbb{Z}_2\rangle\!\rangle
		+
		\sqrt{S^{(3)}_{1,3}}\|\mathcal{A},1,r,3\|1,3;\tilde{\mathcal{B}},\mathbb{Z}_2\rangle\!\rangle
	\right),
\end{equation}
where
\begin{equation}
	|t,s;\tilde{\mathcal{B}},\mathbb{Z}_2\rangle\!\rangle=|t,1,1,s;\mathcal{B},\mathbb{Z}_2\rangle\!\rangle+|t,2,2,s;\mathcal{B},\mathbb{Z}_2\rangle\!\rangle.
\end{equation}
The corresponding branching rules are
\begin{align}
	\phi^{\mathcal{B}}_{[1,1,1,1]}&=\phi^{\text{Ising}}_{[1,1;1]}\otimes\phi^{\text{TIM}}_{[1,1;1]}\oplus \phi^{\text{Ising}}_{[1,1;3]}\otimes\phi^{\text{TIM}}_{[3,1;1]}
	=\phi^{\mathcal{{U}}}_{0}\oplus \phi^{\mathcal{{U}}}_{3}\oplus\phi^{\mathcal{{U}}}_{(2;(0,0))}\oplus \phi^{\mathcal{{U}}}_{(2;(\frac12,\frac32))}
	\\
	\phi^{\mathcal{B}}_{[1,1,1,3]}&=\phi^{\text{Ising}}_{[1,1;1]}\otimes\phi^{\text{TIM}}_{[1,1;3]}\oplus \phi^{\text{Ising}}_{[1,1;3]}\otimes\phi^{\text{TIM}}_{[3,1;3]}
	=\phi^{\mathcal{{U}}}_{\frac85}\oplus \phi^{\mathcal{{U}}}_{\frac{13}{5}}\oplus \phi^{\mathcal{{U}}}_{(\frac35;(0,\frac35))}\oplus \phi^{\mathcal{{U}}}_{(\frac35;(\frac12,\frac1{10}))}
	\\
	\phi^{\mathcal{B}}_{[1,2,2,1]}&=\phi^{\text{Ising}}_{[1,2;2]}\otimes\phi^{\text{TIM}}_{[2,2;1]}=
	\phi^{\mathcal{{U}}}_{(\frac12;(\frac1{16},\frac7{16}))}\oplus \phi^{\mathcal{{U}}}_{(\frac32;(\frac1{16},\frac7{16}))}
	\\
	\phi^{\mathcal{B}}_{[1,2,2,3]}&=\phi^{\text{Ising}}_{[1,2;2]}\otimes\phi^{\text{TIM}}_{[2,2;3]}=
	\phi^{\mathcal{{U}}}_{(\frac1{10};(\frac1{16},\frac3{80}))}\oplus \phi^{\mathcal{{U}}}_{(\frac{11}{10};(\frac1{16},\frac3{80}))}
\end{align}

We first consider the multiplicity space of $h=2$ primaries. Since $\Omega_{\hat\rho}$ is completely determined by its action on the lowest weight states of the corresponding Ishibashi states, our strategy is to use the embedding into $V_{\sqrt{2}A_2}$ to explicitly identify and construct these states from the branching rules of $\mathcal{A}$ into $\mathcal{U}$, and then extract the corresponding action matrix by applying $\hat{\rho}$. Recall that around~\eqref{eq:inclusion}, we argued that
\begin{equation}
	\Ish{2;(0,0)}_\mathcal{U}\supset \ket{T'}\otimes U\ket{\bar T'},\quad \ket{T'}:=\frac{1}{\sqrt{21}}\left(7 L^{\text{Ising}}_{-2}-5L^{\text{TIM}}_{-2}\right) \ket{0}.
\end{equation}
And from $\ket{\frac12,\frac32}_\mathcal{A}\propto J\ket{0}$, we obtain
\begin{equation}
	\label{eq:embeddingJ}
	\Ish{2;(\frac12,\frac32)}_\mathcal{U}\supset \ket{J}\otimes U\ket{\bar J},\quad \ket{J}:=\frac{1}{6\sqrt{7}}J_{-2}\ket{0}.
\end{equation}
The Potts factors are chosen to be $\ket{0}$ and the normalization is determined by the following two point function:
\begin{equation}
	\langle w(\alpha_i)(z)w(\alpha_j)(0)\rangle
	=\frac1{z^4}
	\begin{cases}
		4,&i=j\\
		\frac12,&i\ne j
	\end{cases}
\end{equation}
Using the embedding~\eqref{eq:embbeding}, we can rewrite
\begin{equation}
	\ket{T'}=\frac1{\sqrt{21}}
	\bigl(2w(\alpha_1)-w(\alpha_0)-w(\alpha_2)\bigr)_{-2}|0\rangle,\quad 
	\ket{J}=\frac{1}{\sqrt{7}}\bigl(w(\alpha_2)-w(\alpha_0)\bigr)_{-2}\ket{0}.
\end{equation}
Acting on these states with $\hat\rho=\theta\rho$, we obtain
\begin{equation}
	\label{eq:Vrho2}
	\Omega_{\hat\rho} 
	\begin{pmatrix}
		\ket{T'}\\
		\ket{J}
	\end{pmatrix}
	=
	\begin{pmatrix}
		-\frac12 & \frac{\sqrt{3}}{2}\\
		\frac{\sqrt{3}}{2}&\frac12
	\end{pmatrix}
	\begin{pmatrix}
		\ket{T'}\\
		\ket{J}
	\end{pmatrix}
\end{equation}
Similarly, for the multiplicity space of $h=\frac35$, if we choose the Potts factor to be $\ket{\frac25}$, the two dimensional lowest weight space is spanned by $\alpha_1\cdot \boldsymbol{j}$ and $\alpha_2\cdot \boldsymbol{j}$. We have
\begin{align}
		\Ish{\frac35;(0,\frac35)}_\mathcal{U}\supset & \ket{Z_1}\otimes U\ket{\bar Z_1},\quad \ket{Z_1}:=\frac{1}{\sqrt{6}}\left(\alpha_1\cdot \boldsymbol{j}+2\alpha_2\cdot \boldsymbol{j}\right)_{-1} \ket{0},\\
		\Ish{\frac35;(\frac12,\frac1{10})}_\mathcal{U}\supset & \ket{Z_2}\otimes U\ket{\bar Z_2},\quad \ket{Z_2}:=\frac{1}{\sqrt{2}}\left(\alpha_1\cdot \boldsymbol{j}\right)_{-1}\ket{0}.
\end{align}
The following OPEs can be used to check the conformal weights and normalizations:
	\begin{align}
		j^{(a)}(z)j^{(b)}(w)\sim{}&\frac{\delta^{ab}}{(z-w)^2},\\
		w(\alpha_i)(z)j^a(w)\sim{}&
		\frac{\alpha_i^a}{(z-w)^2}(\alpha_i\cdot\boldsymbol{j})(w)
		+\frac{w(\alpha_i)_{-1}j^a}{z-w},
		\\
		w(\alpha_i)_{-1}j^a={}&\alpha_i^a\left[\alpha_i\cdot\partial\boldsymbol{j}+\sqrt{2}\left(:e^{i\sqrt{2}\alpha_i\cdot\boldsymbol{\phi}}:-:e^{-i\sqrt{2}\alpha_i\cdot\boldsymbol{\phi}}:\right)\right].
	\end{align}
Acting with $\hat{\rho}$, we obtain 
\begin{equation}
	\Omega_{\hat\rho} 
	\begin{pmatrix}
		\ket{Z_1}\\
		\ket{Z_2}
	\end{pmatrix}
	=
	\begin{pmatrix}
		-\frac12 & -\frac{\sqrt{3}}{2}\\
		-\frac{\sqrt{3}}{2}&\frac12
	\end{pmatrix}
	\begin{pmatrix}
		\ket{Z_1}\\
		\ket{Z_2}
	\end{pmatrix}
\end{equation}
However, note that in this case the difference between $\rho$ and $\hat{\rho}$ matters, acting only with $\rho$ gives the wrong sign. 

The analysis of multiplicity free states is similar but simpler. The lowest weight state in $\Ish{0}_{\mathcal{U}}$ is the vacuum, which is invariant under $\hat\rho$. Since the two $\mathcal{U}$-modules corresponding to $\Ish{0}_{\mathcal{U}}$ and $\Ish{3}_{\mathcal{U}}$ arise from the splitting of the $\mathcal{W}_3$ vacuum module, their lowest weight states are $\ket{0}$ and $W_{-3}\ket{0}$, respectively. Since $\hat\rho$ acts on the spin-3 current as $W\to -W$, the latter state is odd under ${\hat\rho}$. Consequently, $\Omega_{\hat\rho}$ acts on $\Ish{3}_{\mathcal{U}}$ with a minus sign. Similarly, according to the discussion around~\eqref{eq:wVOPE}, the lowest weight state of $\Ish{\frac12;(\frac{1}{16},\frac{7}{16})}$ is $V_{\alpha_0/\sqrt{2}}-V_{-\alpha_0/\sqrt{2}}$ and is therefore odd under $\Omega_{\hat\rho}$. Its partner in $\mathcal{W}_3$, $\Ish{\frac32;(\frac{1}{16},\frac{7}{16})}$, is even under $\Omega_{\hat \rho}$. We list below the explicit constructions of the lowest weight states for the remaining multiplicity free Ishibashi states:
\begin{align}
	\Ish{\frac85}\supset& \ket{X}\otimes U\ket{\bar X},\quad  \Ish{\frac{13}5}\supset W_{-1}\ket{X}\otimes U\bar W_{-1}\ket{\bar X},\\
	\Ish{\frac1{10};(\frac1{16},\frac{3}{80})}\supset& \ket{E}\otimes U\ket{\bar E},\quad  \Ish{\frac{11}{10};(\frac1{16},\frac{3}{80})}\supset W_{-1}\ket{E}\otimes U\bar W_{-1}\ket{\bar E}.
\end{align}
Here, the Potts factors are $\ket{\frac25}$ and we write
\begin{equation}
	\label{eq:EX}
	\ket{X}\propto\sum_{i=0}^2\left(V_{\sqrt{2}\alpha_i}-V_{-\sqrt{2}\alpha_i}\right),\quad
	\ket{E}\propto V_{\alpha_0/\sqrt{2}}+V_{-\alpha_0/\sqrt{2}}.
\end{equation}
As in the calculation of $\Ish{\frac12;(\frac1{16},\frac7{16})}$, there is a subtlety here. The state $\ket{\frac1{16}}\otimes\ket{\frac3{80}}$ can also be realized by $V_{\alpha_2/\sqrt{2}}+V_{-\alpha_2/\sqrt{2}}$. However, the corresponding module is not invariant under $\hat\rho$. Since the bulk CFT with chiral algebra $\mathcal{B}$ is diagonal, a $\hat\rho$-twisted Ishibashi state can only arise from a $\hat\rho$-invariant module. Therefore, this realization must be excluded.

From~\eqref{eq:EX}, it is clear that $\ket{X}$ is odd under $\Omega_{\hat\rho}$ and $\ket{E}$ is even under $\Omega_{\hat\rho}$. Collecting all the above results for $\Omega_{\hat\rho}$, we obtain
\begin{equation}
	\begin{aligned}
	\Ish{1,1,1,1; \mathcal{B},\mathbb{Z}_2}=&
	\Ish{0}_{\mathcal{U}}-\Ish{3}_{\mathcal{U}}
	-\frac12\Ish{2;(0,0)}_{\mathcal{U}}+\frac{1}{2}\Ish{2;(\frac12,\frac32)}_{\mathcal{U}}\\
	&+\frac{\sqrt{3}}{2}\Ish{2;(0,0),(\frac12,\frac32)}_{\mathcal{U}}+\frac{\sqrt{3}}{2}\Ish{2;(\frac12,\frac32),(0,0)}_{\mathcal{U}}
\end{aligned}
\end{equation}
\begin{equation}
	\begin{aligned}
		\Ish{1,1,1,3; \mathcal{B},\mathbb{Z}_2}=&
		-\Ish{\frac{8}{5}}_{\mathcal{U}}+\Ish{\frac{13}{5}}_{\mathcal{U}}
		-\frac12\Ish{\frac35;(0,\frac35)}_{\mathcal{U}}+\frac{1}{2}\Ish{\frac35;(\frac12,\frac1{10})}_{\mathcal{U}}\\
		&-\frac{\sqrt{3}}{2}\Ish{\frac35;(0,\frac35),(\frac12,\frac1{10})}_{\mathcal{U}}-\frac{\sqrt{3}}{2}\Ish{\frac35;(\frac12,\frac1{10}),(0,\frac35)}_{\mathcal{U}}
	\end{aligned}
\end{equation}
\begin{equation}
	\Ish{1,2,2,1; \mathcal{B},\mathbb{Z}_2}=-\Ish{\frac12;(\frac1{16},\frac7{16})}_{\mathcal{U}}+\Ish{\frac32;(\frac1{16},\frac7{16})}_{\mathcal{U}}
\end{equation}
\begin{equation}
	\Ish{1,2,2,3; \mathcal{B},\mathbb{Z}_2}=\Ish{\frac1{10};(\frac1{16},\frac3{80})}_{\mathcal{U}}-\Ish{\frac{11}{10};(\frac1{16},\frac3{80})}_{\mathcal{U}}
\end{equation}
The projector $\|\mathcal{A},t,r,s\|$ projects out these
Ishibashi states involving multiplicity mixing.
Consequently, the RG domain wall can be identified with
one of the symmetry breaking boundary states listed in~\eqref{eq:B1}--\eqref{eq:B6}. Using~\eqref{eq:SWZW}, we obtain
\begin{equation}
	S^{(1)}_{1,1}=S^{(2)}_{1,2}=\frac1{\sqrt2},\quad
	S^{(2)}_{1,1}=S^{(2)}_{1,3}=\frac12,\quad
	S^{(3)}_{1,1}=\frac{1}{2} \sqrt{1-\frac{1}{\sqrt{5}}},\quad
	S^{(3)}_{1,3}=\frac{1}{2} \sqrt{1+\frac{1}{\sqrt{5}}}.
\end{equation}
Noting that $\|\mathcal{A},1,r,s\|$ selects the Ishibashi states without multiplicity mixing that originate from
\begin{equation}
	\left.\left|{h_{\text{Ising}}=\frac{(4 - 3 r)^2 - 1}{48},h_{\text{TIM}}=\frac{(5r - 4 s)^2 - 1}{80}}\right\rangle\!\right\rangle_\mathcal{A},
\end{equation}
we obtain
\begin{align}
		\ket{RG}
		={}&
		\sum_{s\in\{1,3\}}\sum_{r=1}^{3}
		\frac{\sqrt{S^{(1)}_{1,1}S^{(3)}_{1,s}}}{S^{(2)}_{1,r}}
		\|\mathcal A,1,r,s\|
		\left(
		\Ish{1,1,1,s;\mathcal B,\mathbb Z_2}
		+\Ish{1,2,2,s;\mathcal B,\mathbb Z_2}
		\right)
		\notag\\
		={}&
		\sqrt{S^{(1)}_{1,1}S^{(3)}_{1,1}}
		\Biggl[
		\frac{
			\Ish{0}_{\mathcal U}
			-\frac12\Ish{2;(0,0)}_{\mathcal U}
		}{S^{(2)}_{1,1}}
		\notag\\
		&\hspace{28mm}
		+\frac{
			-\Ish{\frac12;(\frac1{16},\frac7{16})}_{\mathcal U}
			+\Ish{\frac32;(\frac1{16},\frac7{16})}_{\mathcal U}
		}{S^{(2)}_{1,2}}
		+\frac{
			-\Ish{3}_{\mathcal U}
			+\frac12\Ish{2;(\frac12,\frac32)}_{\mathcal U}
		}{S^{(2)}_{1,3}}
		\Biggr]
		\notag\\
		&+
		\sqrt{S^{(1)}_{1,1}S^{(3)}_{1,3}}
		\Biggl[
		\frac{
			\Ish{\frac{13}{5}}_{\mathcal U}
			-\frac12\Ish{\frac35;(0,\frac35)}_{\mathcal U}
		}{S^{(2)}_{1,1}}
		\notag\\
		&\hspace{28mm}
		+\frac{
			\Ish{\frac1{10};(\frac1{16},\frac3{80})}_{\mathcal U}
			-\Ish{\frac{11}{10};(\frac1{16},\frac3{80})}_{\mathcal U}
		}{S^{(2)}_{1,2}}
		+\frac{
			-\Ish{\frac85}_{\mathcal U}
			+\frac12\Ish{\frac35;(\frac12,\frac1{10})}_{\mathcal U}
		}{S^{(2)}_{1,3}}
		\Biggr]
		\notag\\
		={}&
		\frac{\sqrt2}{(1+\varphi^2)^{1/4}}
		\Biggl\{
		\Ish{0}_{\mathcal U}
		-\Ish{3}_{\mathcal U}
		-\frac12\Ish{2;(0,0)}_{\mathcal U}
		+\frac12\Ish{2;(\frac12,\frac32)}_{\mathcal U}
		\notag\\
		&\hspace{16mm}
		+\frac1{\sqrt2}
		\left(
		-\Ish{\frac12;(\frac1{16},\frac7{16})}_{\mathcal U}
		+\Ish{\frac32;(\frac1{16},\frac7{16})}_{\mathcal U}
		\right)
		\notag\\
		&\hspace{16mm}
		+\sqrt{\varphi}\,
		\Biggl[
		-\Ish{\frac85}_{\mathcal U}
		+\Ish{\frac{13}{5}}_{\mathcal U}
		-\frac12\Ish{\frac35;(0,\frac35)}_{\mathcal U}
		+\frac12\Ish{\frac35;(\frac12,\frac1{10})}_{\mathcal U}
		\notag\\
		&\hspace{42mm}
		+\frac1{\sqrt2}
		\left(
		\Ish{\frac1{10};(\frac1{16},\frac3{80})}_{\mathcal U}
		-\Ish{\frac{11}{10};(\frac1{16},\frac3{80})}_{\mathcal U}
		\right)
		\Biggr]
		\Biggr\}
		\notag\\
		={}&\ket{B_2}.
\end{align}

\section{Conclusions and Discussion}
\label{sec:4}
In this work, we have investigated the classification of conformal interfaces between the Ising and tricritical Ising models. Using the folding trick, we reformulate this problem as the classification of conformal boundary states in the Ising $\times$ TIM theory.

Our starting point is the chiral algebra $U_{2A}$, obtained as a simple-current extension of $\mathcal{A}=\mathsf{Vir}_{c=1/2}\otimes\mathsf{Vir}_{c=7/10}$. Taking the invariant subalgebra of $U_{2A}$ under its $S_3$ automorphism group, we obtain $\mathcal{U}=\mathcal{W}_3^{\mathbb{Z}_2}$, which is also a subalgebra of $\mathcal{A}$. Since the bulk CFT remains rational with respect to $\mathcal{U}$, we can systematically classify boundary states preserving this chiral algebra using the framework of~\cite{Petkova:2000dv}. By solving the Cardy condition, we obtain a complete set of 24 fundamental boundary states, of which 18 are the factorized Cardy states preserving the full chiral algebra $\mathcal{A}$, while the remaining six are new symmetry breaking boundary states. Furthermore, we identify one of these six states with the RG domain wall between the Ising and TIM models proposed in~\cite{Gaiotto:2012np}.

Remarkably, we find that all six symmetry breaking boundary states have the same transmission coefficient. This observation leads us to establish that they are all related by topological defect line dressings. Since the RG domain wall has already been studied in lattice realizations~\cite{Cogburn:2023xzw,Antunes:2026kqb}, and the relevant topological defect lines also admit lattice realizations~\cite{Aasen:2016dop,Sinha:2025jhh}, we expect that the remaining five symmetry-breaking interfaces can likewise be realized on the lattice.

Nevertheless, the complete classification of conformal interfaces between the Ising and TIM models remains an open problem. In this work, we have restricted our attention to boundary states preserving $\mathcal{U}$, without considering interfaces that break this symmetry further. A complete classification would require a systematic understanding of the vertex operator subalgebras of $\mathcal{U}$. More generally, classifying boundary states preserving only the diagonal Virasoro algebra $\mathsf{Vir}_{c=6/5}^{\mathrm{diag}}$ remains particularly challenging, since the folded theory is no longer rational with respect to this smaller chiral algebra.

Even within the class of boundary states preserving $\mathcal{U}$, our analysis has been restricted to the Cardy condition. We have not solved the full Cardy--Lewellen sewing constraints~\cite{Cardy:1991tv,Lewellen:1991tb}, which are required to determine consistent boundary OPE coefficients. The uniqueness of the NIM-rep established in the appendix leads us to expect that at least one consistent solution to the full sewing constraints exists. This expectation is further supported by the identification of the symmetry breaking boundary states as TDL dressings of the RG domain wall. Explicitly constructing the corresponding boundary OPE coefficients and verifying the full sewing constraints remain interesting directions for future work. For general discussion of the existence of boundaries consistent with a given bulk RCFT, we refer the reader to~\cite{Fjelstad:2005ua,Kong:2009inh,Kong:2013gca}.

\acknowledgments

The author thanks Hosho Katsura, Jiaxin Qiao, Masataka Watanabe, Hiroshi Yamauchi, and Masahito Yamazaki for helpful discussions. The author acknowledges support from the Global Science Graduate Course (GSGC) at the University of Tokyo, as well as AI tool support provided by the Center for Data-Driven Discovery (CD3) at Kavli IPMU. ChatGPT-6.1 was used to assist with the NIM-rep computations and manuscript preparation.

\paragraph{Note added.} 
We would like to draw the reader's attention to the independent work~\cite{Antunes:2026}, which investigates closely related questions and is being submitted to arXiv concurrently with our manuscript. We thank the authors for coordinating the submissions.

\appendix

\section{Construction and uniqueness of the NIM-rep}
In classifying the boundary states in the main text, we need to construct
a NIM-rep of a rank-24 fusion ring. Although finding NIM-reps is a
finite-dimensional search problem, its computational complexity grows
rapidly with the rank of the fusion ring~\cite{Gannon:2001ki}.
In this appendix, we describe a concrete algorithm that reduces
the final search to a size that can be efficiently handled by a computer.

Using the factorization of the fusion ring given in~\eqref{eq:fusion}
and the basis convention in~\eqref{eq:fibbasis},
we can decompose the desired NIM-rep into
$\mathsf{X}_a:=n_{a,1}$, $\mathsf{T}:=n_{1,\tau}$,
and $n_{a,\tau}$.
Hereafter, we denote the set of possible values of $a$ by
$\mathcal{I}^M$, and define
$\mathcal{I}^{\mathsf{Fib}}=\{1,\tau\}$.
The fusion rules immediately imply
\begin{equation}
	n_{a,\tau}=\mathsf{X}_a\mathsf{T},\quad
	\mathsf{X}_a\mathsf{X}_b
	=\sum_c {M_{ab}}^c\mathsf{X}_c,\quad
	\mathsf{T}^2=\mathbb{I}+\mathsf{T},\quad
	[\mathsf{X}_a,\mathsf{T}]=0.
\end{equation}
Therefore, the problem of finding a NIM-rep of
${\mathcal{N}_{ij}}^k$ is reduced to finding
commuting NIM-reps of ${M_{ab}}^c$ and $\mathsf{Fib}$.

First, up to simultaneous permutations of rows and columns,
$n_i\sim qn_iq^{-1}$, the NIM-rep of $\mathsf{Fib}$
must take the following form~\cite{Davydov:2011pp}:
\begin{equation}
	\label{eq:FibT}
	\mathsf{T}=
	\begin{pmatrix}
		0&\mathbb{I}_{12}\\
		\mathbb{I}_{12}&\mathbb{I}_{12}
	\end{pmatrix}.
\end{equation}
Once $\mathsf{T}$ is fixed to this form,
the remaining permutation symmetry is reduced
from $S_{24}$ to
\begin{equation}
	\label{eq:PS12}
	q=
	\begin{pmatrix}
		p&0\\
		0&p
	\end{pmatrix},
	\quad p\in S_{12}.
\end{equation}
Furthermore, the commutativity of $\mathsf{X}_a$ and
$\mathsf{T}$ implies that $\mathsf{X}_a$
must have the following block structure:
\begin{equation}
	\mathsf{X}_a=
	\begin{pmatrix}
		\mathsf{A}_a&\mathsf{B}_a\\
		\mathsf{B}_a&\mathsf{A}_a+\mathsf{B}_a
	\end{pmatrix},
\end{equation}
where $\mathsf{A}_a$ and $\mathsf{B}_a$ are
$12\times12$ matrices.
Note that the permutations in~\eqref{eq:PS12}
do not mix $\mathsf{A}_a$ and $\mathsf{B}_a$.

The two eigenspaces of $\mathsf{T}$ can be parametrized
by $v\in\mathbb{C}^{12}$ as
\begin{equation}
	\mathsf{T}
	\begin{pmatrix}
		v\\
		\varphi_\pm v
	\end{pmatrix}
	=
	\varphi_\pm
	\begin{pmatrix}
		v\\
		\varphi_\pm v
	\end{pmatrix},
	\qquad
	\varphi_\pm=\frac{1\pm\sqrt{5}}{2}.
\end{equation}
Both eigenspaces are invariant under $\mathsf{X}_a$,
and the restrictions of $\mathsf{X}_a$ to these
subspaces are represented by
$\mathsf{A}_a+\varphi_\pm\mathsf{B}_a$, respectively.
Consequently, we obtain the spectral decomposition
\begin{equation}
	\operatorname{Spec}(n_{a,1})
	=\operatorname{Spec}(\mathsf{X}_a)
	=\operatorname{Spec}(\mathsf{A}_a+\varphi_+\mathsf{B}_a)
	\cup
	\operatorname{Spec}(\mathsf{A}_a+\varphi_-\mathsf{B}_a).
\end{equation}
On the other hand, according to~\eqref{eq:nidecomp},
we have
\begin{equation}
	\label{eq:spec}
		\operatorname{Spec}(n_{a,1})
		=
		\left\{
		\frac{S^{\mathcal{U}}_{(a,1),(b,f)}}
		{S^{\mathcal{U}}_{(1,1),(b,f)}}
		\right\}
		=
		\left\{
		\frac{S^M_{a,b}S^{\mathsf{Fib}}_{1,f}}
		{S^M_{1,b}S^{\mathsf{Fib}}_{1,f}}
		\right\}
		=
		\left\{
		\frac{S^M_{a,b}}{S^M_{1,b}}
		\right\},
		\qquad
		a\in\mathcal{I}^M,\quad
		(b,f)\in\mathcal{E}^\mathcal{U}.
\end{equation}
Combining this with~\eqref{eq:SM}
and the definitions of the constants in~\eqref{eq:const},
we obtain
\begin{equation}
	\operatorname{Spec}(\mathsf{A}_a+\varphi_\pm\mathsf{B}_a)
	\subset
	\operatorname{Spec}(n_{a,1})
	\subset
	\mathbb{Q}\left(
	\sqrt{2},2\cos\left(\frac{\pi}{9}\right)
	\right)=K.
\end{equation}
A useful algebraic fact is that
$\sqrt{5}\notin K$, and
\begin{equation}
	\sigma\in
	\operatorname{Gal}\left(K(\sqrt{5})/K\right)
	\cong\mathbb{Z}_2,
	\qquad
	\sigma:\sqrt{5}\mapsto-\sqrt{5}.
\end{equation}
Since $\sigma$ exchanges $\varphi_+$ and $\varphi_-$
while acting trivially on $K$,
the matrices $\mathsf{A}_a+\varphi_+\mathsf{B}_a$
and $\mathsf{A}_a+\varphi_-\mathsf{B}_a$
must have identical spectra.
It follows that
\begin{equation}
		0
		=
		\operatorname{Tr}
		(\mathsf{A}_a+\varphi_+\mathsf{B}_a)^2
		-
		\operatorname{Tr}
		(\mathsf{A}_a+\varphi_-\mathsf{B}_a)^2\\
		=
		\sqrt{5}\left(
		2\operatorname{Tr}(\mathsf{A}_a\mathsf{B}_a)
		+\operatorname{Tr}(\mathsf{B}_a^2)
		\right).
\end{equation}
Since $\mathsf{A}_a$ and $\mathsf{B}_a$
are symmetric matrices with nonnegative integer entries,
the above equation immediately implies
\begin{equation}
	\mathsf{B}_a=0,
	\qquad \forall a\in\mathcal{I}^M.
\end{equation}
This shows that the problem of finding a
24-dimensional NIM-rep of ${\mathcal{N}_{ij}}^k$
can be further reduced to finding a
12-dimensional NIM-rep of ${M_{ab}}^c$:
\begin{equation}
	\label{eq:AFusion}
	\mathsf{X}_a=
	\begin{pmatrix}
		\mathsf{A}_a&0\\
		0&\mathsf{A}_a
	\end{pmatrix},
	\quad
	\mathsf{A}_a\mathsf{A}_b
	=\sum_c {M_{ab}}^c\mathsf{A}_c,
	\quad
	\mathsf{A}_1=\mathbb{I}_{12},
	\quad
	\mathsf{A}_a^{\mathsf{T}}=\mathsf{A}_a.
\end{equation}
We now determine this NIM-rep in several steps.
\subsubsection*{Step 1: $\mathsf{A}_{12}$}

We first consider the primary with $h=3$ and denote the corresponding matrix
$\mathsf{A}_{12}$ by $\mathsf{P}$.
Using~\eqref{eq:spec} and
$\operatorname{Spec}(\mathsf{X}_a)
=\operatorname{Spec}(\mathsf{A}_a\oplus\mathsf{A}_a)$,
we obtain
\begin{equation}
	\operatorname{Spec}(\mathsf{P})
	=\{1\}^8\cup\{-1\}^4,\quad
	\mathsf{P}^{\mathsf{T}}=\mathsf{P}
	\quad\Rightarrow\quad
	\operatorname{Tr}(\mathsf{P})=4,\quad
	\mathsf{P}^2=\mathbb{I}_{12}.
\end{equation}
Since all entries of $\mathsf{P}$ are nonnegative integers,
the condition $\mathsf{P}^2=\mathbb{I}_{12}$ implies that
each row and column contains exactly one entry equal to $1$,
with all remaining entries being zero.
Thus, $\mathsf{P}$ is a permutation matrix.
Furthermore, its trace implies that the corresponding permutation
consists of exactly four fixed points and four transpositions.

\subsubsection*{Step 2: $\mathsf{A}_{11}$}

We next consider the primary with $h=2$, which corresponds to
$\mathsf{A}_{11}$.
Applying~\eqref{eq:spec} again, we find
\begin{equation}
	\operatorname{Spec}(\mathsf{A}_{11})
	=\operatorname{Spec}(\mathsf{P})+1.
\end{equation}
Here, the eigenvalues are understood to be ordered according to
a common eigenbasis of $\mathsf{P}$ and $\mathsf{A}_{11}$,
so that the equality holds entrywise.
Since $\mathsf{P}$ and $\mathsf{A}_{11}$ can be simultaneously diagonalized,
it follows that
\begin{equation}
	\mathsf{A}_{11}=\mathbb{I}_{12}+\mathsf{P}.
\end{equation}

\subsubsection*{Step 3: $\mathsf{A}_{3,6,8}$}

Following the same argument as in the previous step,
we consider the primaries with $h=1/9,7/9,13/9$,
corresponding to $\mathsf{A}_3,\mathsf{A}_6,\mathsf{A}_8$,
respectively.
Evaluating their spectra using~\eqref{eq:spec},
we find that their eigenvalues coincide entrywise
when ordered according to a common eigenbasis:
\begin{equation}
	\operatorname{Spec}(\mathsf{A}_3)
	=\operatorname{Spec}(\mathsf{A}_6)
	=\operatorname{Spec}(\mathsf{A}_8)=\{4\}^2\cup\{-2\}^2\cup\{0\}^8.
\end{equation}
Since the fusion ring is commutative, these matrices can be
simultaneously diagonalized.
By the same argument as above, we therefore obtain
\begin{equation}
	\mathsf{A}_3=\mathsf{A}_6=\mathsf{A}_8=:\mathsf{R}.
\end{equation}

\subsubsection*{Step 4: Fusion relations}

We now make use of the fusion rules satisfied by
$\mathsf{A}_a$ in~\eqref{eq:AFusion}.
Computing the fusion coefficients ${M_{ab}}^c$
directly from~\eqref{eq:SM}, we find the following relations:
\begin{equation}
	\mathsf{A}_5=\mathsf{P}\mathsf{A}_9=\mathsf{P}\mathsf{F},
	\quad
	\mathsf{A}_{10}=\mathsf{P}\mathsf{A}_2=\mathsf{P}\mathsf{T}_0,
	\quad
	\mathsf{A}_7=\mathsf{P}\mathsf{A}_4=\mathsf{P}\mathsf{T}_1,
\end{equation}
where we have introduced
$\mathsf{F}:=\mathsf{A}_9$,
$\mathsf{T}_0:=\mathsf{A}_2$, and
$\mathsf{T}_1:=\mathsf{A}_4$.
Here, $\mathsf{A}_4,\mathsf{A}_5,\mathsf{A}_7,
\mathsf{A}_9,\mathsf{A}_{10}$
correspond to the primaries with
$h=7/16,1/2,15/16,3/2,25/16$, respectively.
At this stage, all 12 matrices are determined by
$\mathsf{P}$ and four additional matrices
$\mathsf{T}_0,\mathsf{T}_1,\mathsf{R},\mathsf{F}$:
\begin{equation}
	\label{eq:12to4}
	\left\{
	\mathbb{I}_{12},
	\mathsf{T}_0,
	\mathsf{R},
	\mathsf{T}_1,
	\mathsf{P}\mathsf{F},
	\mathsf{R},
	\mathsf{P}\mathsf{T}_1,
	\mathsf{R},
	\mathsf{F},
	\mathsf{P}\mathsf{T}_0,
	\mathbb{I}_{12}+\mathsf{P},
	\mathsf{P}
	\right\}.
\end{equation}

\subsubsection*{Step 5: $\mathsf{F}$ and $\mathsf{P}$}

We list below some fusion relations that will be useful
in the subsequent analysis:
\begin{equation}
	\label{eq:FP}
	\mathsf{F}^2
	=2\mathbb{I}_{12}+\mathsf{P}
	+(\mathbb{I}_{12}+\mathsf{P})\mathsf{F},
	\quad
	\mathsf{R}^2=\mathsf{K}+2\mathsf{R},
	\quad
	\mathsf{R}\mathsf{F}=3\mathsf{R},
\end{equation}
where
$\mathsf{K}:=(\mathbb{I}_{12}+\mathsf{P})
(\mathbb{I}_{12}+\mathsf{F})$.
We begin with the commutation relation
$[\mathsf{F},\mathsf{P}]=0$.
Their eigenvalues can be computed using the method described above,
from which we obtain
\begin{equation}
	\mathsf{P}\mathsf{K}=\mathsf{K},
	\qquad
	\operatorname{Spec}(\mathsf{K})
	=\{8\}^4\cup\{0\}^8.
\end{equation}

Suppose that $\mathsf{K}$ can be decomposed into a direct sum
of irreducible matrices by simultaneous permutations of
rows and columns,
$\mathsf{K}=\mathsf{K}_1\oplus\cdots\oplus\mathsf{K}_\ell$.
Each block $\mathsf{K}_i$ is an irreducible nonnegative matrix,
so the Perron--Frobenius theorem guarantees that its
largest eigenvalue is simple.
Since the eigenvalues of each $\mathsf{K}_i$ can only be
$0$ or $8$, and $\mathsf{K}\geq\mathbb{I}_{12}$
holds entrywise by construction, none of the blocks
can be the zero matrix.
Therefore, the largest eigenvalue of each block must be $8$.
It follows that $\mathsf{K}$ decomposes into exactly
four irreducible blocks, each with spectrum
$\{8,0,\ldots,0\}$.
Consequently,
\begin{equation}
	\mathsf{K}=\bigoplus_{i=1}^4\mathsf{K}_i,
	\qquad
	\mathsf{K}_i=\mathbf{v}\mathbf{v}^{\mathsf{T}},
	\qquad
	v_a>0,\quad \sum_a v_a^2=8.
\end{equation}
The condition $\mathsf{K}\geq\mathbb{I}_{12}$
implies $v_a^2\in\mathbb{Z}_{\geq1}$.
Using this condition together with $v_av_b\in\mathbb{Z}_{\geq0}$,
we find only five possible forms of $\mathbf{v}$ by direct enumeration:
\begin{equation}
	(1^{\oplus8}),\quad
	(1^{\oplus4},2),\quad
	(\sqrt{2}^{\oplus4}),\quad
	(2,2),\quad
	(2\sqrt{2}).
\end{equation}
Considering the dimensions of the corresponding blocks
$\mathsf{K}_i$, we find that the decomposition of
$\mathsf{K}$ can only take one of the following four forms:
\begin{equation}
	(8,2,1,1),\quad
	(5,5,1,1),\quad
	(5,4,2,1),\quad
	(4,4,2,2),
\end{equation}
where the entries denote $\dim\mathsf{K}_i$.

We first fix $\mathsf{K}$ to one of these four forms
by simultaneous permutations of rows and columns.
The remaining permutation symmetry is then reduced
from $S_{12}$ in~\eqref{eq:PS12}
to the stabilizer subgroup of $\mathsf{K}$.
Using $\mathsf{P}\mathsf{K}=\mathsf{K}$
together with the possible cycle structure of
$\mathsf{P}$ derived in Step 1,
we can enumerate all possible $\mathsf{P}$
up to the remaining permutations.
Furthermore, for any fixed point $a$ of $\mathsf{P}$,
i.e. $\mathsf{P}_{aa}=1$, we have
$\mathsf{K}_{aa}=2(1+\mathsf{F}_{aa})\geq2$.
Taking this additional constraint into account,
we obtain eight possible pairs $(\mathsf{K},\mathsf{P})$:
\begin{equation}
	\begin{alignedat}{3}
		\mathsf{K}&=(8,2,1,1):\qquad
		&\mathsf{P}&=&&(12)(34)(56)(78),\\
		\mathsf{K}&=(5,5,1,1):
		&\mathsf{P}&=&&(12)(34)(67)(89),\\
		\mathsf{K}&=(5,4,2,1):
		&\mathsf{P}&=&&(12)(34)(67)(89),\
		(12)(34)(67)(10,11),\\
		\mathsf{K}&=(4,4,2,2):
		&\mathsf{P}&=&&(12)(34)(56)(78),\
		(12)(34)(56)(9,10),\\
		&&&&&(12)(34)(9,10)(11,12),\
		(12)(56)(9,10)(11,12).
	\end{alignedat}
\end{equation}
Note that the remaining permutation symmetry is now
$\operatorname{Stab}_{S_{12}}(\mathsf{K},\mathsf{P})$.

\subsubsection*{Step 6: $\mathsf{R}$}

Once $(\mathsf{K},\mathsf{P})$ is fixed,
the following equations for $\mathsf{F}$ and $\mathsf{R}$
become linear:
\begin{equation}
	\begin{gathered}
		\mathsf{F}=\mathsf{F}^{\mathsf{T}},
		\quad
		\mathsf{R}=\mathsf{R}^{\mathsf{T}},
		\quad
		\mathsf{F}\mathsf{P}=\mathsf{P}\mathsf{F},
		\quad
		\mathsf{K}\mathsf{R}=8\mathsf{R},\\
		\mathsf{K}=(\mathbb{I}_{12}+\mathsf{P})
		(\mathbb{I}_{12}+\mathsf{F}),
		\quad
		\operatorname{Tr}(\mathsf{F})
		=2\operatorname{Tr}(\mathsf{R})=8.
	\end{gathered}
\end{equation}
We first solve these linear equations to reduce the
number of unknowns, and then enumerate the remaining
possibilities subject to~\eqref{eq:FP}.
Note that the matrix elements of a NIM-rep $n_i$
cannot exceed the quantum dimension of the corresponding
primary, $d_i=S_{0i}/S_{00}$~\cite{Gannon:2001ki}.
Therefore, it suffices to restrict the search to
\begin{equation}
	0\leq\mathsf{R}_{\alpha\beta}\leq4,
	\qquad
	0\leq\mathsf{F}_{\alpha\beta}\leq3.
\end{equation}
Finally, after quotienting by the residual permutation
symmetry $\operatorname{Stab}_{S_{12}}(\mathsf{K},\mathsf{P})$,
we obtain the possible triples
$(\mathsf{P},\mathsf{F},\mathsf{R})$. In practice, 22 triples $(\mathsf{P},\mathsf{F},\mathsf{R})$ remain after the search.

\subsubsection*{Final Step: $\mathsf{T}_0$ and $\mathsf{T}_1$}

Having fixed $(\mathsf{P},\mathsf{F},\mathsf{R})$,
we first use
\begin{equation}
	\operatorname{Tr}(\mathsf{T}_0)
	=\operatorname{Tr}(\mathsf{T}_1)
	=\operatorname{Tr}(\mathsf{P}\mathsf{T}_0)
	=\operatorname{Tr}(\mathsf{P}\mathsf{T}_1)=0,
\end{equation}
to obtain
\begin{equation}
	(\mathsf{T}_a)_{ii}
	=(\mathsf{T}_a)_{i,\mathsf{P}(i)}=0,
	\qquad a=0,1.
\end{equation}
We then solve the following linear equations to reduce the number of remaining variables:
\begin{equation}
	\begin{aligned}
		\mathsf{F}\mathsf{T}_0
		&=\mathsf{P}\mathsf{T}_0
		+(\mathbb{I}_{12}+\mathsf{P})\mathsf{T}_1,
		\quad
		\mathsf{F}\mathsf{T}_1
		=\mathsf{T}_1+(\mathbb{I}_{12}+\mathsf{P})\mathsf{T}_0,\\
		\mathsf{R}\mathsf{T}_0
		&=(\mathbb{I}_{12}+\mathsf{P})(\mathsf{T}_0+\mathsf{T}_1),
		\quad
		\mathsf{R}\mathsf{T}_1
		=(\mathbb{I}_{12}+\mathsf{P})(\mathsf{T}_0+\mathsf{T}_1),\\
		[\mathsf{T}_a,\mathsf{P}]
		&=[\mathsf{T}_a,\mathsf{F}]
		=[\mathsf{T}_a,\mathsf{R}]=0,
		\qquad a=0,1.
	\end{aligned}
\end{equation}
We further impose the quadratic relations
\begin{equation}
		\mathsf{T}_0^2
		=2\mathbb{I}_{12}+\mathsf{P}
		+3\mathsf{R}+\mathsf{P}\mathsf{F},\quad
		\mathsf{T}_1^2
		=2\mathbb{I}_{12}+\mathsf{P}
		+3\mathsf{R}+\mathsf{F},\quad
		\mathsf{T}_0\mathsf{T}_1
		=3\mathsf{R}+\mathsf{F}+\mathsf{P}\mathsf{F}.
\end{equation}
We then enumerate the remaining possibilities within
the bounds imposed by the quantum dimensions,
\begin{equation}
	0\leq(\mathsf{T}_a)_{\alpha\beta}\leq4,
	\qquad a=0,1.
\end{equation}
After quotienting by the residual permutation symmetry
$\operatorname{Stab}_{S_{12}}
(\mathsf{K},\mathsf{P},\mathsf{F},\mathsf{R})$,
we find a unique admissible tuple
$(\mathsf{P},\mathsf{F},\mathsf{R},\mathsf{T}_0,\mathsf{T}_1)$.
Finally, using~\eqref{eq:12to4}, \eqref{eq:AFusion},
and~\eqref{eq:FibT}, we reconstruct the complete
NIM-rep of ${\mathcal{N}_{ij}}^k$. This is the NIM-rep used to obtain the boundary state coefficient matrix in~\eqref{eq:psi}.

\bibliographystyle{JHEP}
\bibliography{refs}

@article{Andrei:2018die,
	author = "Andrei, N. and others",
	title = "{Boundary and Defect CFT: Open Problems and Applications}",
	eprint = "1810.05697",
	archivePrefix = "arXiv",
	primaryClass = "hep-th",
	doi = "10.1088/1751-8121/abb0fe",
	journal = "J. Phys. A",
	volume = "53",
	number = "45",
	pages = "453002",
	year = "2020"
}

@article{Chen:2026byx,
	author = "Chen, Jin",
	title = "{Renormalization Group Domain Walls from Phantom Symmetries}",
	eprint = "2610.23529",
	archivePrefix = "arXiv",
	primaryClass = "hep-th",
	reportNumber = "USTC-ICTS/PCFT-26-66",
	month = "10",
	year = "2026"
}

@article{Gaiotto:2012np,
	author = "Gaiotto, Davide",
	title = "{Domain Walls for Two-Dimensional Renormalization Group Flows}",
	eprint = "1201.0767",
	archivePrefix = "arXiv",
	primaryClass = "hep-th",
	doi = "10.1007/JHEP12(2012)103",
	journal = "JHEP",
	volume = "12",
	pages = "103",
	year = "2012"
}

@article{Oshikawa:1996dj,
	author = "Oshikawa, Masaki and Affleck, Ian",
	title = "{Boundary conformal field theory approach to the critical two-dimensional Ising model with a defect line}",
	eprint = "cond-mat/9612187",
	archivePrefix = "arXiv",
	doi = "10.1016/S0550-3213(97)00219-8",
	journal = "Nucl. Phys. B",
	volume = "495",
	pages = "533--582",
	year = "1997"
}

@article{Oshikawa:1996ww,
	author = "Oshikawa, Masaki and Affleck, Ian",
	title = "{Defect lines in the Ising model and boundary states on orbifolds}",
	eprint = "hep-th/9606177",
	archivePrefix = "arXiv",
	doi = "10.1103/PhysRevLett.77.2604",
	journal = "Phys. Rev. Lett.",
	volume = "77",
	pages = "2604--2607",
	year = "1996"
}

@article{Cardy:1989ir,
	author = "Cardy, John L.",
	title = "{Boundary Conditions, Fusion Rules and the Verlinde Formula}",
	reportNumber = "UCSB-TH-89-06",
	doi = "10.1016/0550-3213(89)90521-X",
	journal = "Nucl. Phys. B",
	volume = "324",
	pages = "581--596",
	year = "1989"
}

@article{Behrend:1999bn,
	author = "Behrend, Roger E. and Pearce, Paul A. and Petkova, Valentina B. and Zuber, Jean-Bernard",
	title = "{Boundary conditions in rational conformal field theories}",
	eprint = "hep-th/9908036",
	archivePrefix = "arXiv",
	reportNumber = "SACLAY-SPH-T-99-085",
	doi = "10.1016/S0550-3213(99)00592-1",
	journal = "Nucl. Phys. B",
	volume = "570",
	pages = "525--589",
	year = "2000"
}

@inproceedings{Petkova:2000dv,
	author = "Petkova, Valentina B. and Zuber, Jean-Bernard",
	title = "{Conformal boundary conditions and what they teach us}",
	booktitle = "{Eotvos Summer School in Physics: Nonperturbative QFT Methods and Their Applications}",
	eprint = "hep-th/0103007",
	archivePrefix = "arXiv",
	doi = "10.1142/9789812799968_0001",
	pages = "1--35",
	month = "8",
	year = "2000"
}

@article{Frohlich:2006ch,
	author = "Frohlich, Jurg and Fuchs, Jurgen and Runkel, Ingo and Schweigert, Christoph",
	title = "{Duality and defects in rational conformal field theory}",
	eprint = "hep-th/0607247",
	archivePrefix = "arXiv",
	reportNumber = "KCL-MTH-06-08, ZMP-HH-06-11",
	doi = "10.1016/j.nuclphysb.2006.11.017",
	journal = "Nucl. Phys. B",
	volume = "763",
	pages = "354--430",
	year = "2007"
}

@article{Gang:2008sz,
	author = "Gang, Dongmin and Yamaguchi, Satoshi",
	title = "{Superconformal defects in the tricritical Ising model}",
	eprint = "0809.0175",
	archivePrefix = "arXiv",
	primaryClass = "hep-th",
	reportNumber = "SNUTP-08-007",
	doi = "10.1088/1126-6708/2008/12/076",
	journal = "JHEP",
	volume = "12",
	pages = "076",
	year = "2008"
}

@article{Makabe:2017ygy,
	author = "Makabe, Isao and Watts, Gerard M T",
	title = "{Defects in the Tri-critical Ising model}",
	eprint = "1703.09148",
	archivePrefix = "arXiv",
	primaryClass = "hep-th",
	reportNumber = "KCL-MTH-17-01",
	doi = "10.1007/JHEP09(2017)013",
	journal = "JHEP",
	volume = "09",
	pages = "013",
	year = "2017"
}

@article{Antunes:2026kqb,
	author = "Antunes, Ant{\'o}nio and Rong, Junchen",
	title = "{On the Ising/tricritical-Ising interface: From ${W}_3$ symmetry to Rydberg atoms}",
	eprint = "2605.23848",
	archivePrefix = "arXiv",
	primaryClass = "hep-th",
	month = "5",
	year = "2026"
}

@article{Fuchs:1999xn,
	author = "Fuchs, Jurgen and Schweigert, Christoph",
	title = "{Symmetry breaking boundaries. 2. More structures: Examples}",
	eprint = "hep-th/9908025",
	archivePrefix = "arXiv",
	reportNumber = "ETH-TH-99-04",
	doi = "10.1016/S0550-3213(99)00669-0",
	journal = "Nucl. Phys. B",
	volume = "568",
	pages = "543--593",
	year = "2000"
}

@article{Fuchs:1999zi,
	author = "Fuchs, Jurgen and Schweigert, Christoph",
	title = "{Symmetry breaking boundaries. 1. General theory}",
	eprint = "hep-th/9902132",
	archivePrefix = "arXiv",
	reportNumber = "CERN-TH-99-35, ETH-TH-99-03",
	doi = "10.1016/S0550-3213(99)00406-X",
	journal = "Nucl. Phys. B",
	volume = "558",
	pages = "419--483",
	year = "1999"
}

@article{Fuchs:1998xx,
	author = "Fuchs, Jurgen and Schweigert, Christoph",
	title = "{Orbifold analysis of broken bulk symmetries}",
	eprint = "hep-th/9811211",
	archivePrefix = "arXiv",
	reportNumber = "CERN-TH-98-371",
	doi = "10.1016/S0370-2693(98)01598-6",
	journal = "Phys. Lett. B",
	volume = "447",
	pages = "266--276",
	year = "1999"
}

@article{Schellekens:1989am,
	author = "Schellekens, A. N. and Yankielowicz, S.",
	title = "{Extended Chiral Algebras and Modular Invariant Partition Functions}",
	reportNumber = "CERN-TH-5344/89",
	doi = "10.1016/0550-3213(89)90310-6",
	journal = "Nucl. Phys. B",
	volume = "327",
	pages = "673--703",
	year = "1989"
}

@article{Schellekens:1990xy,
	author = "Schellekens, A. N. and Yankielowicz, S.",
	title = "{Simple Currents, Modular Invariants and Fixed Points}",
	reportNumber = "CERN-TH-5622-90",
	doi = "10.1142/S0217751X90001367",
	journal = "Int. J. Mod. Phys. A",
	volume = "5",
	pages = "2903--2952",
	year = "1990"
}

@article{Fuchs:1996dd,
	author = "Fuchs, J. and Schellekens, A. N. and Schweigert, C.",
	title = "{A Matrix S for all simple current extensions}",
	eprint = "hep-th/9601078",
	archivePrefix = "arXiv",
	reportNumber = "IHES-P-96-8, NIKHEF-96-001, DESY-96-008",
	doi = "10.1016/0550-3213(96)00247-7",
	journal = "Nucl. Phys. B",
	volume = "473",
	pages = "323--366",
	year = "1996"
}

@article{Mathieu:2007fx,
	author = "Mathieu, Pierre and Ridout, David",
	title = "{The Extended Algebra of the Minimal Models}",
	eprint = "hep-th/0701250",
	archivePrefix = "arXiv",
	doi = "10.1016/j.nuclphysb.2007.03.030",
	journal = "Nucl. Phys. B",
	volume = "776",
	pages = "365--404",
	year = "2007"
}

@article{dong2003,
	title = {{$\mathbb{Z}_3$ symmetry and $W_3$ algebra in lattice vertex operator algebras}},
	author = {Dong, Chongying and Lam, Ching Hung and Tanabe, Kenichiro and Yamada, Hiromichi and Yokoyama, Kazuhiro},
	year = {2004},
	eprint = {math/0302314},
	archiveprefix = {arXiv},
	primaryclass = {math.QA},
	url = {https://arxiv.org/abs/math/0302314},
	journal = {Pacific J. Math.},
	volume = {215},
	number = {2},
	pages = {245--296},
	doi = {10.2140/pjm.2004.215.245},
}

@article{Blumenhagen:1994wg,
	author = "Blumenhagen, R. and Eholzer, W. and Honecker, A. and Hornfeck, K. and Hubel, R.",
	title = "{Coset realization of unifying W algebras}",
	eprint = "hep-th/9406203",
	archivePrefix = "arXiv",
	reportNumber = "DFTT-25-94, BONN-TH-94-11",
	doi = "10.1142/S0217751X95001157",
	journal = "Int. J. Mod. Phys. A",
	volume = "10",
	pages = "2367--2430",
	year = "1995"
}

@article{milas2020permutation,
	title = {{Permutation orbifolds of $\mathfrak{sl}_2$ vertex operator algebras}},
	author = {Milas, Antun and Penn, Michael},
	journal = {Glas. Mat. Ser. III},
	volume = {55},
	number = {2},
	pages = {277--300},
	year = {2020},
	publisher = {Hrvatsko matemati{\v{c}}ko dru{\v{s}}tvo},
	doi = {10.3336/gm.55.2.08},
}

@article{Al-Ali:2016ndc,
	author = "Al-Ali, Masoumah and Linshaw, Andrew R.",
	title = "{The ${\mathbb{Z}_2}$-Orbifold of the ${\mathcal{W}_3}$-Algebra}",
	eprint = "1608.06255",
	archivePrefix = "arXiv",
	primaryClass = "math.RT",
	doi = "10.1007/s00220-016-2812-7",
	journal = "Commun. Math. Phys.",
	volume = "353",
	number = "3",
	pages = "1129--1150",
	year = "2017"
}

@article{Honecker:1992kz,
	author = "Honecker, A.",
	title = "{Automorphisms of W algebras and extended rational conformal field theories}",
	eprint = "hep-th/9211130",
	archivePrefix = "arXiv",
	reportNumber = "BONN-HE-92-37",
	doi = "10.1016/0550-3213(93)90416-M",
	journal = "Nucl. Phys. B",
	volume = "400",
	pages = "574--596",
	year = "1993"
}

@article{Vafa:1989ih,
	author = "Vafa, Cumrun",
	title = "{Quantum Symmetries of String Vacua}",
	reportNumber = "HUTP-89-A021",
	doi = "10.1142/S0217732389001842",
	journal = "Mod. Phys. Lett. A",
	volume = "4",
	pages = "1615",
	year = "1989"
}

@article{Bhardwaj:2017xup,
	author = "Bhardwaj, Lakshya and Tachikawa, Yuji",
	title = "{On finite symmetries and their gauging in two dimensions}",
	eprint = "1704.02330",
	archivePrefix = "arXiv",
	primaryClass = "hep-th",
	reportNumber = "IPMU-17-0049",
	doi = "10.1007/JHEP03(2018)189",
	journal = "JHEP",
	volume = "03",
	pages = "189",
	year = "2018"
}

@article{hanaki1997,
	title = {{Quantum Galois theory for finite groups}},
	author = {Hanaki, Akihide and Miyamoto, Masahiko and Tambara, Daisuke},
	year = {1999},
	eprint = {q-alg/9709041},
	archiveprefix = {arXiv},
	primaryclass = {q-alg},
	url = {https://arxiv.org/abs/q-alg/9709041},
	journal = {Duke Math. J.},
	volume = {97},
	number = {3},
	pages = {541--544},
	doi = {10.1215/S0012-7094-99-09720-X},
}

@article{Ho-Kim:1988ast,
	author = "Ho-Kim, Q. and Zheng, H. B.",
	title = "{Twisted Conformal Field Theories With $Z$(3) Invariance}",
	doi = "10.1016/0370-2693(88)91238-5",
	journal = "Phys. Lett. B",
	volume = "212",
	pages = "71--76",
	year = "1988"
}

@article{Bouwknegt:1992wg,
	author = "Bouwknegt, Peter and Schoutens, Kareljan",
	title = "{W symmetry in conformal field theory}",
	eprint = "hep-th/9210010",
	archivePrefix = "arXiv",
	reportNumber = "CERN-TH-6583-92, ITPO-SB-92-23",
	doi = "10.1016/0370-1573(93)90111-P",
	journal = "Phys. Rept.",
	volume = "223",
	pages = "183--276",
	year = "1993"
}

@article{Birke:1999ik,
	author = "Birke, Lothar and Fuchs, Jurgen and Schweigert, Christoph",
	title = "{Symmetry breaking boundary conditions and WZW orbifolds}",
	eprint = "hep-th/9905038",
	archivePrefix = "arXiv",
	reportNumber = "ETH-TH-99-11",
	doi = "10.4310/ATMP.1999.v3.n3.a8",
	journal = "Adv. Theor. Math. Phys.",
	volume = "3",
	pages = "671--726",
	year = "1999"
}

@article{dong2017orbifold,
	title = {{On orbifold theory}},
	author = {Dong, Chongying and Ren, Li and Xu, Feng},
	journal = {Adv. Math.},
	volume = {321},
	pages = {1--30},
	year = {2017},
	publisher = {Elsevier},
	doi = {10.1016/j.aim.2017.09.032},
}

@article{Ishibashi:1988kg,
	author = "Ishibashi, Nobuyuki",
	title = "{The Boundary and Crosscap States in Conformal Field Theories}",
	reportNumber = "UT-530-TOKYO",
	doi = "10.1142/S0217732389000320",
	journal = "Mod. Phys. Lett. A",
	volume = "4",
	pages = "251",
	year = "1989"
}

@book{Recknagel:2013uja,
	author = "Recknagel, Andreas and Schomerus, Volker",
	title = "{Boundary Conformal Field Theory and the Worldsheet Approach to D-Branes}",
	doi = "10.1017/CBO9780511806476",
	isbn = "978-0-521-83223-6, 978-1-107-49612-5",
	publisher = "Cambridge University Press",
	series = "Cambridge Monographs on Mathematical Physics",
	month = "11",
	year = "2013"
}

@article{Gaberdiel:2001xm,
	author = "Gaberdiel, M. R. and Recknagel, A. and Watts, G. M. T.",
	title = "{The Conformal boundary states for SU(2) at level 1}",
	eprint = "hep-th/0108102",
	archivePrefix = "arXiv",
	reportNumber = "KCL-MTH-01-36",
	doi = "10.1016/S0550-3213(02)00033-0",
	journal = "Nucl. Phys. B",
	volume = "626",
	pages = "344--362",
	year = "2002"
}

@article{Hoback:2026yqj,
	author = "Hoback, Sarah and Jafferis, Daniel L. and Wei, Zixia",
	title = "{Monitoring a de Sitter universe through an anti-de Sitter window}",
	eprint = "2606.31705",
	archivePrefix = "arXiv",
	primaryClass = "hep-th",
	month = "6",
	year = "2026"
}

@article{Fuchs:2002cm,
	author = "Fuchs, Jurgen and Runkel, Ingo and Schweigert, Christoph",
	title = "{TFT construction of RCFT correlators 1. Partition functions}",
	eprint = "hep-th/0204148",
	archivePrefix = "arXiv",
	reportNumber = "PAR-LPTHE-02-25",
	doi = "10.1016/S0550-3213(02)00744-7",
	journal = "Nucl. Phys. B",
	volume = "646",
	pages = "353--497",
	year = "2002"
}

@article{Gannon:2001ki,
	author = "Gannon, Terry",
	title = "{Boundary conformal field theory and fusion ring representations}",
	eprint = "hep-th/0106105",
	archivePrefix = "arXiv",
	doi = "10.1016/S0550-3213(01)00632-0",
	journal = "Nucl. Phys. B",
	volume = "627",
	pages = "506--564",
	year = "2002"
}

@article{Sousa:2002te,
	author = "Sousa, N. and Schellekens, A. N.",
	title = "{Orientation matters for NIMreps}",
	eprint = "hep-th/0210014",
	archivePrefix = "arXiv",
	reportNumber = "NIKHEF-02-009",
	doi = "10.1016/S0550-3213(02)01124-0",
	journal = "Nucl. Phys. B",
	volume = "653",
	pages = "339--368",
	year = "2003"
}

@article{Quella:2006de,
	author = "Quella, Thomas and Runkel, Ingo and Watts, Gerard M. T.",
	title = "{Reflection and transmission for conformal defects}",
	eprint = "hep-th/0611296",
	archivePrefix = "arXiv",
	reportNumber = "KCL-MTH-06-12, NSF-KITP-06-110",
	doi = "10.1088/1126-6708/2007/04/095",
	journal = "JHEP",
	volume = "04",
	pages = "095",
	year = "2007"
}

@article{Davydov:2011pp,
	author = {Booker, Thomas and Davydov, Alexei},
	title = {{Commutative algebras in Fibonacci categories}},
	eprint = {1103.3537},
	archiveprefix = {arXiv},
	primaryclass = {math.CT},
	year = {2012},
	journal = {J. Algebra},
	volume = {355},
	pages = {176--204},
	doi = {10.1016/j.jalgebra.2011.12.029},
}

@article{hohn2012mckay,
	title = {{McKay's $E_7$ observation on the Baby Monster}},
	author = {H{\"o}hn, Gerald and Lam, Ching Hung and Yamauchi, Hiroshi},
	journal = {Int. Math. Res. Not.},
	volume = {2012},
	number = {1},
	pages = {166--212},
	year = {2012},
	publisher = {OUP},
	doi = {10.1093/imrn/rnr009},
}

@article{Cogburn:2023xzw,
	author = "Cogburn, Cameron V. and Fitzpatrick, Andrew Liam and Geng, Hao",
	title = "{CFT and lattice correlators near an RG domain wall between minimal models}",
	eprint = "2308.00737",
	archivePrefix = "arXiv",
	primaryClass = "hep-th",
	doi = "10.21468/SciPostPhysCore.7.2.021",
	journal = "SciPost Phys. Core",
	volume = "7",
	pages = "021",
	year = "2024"
}

@article{Aasen:2016dop,
	author = "Aasen, David and Mong, Roger S. K. and Fendley, Paul",
	title = "{Topological Defects on the Lattice I: The Ising model}",
	eprint = "1601.07185",
	archivePrefix = "arXiv",
	primaryClass = "cond-mat.stat-mech",
	doi = "10.1088/1751-8113/49/35/354001",
	journal = "J. Phys. A",
	volume = "49",
	number = "35",
	pages = "354001",
	year = "2016"
}

@article{Sinha:2025jhh,
	author = "Sinha, Madhav and Tavares, Thiago Silva and Roy, Ananda and Saleur, Hubert",
	title = "{Integrability and lattice discretizations of all topological defect lines in minimal CFTs}",
	eprint = "2509.04257",
	archivePrefix = "arXiv",
	primaryClass = "hep-th",
	doi = "10.1016/j.nuclphysb.2026.117308",
	journal = "Nucl. Phys. B",
	volume = "1023",
	pages = "117308",
	year = "2026"
}

@article{Cardy:1991tv,
	author = "Cardy, John L. and Lewellen, David C.",
	title = "{Bulk and boundary operators in conformal field theory}",
	reportNumber = "UCSB-TH-90-70, NSF-ITP-91-15",
	doi = "10.1016/0370-2693(91)90828-E",
	journal = "Phys. Lett. B",
	volume = "259",
	pages = "274--278",
	year = "1991"
}

@article{Lewellen:1991tb,
	author = "Lewellen, David C.",
	title = "{Sewing constraints for conformal field theories on surfaces with boundaries}",
	reportNumber = "NSF-ITP-91-32",
	doi = "10.1016/0550-3213(92)90370-Q",
	journal = "Nucl. Phys. B",
	volume = "372",
	pages = "654--682",
	year = "1992"
}

@article{dong1996algebraic,
	title = {{The algebraic structure of relative twisted vertex operators}},
	author = {Dong, Chongying and Lepowsky, James},
	journal = {J. Pure Appl. Algebra},
	volume = {110},
	number = {3},
	pages = {259--295},
	year = {1996},
	publisher = {Elsevier},
	doi = {10.1016/0022-4049(95)00095-X},
}

@article{kitazume2000ternary,
	title = {{Ternary codes and vertex operator algebras}},
	author = {Kitazume, Masaaki and Miyamoto, Masahiko and Yamada, Hiromichi},
	journal = {J. Algebra},
	volume = {223},
	number = {2},
	pages = {379--395},
	year = {2000},
	publisher = {Elsevier},
	doi = {10.1006/jabr.1999.8058},
}

@article{lam2008faithful,
	author = {Lam, Ching-Hung and Sakuma, Shinya},
	title = {{On a class of vertex operator algebras having a faithful $S_{n+1}$-action}},
	journal = {Taiwanese J. Math.},
	volume = {12},
	number = {9},
	pages = {2465--2488},
	year = {2008},
	doi = {10.11650/twjm/1500405190}
}

@book{lepowsky2004introduction,
	title={Introduction to vertex operator algebras and their representations},
	author={Lepowsky, James and Li, Haisheng},
	volume={227},
	year={2004},
	publisher={Springer Science \& Business Media}
}

@article{Kormos:2009sk,
	author = "Kormos, Marton and Runkel, Ingo and Watts, Gerard M. T.",
	title = "{Defect flows in minimal models}",
	eprint = "0907.1497",
	archivePrefix = "arXiv",
	primaryClass = "hep-th",
	reportNumber = "KCL-MTH-09-06",
	doi = "10.1088/1126-6708/2009/11/057",
	journal = "JHEP",
	volume = "11",
	pages = "057",
	year = "2009"
}

@inproceedings{Caldeira:2004jy,
	author = "Caldeira, Alexandre F. and Wheater, John F.",
	title = "{Boundary states and broken bulk symmetries in W A(r) minimal models}",
	booktitle = "{From Fields to Strings: Circumnavigating Theoretical Physics: A Conference in Tribute to Ian Kogan}",
	eprint = "hep-th/0404052",
	archivePrefix = "arXiv",
	reportNumber = "OUTP-04-05P",
	doi = "10.1142/9789812775344_0033",
	pages = "1441--1469",
	month = "4",
	year = "2004"
}

@book{DiFrancesco:1997nk,
	author = "Di Francesco, P. and Mathieu, P. and Senechal, D.",
	title = "{Conformal Field Theory}",
	doi = "10.1007/978-1-4612-2256-9",
	isbn = "978-0-387-94785-3, 978-1-4612-7475-9",
	publisher = "Springer-Verlag",
	address = "New York",
	series = "Graduate Texts in Contemporary Physics",
	year = "1997"
}

@article{Quella:2002ct,
	author = "Quella, Thomas and Schomerus, Volker",
	title = "{Symmetry breaking boundary states and defect lines}",
	eprint = "hep-th/0203161",
	archivePrefix = "arXiv",
	reportNumber = "AEI-2002-024, PAR-LPTHE-02-09",
	doi = "10.1088/1126-6708/2002/06/028",
	journal = "JHEP",
	volume = "06",
	pages = "028",
	year = "2002"
}

@article{Furuta:2025ahl,
	author = "Furuta, Yuma and Kusuki, Yuya and Onagi, Toshiki",
	title = "{Transmission coefficients from phantom currents}",
	eprint = "2511.00356",
	archivePrefix = "arXiv",
	primaryClass = "hep-th",
	reportNumber = "KYUSHU-HET-341, RIKEN-iTHEMS-Report-25",
	doi = "10.1103/fdfb-tkz4",
	journal = "Phys. Rev. D",
	volume = "113",
	number = "4",
	pages = "045008",
	year = "2026"
}

@article{Zamolodchikov:1987ti,
	author = "Zamolodchikov, A. B.",
	title = "{Renormalization Group and Perturbation Theory Near Fixed Points in Two-Dimensional Field Theory}",
	journal = "Sov. J. Nucl. Phys.",
	volume = "46",
	pages = "1090",
	year = "1987"
}

@article{Zamolodchikov:1991vx,
	author = "Zamolodchikov, A. B.",
	title = "{From tricritical Ising to critical Ising by thermodynamic Bethe ansatz}",
	reportNumber = "ENS-LPS-327-1991",
	doi = "10.1016/0550-3213(91)90423-U",
	journal = "Nucl. Phys. B",
	volume = "358",
	pages = "524--546",
	year = "1991"
}

@article{Chang:2018iay,
	author = "Chang, Chi-Ming and Lin, Ying-Hsuan and Shao, Shu-Heng and Wang, Yifan and Yin, Xi",
	title = "{Topological Defect Lines and Renormalization Group Flows in Two Dimensions}",
	eprint = "1802.04445",
	archivePrefix = "arXiv",
	primaryClass = "hep-th",
	reportNumber = "CALT-TH-2017-067, CALT-TH 2017-067, PUPT-2546",
	doi = "10.1007/JHEP01(2019)026",
	journal = "JHEP",
	volume = "01",
	pages = "026",
	year = "2019"
}

@unpublished{Antunes:2026,
	author = "Antunes, Ant{\'o}nio and Chen, Jin and Rong, Junchen",
	title = "{Rational conformal interfaces between the Ising and tricritical Ising models from a tetrality web}",
	note = "to appear on arXiv",
	year = "2026"
}

@article{Fjelstad:2005ua,
	author = "Fjelstad, Jens and Fuchs, Jurgen and Runkel, Ingo and Schweigert, Christoph",
	title = "{TFT construction of RCFT correlators. V. Proof of modular invariance and factorisation}",
	eprint = "hep-th/0503194",
	archivePrefix = "arXiv",
	reportNumber = "AEI-2005-037, LPTHE-05-07",
	journal = "Theor. Appl. Categor.",
	volume = "16",
	pages = "342--433",
	year = "2006"
}

@article{Kong:2009inh,
	author = "Kong, Liang and Runkel, Ingo",
	title = "{Cardy algebras and sewing constraints. I.}",
	eprint = "0807.3356",
	archivePrefix = "arXiv",
	primaryClass = "math.QA",
	reportNumber = "KCL-MTH-08-06",
	doi = "10.1007/s00220-009-0901-6",
	journal = "Commun. Math. Phys.",
	volume = "292",
	pages = "871--912",
	year = "2009"
}

@article{Kong:2013gca,
	author = "Kong, Liang and Li, Qin and Runkel, Ingo",
	title = "{Cardy algebras and sewing constraints, II}",
	eprint = "1310.1875",
	archivePrefix = "arXiv",
	primaryClass = "math.QA",
	doi = "10.1016/j.aim.2014.05.020",
	journal = "Adv. Math.",
	volume = "262",
	pages = "604--681",
	year = "2014"
}

\end{document}